\documentclass[aps,prx,reprint,superscriptaddress,nofootinbib]{revtex4-2}
\usepackage{blindtext}
\usepackage{lipsum}
\usepackage{graphics}
\usepackage{amsmath}
\usepackage{graphicx}
\usepackage{graphics}
\usepackage{amssymb}
\usepackage{verbatim}
\usepackage{physics}
\usepackage{bm}
\usepackage{float}
\usepackage{dsfont}
\usepackage[dvipsnames]{xcolor}
\usepackage{framed}
\definecolor{shadecolor}{RGB}{224,224,224}
\usepackage{amsthm}
\usepackage{amsfonts}
\usepackage{outlines}
\usepackage{enumitem}
\usepackage{booktabs}
\setenumerate[1]{label=\Roman*.}
\setenumerate[2]{label=\Alph*.}
\setenumerate[3]{label=\roman*.}
\setenumerate[4]{label=\alph*.}
\usepackage{mdframed}
\surroundwithmdframed[
  backgroundcolor=shadecolor,
  linecolor=gray!10,
  linewidth=1pt
]{definition}

\theoremstyle{definition}

\newmdtheoremenv[
  backgroundcolor=shadecolor, 
  linecolor=gray!10         
]{theorem}{Theorem}
\usepackage{url}
\usepackage{chngcntr}
\counterwithout{equation}{section}

\makeatletter
\def\l@subsubsection#1#2{}
\makeatother

\usepackage[dvipsnames]{xcolor}
\usepackage[normalem]{ulem}
\usepackage{wasysym} 
\usepackage[export]{adjustbox}

\makeatletter
\DeclareFontFamily{OMX}{MnSymbolE}{}
\DeclareSymbolFont{MnLargeSymbols}{OMX}{MnSymbolE}{m}{n}
\SetSymbolFont{MnLargeSymbols}{bold}{OMX}{MnSymbolE}{b}{n}
\DeclareFontShape{OMX}{MnSymbolE}{m}{n}{
    <-6>  MnSymbolE5
   <6-7>  MnSymbolE6
   <7-8>  MnSymbolE7
   <8-9>  MnSymbolE8
   <9-10> MnSymbolE9
  <10-12> MnSymbolE10
  <12->   MnSymbolE12
}{}
\DeclareFontShape{OMX}{MnSymbolE}{b}{n}{
    <-6>  MnSymbolE-Bold5
   <6-7>  MnSymbolE-Bold6
   <7-8>  MnSymbolE-Bold7
   <8-9>  MnSymbolE-Bold8
   <9-10> MnSymbolE-Bold9
  <10-12> MnSymbolE-Bold10
  <12->   MnSymbolE-Bold12
}{}

\let\llangle\@undefined
\let\rrangle\@undefined
\DeclareMathDelimiter{\llangle}{\mathopen}%
                     {MnLargeSymbols}{'164}{MnLargeSymbols}{'164}
\DeclareMathDelimiter{\rrangle}{\mathclose}%
                     {MnLargeSymbols}{'171}{MnLargeSymbols}{'171}
\makeatother

\usepackage{tikz, ifthen}
\usepackage{tikz-cd}
\usetikzlibrary{arrows}
\usetikzlibrary{snakes}
\usetikzlibrary{intersections}
\usetikzlibrary{shapes.geometric}
\usetikzlibrary{decorations.pathmorphing, patterns,shapes}
\usetikzlibrary{decorations.markings}
\usetikzlibrary{calc}

\tikzset{
	partial ellipse/.style args={#1:#2:#3}{
		insert path={+ (#1:#3) arc (#1:#2:#3)}
	}
}

\tikzset{
	mid arrow/.style={postaction={decorate,decoration={
				markings,
				mark=at position .575 with {\arrow[#1]{stealth}}
	}}},
	near arrow/.style={postaction={decorate,decoration={
				markings,
				mark=at position .275 with {\arrow[#1]{stealth}}
	}}},
	far arrow/.style={postaction={decorate,decoration={
				markings,
				mark=at position .800 with {\arrow[#1]{stealth}}
	}}},
}

\pgfdeclarepatternformonly{south west lines}{\pgfqpoint{-0pt}{-0pt}}{\pgfqpoint{3pt}{3pt}}{\pgfqpoint{3pt}{3pt}}{
	\pgfsetlinewidth{0.4pt}
	\pgfpathmoveto{\pgfqpoint{0pt}{0pt}}
	\pgfpathlineto{\pgfqpoint{3pt}{3pt}}
	\pgfpathmoveto{\pgfqpoint{2.8pt}{-.2pt}}
	\pgfpathlineto{\pgfqpoint{3.2pt}{.2pt}}
	\pgfpathmoveto{\pgfqpoint{-.2pt}{2.8pt}}
	\pgfpathlineto{\pgfqpoint{.2pt}{3.2pt}}
	\pgfusepath{stroke}}

\newcommand{\lw}{0.5}
\newcommand{\qubitsize}{1.5}
\newcommand{\qubitsizedos}{1.2}
\newcommand{\Cdelta}{-0.15}
\newcommand{\aAdeltax}{0.2}
\newcommand{\aAdeltay}{0.1}
\newcommand{\CZlwtwo}{0.65}

\newcommand{\oC}[1]{\textcolor{orange(ryb)}{#1}}
\newcommand{\bB}[1]{\textcolor{dodgerblue}{#1}}
\newcommand{\AR}[1]{\textcolor{red}{#1}}
\newcommand{\aP}[1]{\textcolor{goodpurple}{#1}}
\newcommand{\cupp}{\smallsmile}
\newcommand{\capp}{\smallfrown}
\newcommand{\bgamma}{\boldsymbol{\gamma}}

\definecolor{goodpurple}{HTML}{E771DF}

\colorlet{cz1color}{PineGreen}
\colorlet{cz2color}{LimeGreen}

\definecolor{orange(ryb)}{HTML}{FFA500}
\definecolor{lightorange(ryb)}{HTML}{FFB300}
\definecolor{dodgerblue}{HTML}{1E90FF}
\definecolor{shamrock}{HTML}{28965A}
\definecolor{lightdodgerblue}{HTML}{4dbff7}
\definecolor{crimson}{HTML}{FF4C4C}
\definecolor{pinkerton}{HTML}{EC368D}
\definecolor{forest}{HTML}{6DD189}
\definecolor{lightishgray}{HTML}{DFDFDF}
\definecolor{error-red}{HTML}{EFB2B6}
\definecolor{lightcrimson}{HTML}{FF6F6F}
\definecolor{maroon}{HTML}{800000}
\definecolor{navy}{HTML}{000080}

\usepackage[colorlinks=true,citecolor=dodgerblue,linkcolor=dodgerblue,urlcolor=dodgerblue,pdftitle={Non-Abelian LDPC}]{hyperref}

\newcommand{\exd}{\text{d}}
\newcommand{\BandC}{$\textcolor{dodgerblue}{B}$ and $\textcolor{orange(ryb)}{C}\text{ }$}

\def \beq {\begin{equation}}
\def \eeq {\end{equation}}
\def \beqa {\begin{eqnarray}}
\def \eeqa {\end{eqnarray}}
\def \bseq {\begin{subequations}}
\def \eseq {\end{subequations}}
\newcommand{\phii}{\varphi}

\newcommand \M {\mathcal{M}}

\begin{document}

\title{Non-Abelian Quantum Low-Density Parity Check Codes and Non-Clifford Operations\\ from Gauging Logical Gates via Measurements}

\author{Maine Christos}
\affiliation{Department of Physics and Institute for Quantum Information and Matter,
California Institute of Technology, Pasadena, CA 91125, USA}
\author{Chiu Fan Bowen Lo}
\affiliation{Department of Physics, Harvard University, Cambridge, MA 02138}
\author{Vedika Khemani}
\affiliation{Department of Physics, Stanford University, Stanford, CA 94305, USA}
\author{Rahul Sahay}
\affiliation{Department of Physics, Harvard University, Cambridge, MA 02138}

\begin{abstract}

In this work, we introduce constructions for non-Abelian qLDPC codes obtained by gauging transversal Clifford gates using measurement and feedback.
In particular, we identify two qualitatively different approaches to gauging qLDPC codes to obtain their non-Abelian counterparts.
The first approach applies to codes that exhibit a generalized form of Poincar\'e duality and leads to a qLDPC non-Abelian Clifford stabilizer code, whose stabilizers are reminiscent of the action of a Type-III twisted quantum double.
Our second approach applies to general qLDPC codes, and uses a graph of ancilla qubits which may be tailored to properties of the input codes to gauge a single transversal gate.
For both constructions, the resulting gauged codes are shown to have properties analogous to 2D non-Abelian topological order---e.g. the analog of a single anyon on a torus. 
We conclude by demonstrating that our gauging procedures enable magic state preparation via the measurement of logical Clifford gates.
Consequently, our gauging constructions offer a protocol for performing non-Clifford operations on any qLDPC code.

\end{abstract}

\maketitle

\section{Introduction}

The physical understanding of both Abelian and non-Abelian topologically ordered phases of matter \cite{wen_topological_1990,RevModPhys.89.041004,Bravyi_2010} has been a driving force of progress in quantum information science for decades.
Indeed, Abelian topological codes such as the toric code have been invaluable for our conceptual understanding of quantum error correction and have paved the way towards the first experimental demonstrations of error corrected quantum computation~\cite{bluvstein_fault-tolerant_2026,bluvstein_logical_2024,google2025quantum}.
Beyond this, the physical properties of these Abelian phases—e.g. their anyonic excitations, exchange and braiding statistics, automorphisms, defects—have inspired numerous routes towards fault-tolerant quantum computation \cite{koenig_quantum_2010,yoshida_topological_2015,breuckmann_hyperbolic_2017,bridgeman_tensor_2017,kesselring_boundaries_2018,lavasani_universal_2019,zhu_instantaneous_2020,kesselring_anyon_2024,davydova_quantum_2024}.
In a complementary vein, non-Abelian phases underlie the entire paradigm of topological quantum computation \cite{Kitaev_2003,freedman_modular_2002,RevModPhys.80.1083,Bonderson_2008,levaillant_universal_2015,cui_universal_2015,beverland_protected_2016,cong_universal_2017,chen_universal_2025,lo_universal_2025} and recently have even provided a means to facilitate non-Clifford logical operations on Abelian codes \cite{bravyi_universal_2006,laubscher_universal_2019,margaritaknots,sajithcodes,kobayashi_clifford_2025,huang_hybrid_2025,warman_transversal_2025, hsin2025automorphismgaugetheorieshigher, hsin_non-abelian_2025}. 

While these topological codes are especially well suited to experimental platforms with local entangling operations, recent technological breakthroughs in engineering non-local connectivity \cite{Periwal_2021, bluvstein_logical_2024,chiu_continuous_2025,sinclair_fault-tolerant_2025,bluvstein_fault-tolerant_2026,moses_race-track_2023,ransford_helios_2025} have motivated the study of more general quantum low-density parity check (qLDPC) codes \cite{MacKay_2004,Tillich_2014,Breuckmann_2021_rev,panteleev_asymptotically_2022,Panteleev_2021,Breuckmann_2021,Hastings_2021,Leverrier_2015,leverrier2022quantumtannercodes,SipserSpiel,Bravyi_2024,Kovalev_2013}. 
This class of codes includes topological codes and is characterized by commuting Hamiltonians whose stabilizers involve a finite number of qubits and whose qubits participate in a finite number of stabilizers.
However, unlike the toric code, these codes generically have no finite-dimensional Euclidean embedding.
Consequently, they can exhibit properties that cannot exist in their finite dimensional counterparts. 
This is exemplified by so-called ``good'' qLDPC codes, which host an extensive number of logical qubits while maintaining an extensive code distance \cite{panteleev_asymptotically_2022,dinur_good_2023,leverrier2022quantumtannercodes,Breuckmann_2021}.
These properties make qLDPC codes ideal for the efficient and robust storage of quantum information.

Despite the potential that qLDPC codes hold as quantum memories, utilizing them for quantum computation---i.e. determining how one fault tolerantly manipulates their stored information---remains a persistent challenge.
This challenge has been met with a plethora of interesting theoretical approaches in recent years, with a wide variety of ``gadgets'' having been developed to perform both Clifford and non-Clifford operations in both precisely tailored and general classes of qLDPC codes \cite{xu2023constantoverheadfaulttolerantquantumcomputation,xu2024fastparallelizablelogicalcomputation,xu2025batchedhighratelogicaloperations,zheng2025highratesurgeryconstantoverheadlogical,berthusen2025automorphismgadgetshomologicalproduct,cross2025improvedqldpcsurgerylogical,swaroop2025universaladaptersquantumldpc,he2025extractorsqldpcarchitecturesefficient}.
Nevertheless, a systematic understanding of the manipulation of information in this setting is lacking.

These challenges are exacerbated by our limited understanding of the physics of qLDPC codes.
Several pioneering works have made progress in this direction by establishing these codes as gapped quantum phases of matter \cite{rakovszky2023physicsgoodldpccodes,rakovszky2024physicsgoodldpccodes,GappedLDPC,yin_low-density_2025,Tan_2025}, recasting their properties in the language of gauge theory, and discovering new phenomena unique to these codes, e.g. the ``no low-energy trivial states'' (NLTS) phenomena \cite{Anshu_2023} and topological spin glasses \cite{placke_topological_2024}.
However, basic questions such as \textit{What is the landscape of qLDPC quantum phases?}, \textit{What is the nature of excitations above these codes?}, \textit{Is it possible to define a notion of exchange and braiding statistics in this context?} among others, remain unanswered.\footnote{Answering some of these questions will be the subject of forthcoming work \cite{homologicalorder}.}

Our work advances both the physical understanding of qLDPC codes and the ability to manipulate their logical information by developing constructions of non-Abelian qLDPC codes and showing that these constructions yield a systematic procedure for performing non-Clifford logical operations on any qLDPC code.
Crucially, our non-Abelian codes expand the landscape of known qLDPC phases of matter while simultaneously providing new avenues towards universal quantum computation.
Concretely, our key results are three fold.

First, we demonstrate how gauging transversal Clifford logical gates between copies of Abelian qLDPC codes can be used to prepare non-Abelian qLDPC codes. 
This generalizes the pioneering measurement-based gauging protocol of Tantivasadakarn et al. (reviewed in Sec.~\ref{sec:prelim}) originally developed to prepare 2D non-Abelian order~\cite{Hierarchy}, e.g. by gauging a transversal $\mathsf{CZ}$ gate between two copies of the toric code.
In doing so, we find two qualitatively distinct approaches to gauging---which reduce to one another in the 2D case---naturally appear in the qLDPC setting (discussed in Section~\ref{sec:gauging-theory}). 
The first approach, dubbed ``homological gauging'', applies to quantum codes which have an algebraic structure analogous to Poincar\'e duality present in the homology of manifolds.
In this approach, the structure of \textit{global redundancies} of the code's stabilizers is tied to the structure of its logical Clifford gates. 
As we will see, we can gauge these gates to prepare a non-Abelian qLDPC code, whose stabilizers are analogous to a type-III twisted quantum double.

In contrast, the second approach, which we term ``graph gauging", applies to general qLDPC codes.
In this approach, we introducing a graph of ancilla qubits---whose structure can be tailored based on desired properties of the resulting code---to gauge a global symmetry of an input qLDPC code. 
Importantly, this construction does not rely on an assumptions of any additional structure of the input codes and consequently can be applied to any qLDPC code.
We provide explicit lattice examples of codes achieved from both forms of gauging in Sec.~\ref{sec:gauging-examples}.
Here, we highlight novel phenomena that can occur in the qLDPC context, including hybrid Abelian non-Abelian codes---where upon gauging an \textit{addressable} logical gate\footnote{By addressable, we mean that the gate acts non-trivially on an $\mathcal{O}(1)$ number of logical qubits.
This is especially notable in qLDPC codes where the number of logical qubits can scale with the number of physical qubits.
We remark that the systematic construction of this addressable gate is the subject of a forthcoming work~\cite{addressablegates}.}, one part of a qLDPC code takes on a non-Abelian character while the rest of the code remains Abelian.

As our second main result, in Sec.~\ref{sec:non-Abelian-properties} we determine general physical properties exhibited by our non-Abelian qLDPC codes.
In particular, we derive the structure of their ground states and logical operators.
Strikingly, we demonstrate that the notion of non-Abelian braiding persists even in the qLDPC context, despite the absence of an underlying spatial manifold. In particular, we show that repeated action of logical operators can trap the qLDPC analog of a ``single anyon on a torus,'' a hallmark of non-Abelian topological order~\cite{bombin_family_2008,Iqbal_2024}.

Third, we demonstrate in Section~\ref{sec:non-Abelian-computation} that the gauging procedures underlying our constructions enable preparing magic states in any qLDPC code---a key component of universal quantum computation.
Our concrete protocol enables the measurement of operators associated with transversal gates. 
When the transversal gates of a qLDPC code are Clifford, measuring them can be shown to be a non-Clifford operation---e.g. measuring the $+1$ eigenvalue of $\mathsf{CZ}$ on the state $\ket{++}$ yields the magic state $\frac{1}{\sqrt{3}} (\ket{00} + \ket{10} + \ket{01})$.
This generalizes the work of Williamson and Yoder \cite{williamson_tanner}, which showed that a gauging construction can be used to measure logical Pauli operators, to the case of general Clifford gates.

Overall, our work lies at the intersection of new frontiers in both condensed matter physics and quantum information science.
Seen in the light of condensed matter, it proposes an example of a quantum phase of matter which can be realized in settings where usual notions of geometric locality in Euclidean space are lost.
Viewed through the lens of quantum information, our construction offers a protocol for implementing non-Clifford operations on quantum codes with arbitrary, sparse connectivity.

\tableofcontents

\section{Preliminaries: Non-Abelian Topological Order in 2D}\label{sec:prelim}

Before introducing our constructions for non-Abelian qLDPC codes, we review the gauging procedure for preparing the $\mathcal{D}(D_4)$ topological order introduced in Refs.~\cite{verresen_efficiently_2022, Tantivasadakarn_2024_LRE, tantivasadakarn_shortest_2023, Hierarchy}.
When we discuss non-Abelian qLDPC codes, we will present explicit constructions that follow the same sequence of steps as this two dimensional gauging procedure.

As a concrete example, we consider a square lattice model for a $\mathcal{D}(D_4)$ topological code in two dimensions.
The construction begins with two copies of the square lattice toric code, which we label $\textcolor{dodgerblue}{B}$ and $\textcolor{orange(ryb)}{C}$. Their stabilizer group is generated by: 
\begin{equation} \label{eq-TC}
    \begin{tikzpicture}[scale = 0.45, baseline = {([yshift=-.5ex]current bounding box.center)}]
    \draw[lightdodgerblue] (1.5,0) -- (-1.5, 0);
    \draw[lightdodgerblue] (0,1.5) -- (0,-1.5);
    \draw[lightorange(ryb)] (1.5 + 2*\Cdelta,0 + 2*\Cdelta) -- (-1.5 + 2*\Cdelta, 0 + 2*\Cdelta);
    \draw[lightorange(ryb)] (0 + 2*\Cdelta,1.5 + 2*\Cdelta) -- (0 + 2*\Cdelta,-1.5 + \Cdelta);
    \node at (0.75, 0) {\normalsize $\bB{X}$};
    \node at (-0.75, 0) {\normalsize $\bB{X}$};
    \node at (0, 0.75) {\normalsize $\bB{X}$};
    \node at (0, -0.75) {\normalsize $\bB{X}$};
\end{tikzpicture},\quad  \begin{tikzpicture}[scale = 0.4, baseline={([yshift=-.5ex]current bounding box.center)}]
\draw[lightdodgerblue] (-1, -1) -- (-1, 1) -- (1, 1) -- (1, -1) -- cycle;
\draw[lightorange(ryb)] (-1 + 2*\Cdelta, -1 + 2*\Cdelta) -- (-1 + 2*\Cdelta, 1 + 2*\Cdelta) -- (1 + 2*\Cdelta, 1 + 2*\Cdelta) -- (1 + 2*\Cdelta, -1 + 2*\Cdelta) -- cycle;
\node at (0.0, -1) {\normalsize $\bB{Z}$};
\node at (0.0, 1) {\normalsize $\bB{Z}$};
\node at (-1, 0.0) {\normalsize $\bB{Z}$};
\node at (1, 0.0) {\normalsize $\bB{Z}$};
\end{tikzpicture}, \quad \begin{tikzpicture}[scale = 0.45, baseline = {([yshift=-.5ex]current bounding box.center)}]
    \draw[lightdodgerblue] (1.5,0) -- (-1.5, 0);
    \draw[lightdodgerblue] (0,1.5) -- (0,-1.5);
    \draw[lightorange(ryb)] (1.5 + 2*\Cdelta,0 + 2*\Cdelta) -- (-1.5 + 2*\Cdelta, 0 + 2*\Cdelta);
    \draw[lightorange(ryb)] (0 + 2*\Cdelta,1.5 + 2*\Cdelta) -- (0 + 2*\Cdelta,-1.5 + \Cdelta);
    \node at (0.75 + 2*\Cdelta, 0 + 2*\Cdelta) {\normalsize $\oC{X}$};
    \node at (-0.75 + 2*\Cdelta, 0 + 2*\Cdelta) {\normalsize $\oC{X}$};
    \node at (0 + 2*\Cdelta, 0.75 + 2*\Cdelta) {\normalsize $\oC{X}$};
    \node at (0 + 2*\Cdelta, -0.75 + 2*\Cdelta) {\normalsize $\oC{X}$};
\end{tikzpicture}, \quad
\begin{tikzpicture}[scale = 0.4, baseline={([yshift=-.5ex]current bounding box.center)}]
\draw[lightdodgerblue] (-1, -1) -- (-1, 1) -- (1, 1) -- (1, -1) -- cycle;
\draw[lightorange(ryb)] (-1 + 2*\Cdelta, -1 + 2*\Cdelta) -- (-1 + 2*\Cdelta, 1 + 2*\Cdelta) -- (1 + 2*\Cdelta, 1 + 2*\Cdelta) -- (1 + 2*\Cdelta, -1 + 2*\Cdelta) -- cycle;
\node at (0.0 + 2*\Cdelta, -1 + 2*\Cdelta) {\normalsize $\oC{Z}$};
\node at (0.0 + 2*\Cdelta, 1 + 2*\Cdelta) {\normalsize $\oC{Z}$};
\node at (-1 + 2*\Cdelta, 0.0 + 2*\Cdelta) {\normalsize $\oC{Z}$};
\node at (1 + 2*\Cdelta, 0.0 + 2*\Cdelta) {\normalsize $\oC{Z}$};
\end{tikzpicture}.
\end{equation}
A $\mathcal{D}(D_4)$ topological order can be realized by gauging the following global $\mathbb{Z}_2$ ``$\mathsf{CZ}$'' operation which acts as:
\begin{equation}\label{eq:CZ}
    U_{\mathsf{CZ}}= \begin{tikzpicture}[scale = 0.9, baseline={([yshift=-.5ex]current bounding box.center)}]
    \foreach \i in {0, 1, 2, 3}{        
        \draw[color = lightorange(ryb), line width = \lw pt] (-0.5 + \Cdelta, \i + \Cdelta) -- (3.5 + \Cdelta, \i + \Cdelta);
        \draw[color = lightorange(ryb), line width =\lw  pt] (\i + \Cdelta, -0.5 + \Cdelta) -- (\i + \Cdelta, 3.5 + \Cdelta);
        \draw[color = lightdodgerblue, line width =\lw  pt] (-0.5, \i) -- (3.5, \i);
        \draw[color = lightdodgerblue, line width =\lw  pt] (\i, -0.5) -- (\i, 3.5);
    }
    \foreach \i in {0, 1, 2, 3} {
        \foreach \j in {0, 1, 2, 3}{
            \ifthenelse{\i < 4}{\ifthenelse{\j<3}{\draw[color = lightgray, line width = 1 pt] (\i, \j + 0.5) -- (\i + 0.5 + \Cdelta, \j + 1 + \Cdelta)}{}}{};
            \ifthenelse{\i < 3}{\ifthenelse{\j<4}{\draw[color = black, line width = 1 pt] (\i+ 0.5, \j) -- (\i + 1 + \Cdelta, \j + 0.5 + \Cdelta)}{}}{};

            \ifthenelse{\i < 3}{
            \filldraw[color = dodgerblue] (\i + 0.5, \j) circle (\qubitsize pt); 
            \filldraw[color = orange(ryb)] (\i + 0.5 + \Cdelta, \j + \Cdelta) circle (\qubitsize pt); 
            }{}
            \ifthenelse{\j < 3}{
            \filldraw[color = dodgerblue] (\i, \j + 0.5) circle (\qubitsize pt); 
            \filldraw[color = orange(ryb)] (\i + \Cdelta, \j + \Cdelta + 0.5) circle (\qubitsize pt); 
            }{}
            
        };
    };
    \end{tikzpicture}
\end{equation}
where the black (gray) lines are $\mathsf{CZ}$ gates which pair a horizontal (vertical) edge of the toric code on the blue sublattice with an adjacent vertical (horizontal) edge on the orange sublattice.
The action of $U_{\mathsf{CZ}}$ as defined above will map $X$-stabilizers of the $\bB{B}$ code to a products of themselves with other $Z$-stabilizers of the $\textcolor{orange(ryb)}{C}$ code (and vice versa).
Consequently, $U_{\mathsf{CZ}}$ will not leave the toric code Hamiltonian invariant, rather, it can be thought of as an emergent symmetry of the ground state subspace.

\subsection{Non-Abelian Anyons from Gauging}
The statement that gauging this $\mathsf{CZ}$ symmetry yields a $\mathcal{D}(D_4)$ topological order follows from the fact that $U_{\mathsf{CZ}}$ (taken as an operation on the ground state subspace) implements a logical $\mathsf{CZ}$ gate between the logical qubits of the $\bB{B}$ and $\oC{C}$ toric codes.
Under this interpretation of $U_{\mathsf{CZ}}$, a ``patch'' operator of the $\mathsf{CZ}$ symmetry will act non-trivially on the toric code anyons. When an anyon moves through a patch, it will undergo an anyon automorphism, e.g.
\begin{equation}
    \includegraphics[valign = c]{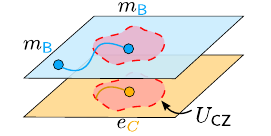}
\end{equation}
Specifically, the $m_{\bB{\mathsf{B}}}$ anyon will be transmuted to the $m_{\bB{\mathsf{B}}}e_{\oC{\mathsf{C}}}$ anyon and the $m_{\oC{\mathsf{C}}}$ anyon will be transmuted to  the $m_{\oC{\mathsf{C}}}e_{\bB{\mathsf{B}}}$ anyon.
To gauge this $\mathsf{CZ}$ symmetry, one must introduce ancillary gauge fields living on a distinct ``$\textcolor{red}{A}$'' sublattice.
Upon doing so, symmetry patch operators of the above form are encoded in loops of $\textcolor{red}{A}$ gauge fields along the symmetry patch boundary: when anyons cross these gauge fields, they are acted on by the corresponding anyon automorphism [Eq.~\eqref{eq:anyonpermute}, left].
Moreover, these $\AR{A}$ loop operators can be ``opened up'' into strings with new anyon excitations at their endpoints.
The process of braiding an $m_{\textcolor{dodgerblue}{B}}$ or $m_{\textcolor{orange(ryb)}{C}}$ anyon with an $m_{\textcolor{red}{A}}$ anyon implements a symmetry automorphism on the $m_{\textcolor{dodgerblue}{B}}$ and $m_{\textcolor{orange(ryb)}{C}}$ anyons of the original toric code, leading to non-Abelian braiding statistics [Eq.~\eqref{eq:anyonpermute}, right].

\begin{equation}\label{eq:anyonpermute}
    \includegraphics[valign = c]{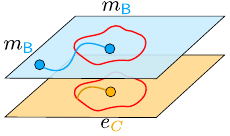} \to \includegraphics[valign = c]{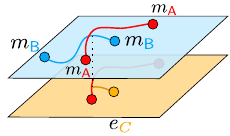}
\end{equation}
This intertwining of the symmetry action on the original codes with additional gauge fields and their anyons is the origin of the non-Abelian character of the resulting state.
After gauging, the $e$ and $m$ anyons related by the $\mathsf{CZ}$ symmetry of the original codes become internal states of a new non-Abelian anyon charged under the $\mathsf{CZ}$ gauge symmetry~\cite{barkeshli2019}.

Perhaps this non-Abelian character is made most apparent by winding two pairs of anyons with non-trivial braiding statistics around the two distinct cycles of the torus.
This will result in a state with a single unpaired anyon at the site where the second non-Abelian pair fuses:
\begin{equation}
    \includegraphics[valign = c]{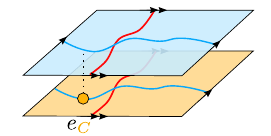}
\end{equation}
This phenomena is a consequence of the non-trivial braiding and fusion structure of the non-Abelian topological order and has even been observed experimentally in Ref.~\cite{Iqbal_2024}. 
Interestingly, we will see a similar phenomena in our discussion of non-Abelian qLDPC codes, even in the absence of an underlying closed manifold.

As a final remark, we note that this intuition extends beyond the example of gauging the $\mathsf{CZ}$ symmetry considered here. In general, the question of which non-Abelian topological order results from gauging a given symmetry of an Abelian topological order is determined by the symmetry action on the anyons of the original Abelian theory. This classification can be made precise in the language of group extensions \cite{barkeshli2019}. The particular example of $\mathcal{D}(D_4)$ topological order we consider here corresponds to an extension of $\mathbb{Z}_2\times\mathbb{Z}_2$ (the gauge group of the input Abelian codes \BandC) by $\mathbb{Z}_2$ (our global symmetry group) \cite{Hierarchy}, mirroring how $D_4$, as a group, can be deconstructed as a semidirect product of $\mathbb{Z}_2\times \mathbb{Z}_2$ with $\mathbb{Z}_2$.

\subsection{Gauging on the Lattice}\label{sec:lattice2D}
Having developed this physical picture of tying symmetry action to gauge fields, we now put this intuition into practice.
Here, we work through the case of gauging the $\mathsf{CZ}$ symmetry of Eq.~\eqref{eq:CZ}, which will result in the $\mathcal{D}(D_4)$ topological order on the square lattice.
First, we introduce two sets of ancilla qubits labeled $\aP{a}$ and $\AR{A}$ initialized in the $\ket{+}$ state, which we place on the vertices and edges of the square lattice respectively (i.e. the two sets of ancillas form a Lieb lattice).
Next, we entangle these ancillas into a $2$D cluster state by performing $\mathsf{CZ}$ gates between neighboring $\aP{a}$ and $\AR{A}$ qubits followed by Hadamard gates on the $\AR{A}$ qubits. The resulting stabilizers are given by: 
\begin{equation} \label{eq-clusterstate}
     \begin{tikzpicture}[scale = 0.45, baseline = {([yshift=-.5ex]current bounding box.center)}]
    \draw[lightcrimson] (1.5,0) -- (-1.5, 0);
    \draw[lightcrimson] (0,1.5) -- (0,-1.5);
    \node at (0.75, 0) {\normalsize $\AR{X}$};
    \node at (-0.75, 0) {\normalsize $\AR{X}$};
    \node at (0, 0.75) {\normalsize $\AR{X}$};
    \node at (0, -0.75) {\normalsize $\AR{X}$};
    \node at (0,0) {\normalsize $\aP{X}$};
\end{tikzpicture}, \quad \begin{tikzpicture}[scale = 0.45, baseline = {([yshift=-.5ex]current bounding box.center)}]
    \draw[lightcrimson] (0,0) -- (1.5, 0);
    \node at (0.75, 0) {\normalsize $\AR{Z}$};
    \node at (1.5, 0) {\normalsize $\aP{Z}$};
    \node at (0,0) {\normalsize $\aP{Z}$};
\end{tikzpicture}, \quad \begin{tikzpicture}[scale = 0.45, baseline = {([yshift=-.5ex]current bounding box.center)}]
    \draw[lightcrimson] (0,0) -- (0, 1.5);
    \node at (0, 0.75) {\normalsize $\AR{Z}$};
    \node at (0, 1.5) {\normalsize $\aP{Z}$};
    \node at (0,0) {\normalsize $\aP{Z}$};
\end{tikzpicture}.
\end{equation}
Note that these stabilizers differ from the usual stabilizers of the cluster state by Hadamard gates acting on the $\AR{A}$ qubits.

We recall the $2$D cluster state on the Lieb lattice is a $\aP{\mathbb{Z}_2^{(0)}} \times \AR{\mathbb{Z}_2^{(1)}}$ symmetry-protected topological (SPT) phase, invariant under a $\aP{\mathbb{Z}^{(0)}_2}$ 0-form symmetry generated by $G_{\aP{a}} = \prod_{v}\aP{ X_{v}}$ and a $\AR{\mathbb{Z}^{(1)}_2}$ $1$-form symmetry $G_{\AR{A}} = \prod_{e \in \gamma} \AR{Z_{e}}$ where $\gamma$ is any closed loop supported on the edges $e$ in the $\AR{A}$ sublattice.
This SPT property means that patches of the global $\mathbb{Z}_2$ symmetry $G_{\aP{a}}$ in a region $\mathcal{R}$ are tied to loop operators on its boundary $\prod_{e \in \partial \mathcal{R}}\AR{X_e}$, as can be seen from the $X$ stabilizer above. These operators are charged (locally anti-commute) under the $\AR{\mathbb{Z}_2^{(1)}}$ symmetry.
As a consequence, the map above can be used to implement the $\mathbb{Z}_2$ gauging map, as first pointed out in Ref.~\cite{Tantivasadakarn_2024_LRE}.

In particular, given a state $\ket{\psi}$ invariant under a global $\mathbb{Z}_2$ symmetry generated by a unitary operator $U$, the general procedure for gauging this state~\cite{Hierarchy} consists of two key steps.
First, we tie the $\mathbb{Z}_2$ symmetry of the state to the $\aP{\mathbb{Z}_2^{(0)}}$ $0$-form symmetry of the cluster state via a ``symmetry-enrichment'' unitary $\Omega$, which satisfies $\Omega G_{\aP{a}} \Omega^{\dagger} = G_{\aP{a}} U$. In what follows, we will see that any $\Omega$ satisfying this takes on the general form of a ``controlled symmetry" gate for the symmetry we wish to gauge.
This symmetry enrichment operator ensures that patches of the combined $G_{\aP{a}} U$ symmetry are equivalent to $\AR{X}$ loop operators at the patch boundaries, as follows from the SPT property of the cluster state on the ancillas.

After the symmetry enrichment, we project the qubits on the $\aP{a}$ sublattice into the $\ket{+}$ state, thereby producing the gauged wavefunction on the remaining sublattices.
As an example, if state $\ket{\psi} = \ket{+}^{\otimes N}$, then the symmetry enrichment map is simply the $\mathsf{CNOT}$ gate between $\ket{\psi}$ and the $\aP{a}$ sublattice; after we project the qubits on the $\aP{a}$ sublattice into the $\ket{+}$ state, Eq.~\eqref{eq-clusterstate} ensures that we prepare the toric code as expected.

To apply this procedure to our case where $U=U_{\mathsf{CZ}}$, we take the following pattern of $\mathsf{CCZ}$ gates as our symmetry enrichment unitary:
\begin{equation}\label{eq:TC-sym-enrich}
    \Omega_{\aP{a}\textcolor{dodgerblue}{B}\textcolor{orange(ryb)}{C}} = \begin{tikzpicture}[scale = 1.2, baseline={([yshift=-.5ex]current bounding box.center)}]
    \foreach \i in {0, 1, 2}{ 
        \draw[color = lightcrimson, line width = \lw pt] (-0.5 +\aAdeltax, \i + \aAdeltay) -- (2.5 +\aAdeltax, \i + \aAdeltay);
        \draw[color = lightcrimson, line width = \lw pt] (\i +\aAdeltax, -0.5 + \aAdeltay) -- (\i +\aAdeltax, 2.5 + \aAdeltay);
        \draw[color = lightorange(ryb), line width = \lw pt] (-0.5 + \Cdelta, \i + \Cdelta) -- (2.5 + \Cdelta, \i + \Cdelta);
        \draw[color = lightorange(ryb), line width = \lw pt] (\i + \Cdelta, -0.5 + \Cdelta) -- (\i + \Cdelta, 2.5 + \Cdelta);
        \draw[color = lightdodgerblue, line width = \lw pt] (-0.5, \i) -- (2.5, \i);
        \draw[color = lightdodgerblue, line width = \lw pt] (\i, -0.5) -- (\i, 2.5);
    }
    \foreach \i in {0, 1, 2} {
        \foreach \j in {0, 1, 2}{
            \ifthenelse{\i < 3}{\ifthenelse{\j<3}{\draw[color = black, line width = 0.8 pt]  (\i +\aAdeltax, \j + \aAdeltay) --(\i, \j + 0.5) -- (\i + 0.5 + \Cdelta, \j + 1 + \Cdelta)}{}}{};
            \ifthenelse{\i < 3}{\ifthenelse{\j<3}{\draw[color = lightgray, line width = 0.8 pt] (\i +\aAdeltax, \j + \aAdeltay) -- (\i+ 0.5, \j) -- (\i + 1 + \Cdelta, \j + 0.5 + \Cdelta)}{}}{};

            \ifthenelse{\i < 2}{
            \filldraw[color = dodgerblue] (\i + 0.5, \j) circle (\qubitsizedos pt); 
            \filldraw[color = orange(ryb)] (\i + 0.5 + \Cdelta, \j + \Cdelta) circle (\qubitsizedos pt); 
            \filldraw[color = crimson] (\i+ 0.5 +\aAdeltax, \j + \aAdeltay) circle (\qubitsizedos pt);
            }{}
            \ifthenelse{\j < 2}{
            \filldraw[color = dodgerblue] (\i, \j + 0.5) circle (\qubitsizedos pt); 
            \filldraw[color = orange(ryb)] (\i + \Cdelta, \j + \Cdelta + 0.5) circle (\qubitsizedos pt); 
            \filldraw[color = crimson] (\i+\aAdeltax, \j + \aAdeltay + 0.5 ) circle (\qubitsizedos pt);
            }{}
            \filldraw[color =goodpurple] (\i +\aAdeltax, \j + \aAdeltay) circle (\qubitsizedos pt);
            
        };
    };
    \end{tikzpicture}
\end{equation}
which has the characteristic property that $\Omega G_{\aP{a}} \Omega^{\dagger} = G_{\aP{a}} U_{\mathsf{CZ}}$.
After applying $\Omega_{\aP{a}\bB{B}\oC{C}}$, the vertex stabilizers of the code are updated as follows: 
\begin{equation} \label{eq-OmegaenrichedX}
    \begin{tikzpicture}[scale = 0.75, baseline = {([yshift=-.5ex]current bounding box.center)}]
    \draw[lightdodgerblue, line width = 0.65 pt] (1.,0) -- (-1., 0);
    \draw[lightdodgerblue, line width = 0.65 pt] (0,1.5) -- (0,-1.);
    \draw[lightdodgerblue, line width = 0.65 pt] (0, 0) -- (1.5, 0) -- (1.5, 1.5) -- (0, 1.5);
    \draw[lightorange(ryb), line width = 0.65 pt] (1.5 + 1.5*\Cdelta,0 + 1.5*\Cdelta) -- (-1. + 1.5*\Cdelta, 0 + 1.5*\Cdelta);
    \draw[lightorange(ryb), line width = 0.65 pt] (0 + 1.5*\Cdelta,1.5 + 1.5*\Cdelta) -- (0 + 1.5*\Cdelta,-1. + \Cdelta);
    \draw[lightorange(ryb), line width = 0.65 pt] (0 + 1.5*\Cdelta, 0 + 1.5*\Cdelta) -- (1.5 + 1.5*\Cdelta, 0 + 1.5*\Cdelta) -- (1.5 + 1.5*\Cdelta, 1.5 + 1.5*\Cdelta) -- (0 + 1.5*\Cdelta, 1.5 + 1.5*\Cdelta);
    \draw[lightcrimson, line width = 0.65 pt] (1.5 + 1.2*\aAdeltax,0 + 2.5*\aAdeltay) -- (-1. + 1.2*\aAdeltax, 0 + 2.5*\aAdeltay);
    \draw[lightcrimson, line width = 0.65 pt] (0 + 1.2*\aAdeltax,1.5 + 2.5*\aAdeltay) -- (0 + 1.2*\aAdeltax,-1. + \aAdeltay);
    \draw[lightcrimson, line width = 0.65 pt] (1.5 + 1.2*\aAdeltax,1.5 + 2.5*\aAdeltay) -- (1.5 + 1.2*\aAdeltax,0 + 2.5*\aAdeltay);
    \draw[lightcrimson, line width = 0.65 pt] (0 + 1.2*\aAdeltax,1.5 + 2.5*\aAdeltay) -- (1.5 + 1.2*\aAdeltax,1.5 + 2.5*\aAdeltay);
    
    \node at (0.5 + 1.2*\aAdeltax, 2.5*\aAdeltay) {\small $\AR{X}$};
    \node at (-0.5 + 1.2*\aAdeltax, 2.5*\aAdeltay) {\small $\AR{X}$};
    \node at (1.2*\aAdeltax, 0.5 + 2.5*\aAdeltay) {\small $\AR{X}$};
    \node at (1.2*\aAdeltax, -0.5 + 2.5*\aAdeltay) {\small $\AR{X}$};
    \node at (1.2*\aAdeltax, 2.5*\aAdeltay) {\small $\aP{X}$};
    \draw[color = black, line width = 0.75 pt](0, 0.8) -- (0.75 + 1.5*\Cdelta, 1.5 + 1.5*\Cdelta);
    \filldraw[color = dodgerblue] (0, 0.8) circle (2 pt);
    \filldraw[color = orange(ryb)] ((0.75 + 1.5*\Cdelta, 1.5 + 1.5*\Cdelta) circle (2 pt);
    \draw[color = lightgray, line width = 0.75 pt](0.8, 0) -- (1.5 + 1.5*\Cdelta, 0.75 + 1.5*\Cdelta);
    \filldraw[color = dodgerblue] (0.8, 0) circle (2 pt);
    \filldraw[color = orange(ryb)] (1.5 + 1.5*\Cdelta, 0.75 + 1.5*\Cdelta) circle (2 pt);
    \end{tikzpicture}
    \qquad 
    \begin{tikzpicture}[scale = 0.75, baseline = {([yshift=-.5ex]current bounding box.center)}]
    \draw[lightdodgerblue, line width = 0.65 pt] (1.,0) -- (-1., 0);
    \draw[lightdodgerblue, line width = 0.65 pt] (0,1.5) -- (0,-1.);
    \draw[lightdodgerblue, line width = 0.65 pt] (0, 0) -- (1.5, 0) -- (1.5, 1.5) -- (0, 1.5);
    \draw[lightorange(ryb), line width = 0.65 pt] (1.5 + 1.5*\Cdelta,0 + 1.5*\Cdelta) -- (-1. + 1.5*\Cdelta, 0 + 1.5*\Cdelta);
    \draw[lightorange(ryb), line width = 0.65 pt] (0 + 1.5*\Cdelta,1.5 + 1.5*\Cdelta) -- (0 + 1.5*\Cdelta,-1. + \Cdelta);
    \draw[lightorange(ryb), line width = 0.65 pt] (0 + 1.5*\Cdelta, 0 + 1.5*\Cdelta) -- (1.5 + 1.5*\Cdelta, 0 + 1.5*\Cdelta) -- (1.5 + 1.5*\Cdelta, 1.5 + 1.5*\Cdelta) -- (0 + 1.5*\Cdelta, 1.5 + 1.5*\Cdelta);
    \draw[lightcrimson, line width = 0.65 pt] (1.5 + 1.2*\aAdeltax,0 + 2.5*\aAdeltay) -- (-1.5 + 1.2*\aAdeltax, 0 + 2.5*\aAdeltay);
    \draw[lightcrimson, line width = 0.65 pt] (0 + 1.2*\aAdeltax,1.5 + 2.5*\aAdeltay) -- (0 + 1.2*\aAdeltax,-1.5 + 2.5*\aAdeltay);
    \node at (0.5 + 0, 0) {\small $\bB{X}$};
    \node at (-0.5 + 0, 0) {\small $\bB{X}$};
    \node at (0, 0.5 + 0) {\small $\bB{X}$};
    \node at (0, -0.5 + 0) {\small $\bB{X}$};
    \draw[color = black, line width = 0.75 pt](0 + 1.2*\aAdeltax, 2.5 *\aAdeltay) -- (0.75 + 1.5*\Cdelta, 1.5 + 1.5*\Cdelta);
    \draw[color = lightgray, line width = 0.75 pt](-1.5 + 1.2*\aAdeltax, 2.5 *\aAdeltay) -- (1.5*\Cdelta, 0.75 + 1.5*\Cdelta);
    \draw[color = black, line width = 0.75 pt](0 + 1.2*\aAdeltax, -1.5 + 2.5 *\aAdeltay) -- (0.75 + 1.5*\Cdelta, 0 + 1.5*\Cdelta);
    \filldraw[color = orange(ryb)] (1.5*\Cdelta, 0.75 + 1.5*\Cdelta) circle (2 pt);
    \filldraw[color = orange(ryb)] (0.75 + 1.5*\Cdelta, 1.5 + 1.5*\Cdelta) circle (2 pt);
    \filldraw[color = orange(ryb)](0.75 + 1.5*\Cdelta, 0 + 1.5*\Cdelta) circle (2 pt);
    \draw[color = lightgray, line width = 0.75 pt] (0 + 1.2*\aAdeltax, 2.5 *\aAdeltay) -- (1.5 + 1.5*\Cdelta, 0.75 + 1.5*\Cdelta);
    \filldraw[color =goodpurple] (0 + 1.2*\aAdeltax, 2.5 *\aAdeltay)  circle (2 pt);
    \filldraw[color =goodpurple] (-1.5 + 1.2*\aAdeltax, 2.5 *\aAdeltay)  circle (2 pt);
    \filldraw[color =goodpurple] (1.2*\aAdeltax, -1.5 + 2.5 *\aAdeltay)  circle (2 pt);
    \filldraw[color = orange(ryb)] (1.5 + 1.5*\Cdelta, 0.75 + 1.5*\Cdelta) circle (2 pt);
    \end{tikzpicture} 
    \qquad 
    \begin{tikzpicture}[scale = 0.75, baseline = {([yshift=-.5ex]current bounding box.center)}]
    \draw[lightdodgerblue, line width = 0.65 pt] (1.,0) -- (-1.5, 0);
    \draw[lightdodgerblue, line width = 0.65 pt] (0,1.) -- (0,-1.5);
    \draw[lightdodgerblue, line width = 0.65 pt] (0, 0) -- (-1.5, 0) -- (-1.5, -1.5) -- (0, -1.5);
    \draw[lightorange(ryb), line width = 0.65 pt] (1. + 1.5*\Cdelta,0 + 1.5*\Cdelta) -- (-1.5 + 1.5*\Cdelta, 0 + 1.5*\Cdelta);
    \draw[lightorange(ryb), line width = 0.65 pt] (0 + 1.5*\Cdelta,1. + 1.5*\Cdelta) -- (0 + 1.5*\Cdelta,-1. + \Cdelta);
    \draw[lightcrimson, line width = 0.65 pt] (1. + 1.2*\aAdeltax,0 + 2.5*\aAdeltay) -- (-1. + 1.2*\aAdeltax, 0 + 2.5*\aAdeltay);
    \draw[lightcrimson, line width = 0.65 pt] (0 + 1.2*\aAdeltax,1. + 2.5*\aAdeltay) -- (0 + 1.2*\aAdeltax,-1. + \aAdeltay);
    \draw[lightcrimson, line width = 0.65 pt] (0 + 1.2*\aAdeltax, 0 + 2.5*\aAdeltay) -- (0 + 1.2*\aAdeltax, -1.5 + 2.5*\aAdeltay)  -- (-1.5 + 1.2*\aAdeltax, -1.5 + 2.5*\aAdeltay) -- (-1.5 + 1.2*\aAdeltax, 0 + 2.5*\aAdeltay)--cycle;
    \node at (0.5 + 1.5*\Cdelta, 1.5*\Cdelta) {\small $\oC{X}$};
    \node at (-0.5 + 1.5*\Cdelta, 1.5*\Cdelta) {\small $\oC{X}$};
    \node at (1.5*\Cdelta, 0.5 + 1.5*\Cdelta) {\small $\oC{X}$};
    \node at (1.5*\Cdelta, -0.5 + 1.5*\Cdelta) {\small $\oC{X}$};
    \draw[color = black, line width = 0.75 pt] (-1.5 + 1.2*\aAdeltax, -1.5 + 2.5 *\aAdeltay) -- (-1.5, -0.75);
    \draw[color = lightgray, line width = 0.75 pt] (-1.5 + 1.2*\aAdeltax, 2.5 *\aAdeltay) -- (-0.75, 0);
    \draw[color = black, line width = 0.75 pt] (1.2*\aAdeltax, -1.5 + 2.5 *\aAdeltay) -- (0, -0.75);
    \filldraw[color = dodgerblue] (-0.75, 0) circle (2 pt);
    \filldraw[color = dodgerblue] (0, -0.75) circle (2 pt);
    \draw[color = lightgray, line width = 0.75 pt] (-1.5 + 1.2*\aAdeltax, -1.5 + 2.5*\aAdeltay) -- (-0.75, -1.5);
    \filldraw[color = dodgerblue] (-0.75, -1.5) circle (2 pt);
    \filldraw[color =goodpurple] (-1.5 + 1.2*\aAdeltax, -1.5 + 2.5 *\aAdeltay)  circle (2 pt);
    \filldraw[color =goodpurple] (1.2*\aAdeltax, -1.5 + 2.5 *\aAdeltay)  circle (2 pt);
    \filldraw[color =goodpurple] (-1.5 + 1.2*\aAdeltax, 2.5 *\aAdeltay)  circle (2 pt);
    \filldraw[color = dodgerblue] (-1.5, -0.75) circle (2 pt);
    \end{tikzpicture},
\end{equation}
whereas the diagonal stabilizers (namely the plaquette stabilizers in Eq.~\eqref{eq-TC} and edge stabilizers on sublattice $\AR{A}$ in Eq.~\eqref{eq-clusterstate}) are unchanged as $\Omega_{\aP{a}\bB{B}\oC{C}}$ is diagonal.

After performing the symmetry enrichment, the only remaining step is to project the $\aP{a}$ sublattice into the $\ket{+}$ state, or equivalently add $X_{\aP{a}}$ to the stabilizer group and determine how the stabilizers are updated.
For some of the stabilizers, this is straightforward.
The $\bB{B}$ and $\oC{C}$ plaquette stabilizers are unaffected by this projection as there is no overlapping support with the $\aP{a}$ qubits.
The diagonal stabilizers on the $\aP{a}\AR{A}$ sublattices [Eq.~\eqref{eq-clusterstate}] anti-commute with measurement of $X_{\aP{a}}$. However, products of these stabilizer around plaquettes are supported only on the $\AR{A}$ sublattice and will survive projection.
The first vertex stabilizer of Eq.~\eqref{eq-OmegaenrichedX} also commutes $X_{\aP{a}}$, and we can therefore simply set $X_{\aP{a}} = 1$ to obtain the updated stabilizer.
The only non-trivial stabilizers to update are the remaining two stabilizers of Eq.~\eqref{eq-OmegaenrichedX} which fail to either commute or anti-commute with $X_{\aP{a}}$.
Using the identity $\mathsf{CZ}_{ve}=Z_v^{n_e}$, where $n_e= (1 - Z_{e})/2$, we can rewrite these as:
\begin{equation}\label{eq-czrewrite}
    \begin{tikzpicture}[scale = 0.75, baseline = {([yshift=-.5ex]current bounding box.center)}]
    \draw[lightdodgerblue, line width = 0.65 pt] (1.,0) -- (-1., 0);
    \draw[lightdodgerblue, line width = 0.65 pt] (0,1.5) -- (0,-1.);
    \draw[lightdodgerblue, line width = 0.65 pt] (0, 0) -- (1.5, 0) -- (1.5, 1.5) -- (0, 1.5);
    \draw[lightorange(ryb), line width = 0.65 pt] (1.5 + 1.5*\Cdelta,0 + 1.5*\Cdelta) -- (-1. + 1.5*\Cdelta, 0 + 1.5*\Cdelta);
    \draw[lightorange(ryb), line width = 0.65 pt] (0 + 1.5*\Cdelta,1.5 + 1.5*\Cdelta) -- (0 + 1.5*\Cdelta,-1. + \Cdelta);
    \draw[lightorange(ryb), line width = 0.65 pt] (0 + 1.5*\Cdelta, 0 + 1.5*\Cdelta) -- (1.5 + 1.5*\Cdelta, 0 + 1.5*\Cdelta) -- (1.5 + 1.5*\Cdelta, 1.5 + 1.5*\Cdelta) -- (0 + 1.5*\Cdelta, 1.5 + 1.5*\Cdelta);
    \draw[lightcrimson, line width = 0.65 pt] (1.5 + 1.2*\aAdeltax,0 + 2.5*\aAdeltay) -- (-1.5 + 1.2*\aAdeltax, 0 + 2.5*\aAdeltay);
    \draw[lightcrimson, line width = 0.65 pt] (0 + 1.2*\aAdeltax,1.5 + 2.5*\aAdeltay) -- (0 + 1.2*\aAdeltax,-1.5 + 2.5*\aAdeltay);
    \draw[lightcrimson, line width = 0.65 pt] (1.5 + 1.2*\aAdeltax,1.5 + 2.5*\aAdeltay) -- (1.5 + 1.2*\aAdeltax,0 + 2.5*\aAdeltay);
    \draw[lightcrimson, line width = 0.65 pt] (0 + 1.2*\aAdeltax,1.5 + 2.5*\aAdeltay) -- (1.5 + 1.2*\aAdeltax,1.5 + 2.5*\aAdeltay);

    \node at  (0.75 + 1.5*\Cdelta, 1.5 + 1.5*\Cdelta) {\scriptsize $1$};
    \node at  (1.5 + 1.5*\Cdelta, 0.75 + 1.5*\Cdelta) {\scriptsize $2$};
    \node at (0.75 + 1.5*\Cdelta, 0 + 1.5*\Cdelta) {\scriptsize $3$};
    \node at (1.5*\Cdelta, 0.75 + 1.5*\Cdelta) {\scriptsize $4$};
    
    
    \node at (0.5+1.2*\aAdeltax, 2.5*\aAdeltay) {\small $\aP{Z}^{n_1+n_2}$};
    \node at (-1.3 + 1.2*\aAdeltax, 2.5*\aAdeltay) {\small $\aP{Z}^{n_4}$};
    \node at (0.2+1.2*\aAdeltax, -1.5 + 2.5*\aAdeltay) {\small $\aP{Z}^{n_3}$};
    
    \end{tikzpicture} 
    \qquad 
    \begin{tikzpicture}[scale = 0.75, baseline = {([yshift=-.5ex]current bounding box.center)}]
    \draw[lightdodgerblue, line width = 0.65 pt] (1.,0) -- (-1.5, 0);
    \draw[lightdodgerblue, line width = 0.65 pt] (0,1.) -- (0,-1.5);
    \draw[lightdodgerblue, line width = 0.65 pt] (0, 0) -- (-1.5, 0) -- (-1.5, -1.5) -- (0, -1.5);
    \draw[lightorange(ryb), line width = 0.65 pt] (1. + 1.5*\Cdelta,0 + 1.5*\Cdelta) -- (-1.5 + 1.5*\Cdelta, 0 + 1.5*\Cdelta);
    \draw[lightorange(ryb), line width = 0.65 pt] (0 + 1.5*\Cdelta,1. + 1.5*\Cdelta) -- (0 + 1.5*\Cdelta,-1. + \Cdelta);
    \draw[lightcrimson, line width = 0.65 pt] (1. + 1.2*\aAdeltax,0 + 2.5*\aAdeltay) -- (-1. + 1.2*\aAdeltax, 0 + 2.5*\aAdeltay);
    \draw[lightcrimson, line width = 0.65 pt] (0 + 1.2*\aAdeltax,1. + 2.5*\aAdeltay) -- (0 + 1.2*\aAdeltax,-1. + \aAdeltay);
    \draw[lightcrimson, line width = 0.65 pt] (0 + 1.2*\aAdeltax, 0 + 2.5*\aAdeltay) -- (0 + 1.2*\aAdeltax, -1.5 + 2.5*\aAdeltay)  -- (-1.5 + 1.2*\aAdeltax, -1.5 + 2.5*\aAdeltay) -- (-1.5 + 1.2*\aAdeltax, 0 + 2.5*\aAdeltay)--cycle;
    \node at (-0.75, 0.05) {\scriptsize $1$};
    \node at (0, -0.75) {\scriptsize $2$};
    \node at (-0.75, -1.5) {\scriptsize $3$};
    \node at (-1 + 1.2*\aAdeltax, 0.05-1.5 + 2.5 *\aAdeltay) {\small $\aP{Z}^{n_3+n_4}$};
    \node at (0.2+1.2*\aAdeltax, -1.5 + 2.5 *\aAdeltay) {\small $\aP{Z}^{n_2}$};
    \node at (-1.3 + 1.2*\aAdeltax, 2.5 *\aAdeltay) {\small $\aP{Z}^{n_1}$};
    \node at (-1.5, -0.75) {\scriptsize $4$};
    \end{tikzpicture}
\end{equation}
The key insight for updating these stabilizers after projecting $X_{\aP{a}} = +1$ is to recognize that the $\bB{B}$ and $\oC{C}$ plaquette stabilizers appear in the $\mathsf{CZ}$ structure above.
To see this, let us note that the first stabilizer can be simplified as:
\begin{equation}
     \begin{tikzpicture}[scale = 0.7, baseline = {([yshift=-.5ex]current bounding box.center)}]
    \draw[lightdodgerblue, line width = 0.65 pt] (1.,0) -- (-1., 0);
    \draw[lightdodgerblue, line width = 0.65 pt] (0,1.5) -- (0,-1.);
    \draw[lightdodgerblue, line width = 0.65 pt] (0, 0) -- (1.5, 0) -- (1.5, 1.5) -- (0, 1.5);
    \draw[lightorange(ryb), line width = 0.65 pt] (1.5 + 1.5*\Cdelta,0 + 1.5*\Cdelta) -- (-1. + 1.5*\Cdelta, 0 + 1.5*\Cdelta);
    \draw[lightorange(ryb), line width = 0.65 pt] (0 + 1.5*\Cdelta,1.5 + 1.5*\Cdelta) -- (0 + 1.5*\Cdelta,-1. + \Cdelta);
    \draw[lightorange(ryb), line width = 0.65 pt] (0 + 1.5*\Cdelta, 0 + 1.5*\Cdelta) -- (1.5 + 1.5*\Cdelta, 0 + 1.5*\Cdelta) -- (1.5 + 1.5*\Cdelta, 1.5 + 1.5*\Cdelta) -- (0 + 1.5*\Cdelta, 1.5 + 1.5*\Cdelta);
    \draw[lightcrimson, line width = 0.65 pt] (1.5 + 1.2*\aAdeltax,0 + 2.5*\aAdeltay) -- (-1.5 + 1.2*\aAdeltax, 0 + 2.5*\aAdeltay);
    \draw[lightcrimson, line width = 0.65 pt] (0 + 1.2*\aAdeltax,1.5 + 2.5*\aAdeltay) -- (0 + 1.2*\aAdeltax,-1.5 + 2.5*\aAdeltay);
    \draw[lightcrimson, line width = 0.65 pt] (1.5 + 1.2*\aAdeltax,1.5 + 2.5*\aAdeltay) -- (1.5 + 1.2*\aAdeltax,0 + 2.5*\aAdeltay);
    \draw[lightcrimson, line width = 0.65 pt] (0 + 1.2*\aAdeltax,1.5 + 2.5*\aAdeltay) -- (1.5 + 1.2*\aAdeltax,1.5 + 2.5*\aAdeltay);

    \node at  (0.75 + 1.5*\Cdelta, 1.5 + 1.5*\Cdelta) {\scriptsize $1$};
    \node at  (1.5 + 1.5*\Cdelta, 0.75 + 1.5*\Cdelta) {\scriptsize $2$};
    \node at (0.75 + 1.5*\Cdelta, 0 + 1.5*\Cdelta) {\scriptsize $3$};
    \node at (1.5*\Cdelta, 0.75 + 1.5*\Cdelta) {\scriptsize $4$};
    
    
    \node at (0.5+1.2*\aAdeltax, 2.5*\aAdeltay) {\small $\aP{Z}^{n_1+n_2}$};
    \node at (-1.3 + 1.2*\aAdeltax, 2.5*\aAdeltay) {\small $\aP{Z}^{n_4}$};
    \node at (0.2+1.2*\aAdeltax, -1.5 + 2.5*\aAdeltay) {\small $\aP{Z}^{n_3}$};
    
    \end{tikzpicture}\  =\  \begin{tikzpicture}[scale = 0.7, baseline = {([yshift=-.5ex]current bounding box.center)}]
    \draw[lightdodgerblue, line width = 0.65 pt] (1.,0) -- (-1., 0);
    \draw[lightdodgerblue, line width = 0.65 pt] (0,1.5) -- (0,-1.);
    \draw[lightdodgerblue, line width = 0.65 pt] (0, 0) -- (1.5, 0) -- (1.5, 1.5) -- (0, 1.5);
    \draw[lightorange(ryb), line width = 0.65 pt] (1.5 + 1.5*\Cdelta,0 + 1.5*\Cdelta) -- (-1. + 1.5*\Cdelta, 0 + 1.5*\Cdelta);
    \draw[lightorange(ryb), line width = 0.65 pt] (0 + 1.5*\Cdelta,1.5 + 1.5*\Cdelta) -- (0 + 1.5*\Cdelta,-1. + \Cdelta);
    \draw[lightorange(ryb), line width = 0.65 pt] (0 + 1.5*\Cdelta, 0 + 1.5*\Cdelta) -- (1.5 + 1.5*\Cdelta, 0 + 1.5*\Cdelta) -- (1.5 + 1.5*\Cdelta, 1.5 + 1.5*\Cdelta) -- (0 + 1.5*\Cdelta, 1.5 + 1.5*\Cdelta);
    \draw[lightcrimson, line width = 0.65 pt] (1.5 + 1.2*\aAdeltax,0 + 2.5*\aAdeltay) -- (-1.5 + 1.2*\aAdeltax, 0 + 2.5*\aAdeltay);
    \draw[lightcrimson, line width = 0.65 pt] (0 + 1.2*\aAdeltax,1.5 + 2.5*\aAdeltay) -- (0 + 1.2*\aAdeltax,-1.5 + 2.5*\aAdeltay);
    \draw[lightcrimson, line width = 0.65 pt] (1.5 + 1.2*\aAdeltax,1.5 + 2.5*\aAdeltay) -- (1.5 + 1.2*\aAdeltax,0 + 2.5*\aAdeltay);
    \draw[lightcrimson, line width = 0.65 pt] (0 + 1.2*\aAdeltax,1.5 + 2.5*\aAdeltay) -- (1.5 + 1.2*\aAdeltax,1.5 + 2.5*\aAdeltay);

    
    
    \node at (0.5+1.2*\aAdeltax, 2.5*\aAdeltay) {\small $\aP{Z}^{\sum_i n_i}$};
    \node at (-0.7 + 1.2*\aAdeltax, 2.5*\aAdeltay) {\small $\AR{Z}^{n_4}$};
    \node at (0.2+1.2*\aAdeltax, -0.9 + 2.5*\aAdeltay) {\small $\AR{Z}^{n_3}$};
    \end{tikzpicture}\ =\  
        \begin{tikzpicture}[scale = 0.7, baseline = {([yshift=-.5ex]current bounding box.center)}]
    \draw[lightdodgerblue, line width = 0.65 pt] (1.,0) -- (-1., 0);
    \draw[lightdodgerblue, line width = 0.65 pt] (0,1.5) -- (0,-1.);
    \draw[lightdodgerblue, line width = 0.65 pt] (0, 0) -- (1.5, 0) -- (1.5, 1.5) -- (0, 1.5);
    \draw[lightorange(ryb), line width = 0.65 pt] (1.5 + 1.5*\Cdelta,0 + 1.5*\Cdelta) -- (-1. + 1.5*\Cdelta, 0 + 1.5*\Cdelta);
    \draw[lightorange(ryb), line width = 0.65 pt] (0 + 1.5*\Cdelta,1.5 + 1.5*\Cdelta) -- (0 + 1.5*\Cdelta,-1. + \Cdelta);
    \draw[lightorange(ryb), line width = 0.65 pt] (0 + 1.5*\Cdelta, 0 + 1.5*\Cdelta) -- (1.5 + 1.5*\Cdelta, 0 + 1.5*\Cdelta) -- (1.5 + 1.5*\Cdelta, 1.5 + 1.5*\Cdelta) -- (0 + 1.5*\Cdelta, 1.5 + 1.5*\Cdelta);
    \draw[lightcrimson, line width = 0.65 pt] (1.5 + 1.2*\aAdeltax,0 + 2.5*\aAdeltay) -- (-1.5 + 1.2*\aAdeltax, 0 + 2.5*\aAdeltay);
    \draw[lightcrimson, line width = 0.65 pt] (0 + 1.2*\aAdeltax,1.5 + 2.5*\aAdeltay) -- (0 + 1.2*\aAdeltax,-1.5 + 2.5*\aAdeltay);
    \draw[lightcrimson, line width = 0.65 pt] (1.5 + 1.2*\aAdeltax,1.5 + 2.5*\aAdeltay) -- (1.5 + 1.2*\aAdeltax,0 + 2.5*\aAdeltay);
    \draw[lightcrimson, line width = 0.65 pt] (0 + 1.2*\aAdeltax,1.5 + 2.5*\aAdeltay) -- (1.5 + 1.2*\aAdeltax,1.5 + 2.5*\aAdeltay);

    
    
    \node at (-0.7 + 1.2*\aAdeltax, 2.5*\aAdeltay) {\small $\AR{Z}^{n_4}$};
    \node at (0.2+1.2*\aAdeltax, -0.9 + 2.5*\aAdeltay) {\small $\AR{Z}^{n_3}$};
    \node at (-0.7+1.5+1.2*\aAdeltax, 2.5*\aAdeltay) {\small $\AR{Z}^{\aP{n}}$};
    \node at (-0.7+1.5+1.2*\aAdeltax, 1.5+2.5*\aAdeltay) {\small $\AR{Z}^{\aP{n}}$};
    \node at (0.1+1.2*\aAdeltax, 0.7+2.5*\aAdeltay) {\small $\AR{Z}^{\aP{n}}$};
    \node at (1.5+0.1+1.2*\aAdeltax, 0.7+2.5*\aAdeltay) {\small $\AR{Z}^{\aP{n}}$};
    \end{tikzpicture} 
\end{equation}
where in the first equality, we used the relation  $\aP{Z}\AR{Z}\aP{Z}=+1$ in Eq.~\eqref{eq-clusterstate}, and in the second equality, we used the fact that $\aP{Z}^{\sum_i\AR{n_i}}=(-1)^{\aP{n}(\sum_{i} \AR{n_i})}=\prod_i \AR{Z}_i^{\aP{n}}$.
As a consequence of the new plaquette stabilizer in the $\AR{A}$ sublattice, we can trivially set $\prod_i \AR{Z}_i^{\aP{n}}=+1$.
As a result, the stabilizer can be simplified to a form that is independent of the $\aP{a}$ sublattice and commutes with the measurement $X_{\aP{a}} = +1$. 
 The stabilizer for the second term in Eq.~\eqref{eq-czrewrite} can also be expressed in a similar form.

As a consequence, after projection, the non-Abelian $\mathcal{D}(D_4)$ code has stabilizers of the form: 
\begin{equation} \label{eq-OmegaenrichedX-final}
\begin{aligned}
    \begin{tikzpicture}[scale = 0.75, baseline = {([yshift=-.5ex]current bounding box.center)}]
    \draw[lightdodgerblue, line width = 0.65 pt] (1.,0) -- (-1., 0);
    \draw[lightdodgerblue, line width = 0.65 pt] (0,1.5) -- (0,-1.);
    \draw[lightdodgerblue, line width = 0.65 pt] (0, 0) -- (1.5, 0) -- (1.5, 1.5) -- (0, 1.5);
    \draw[lightorange(ryb), line width = 0.65 pt] (1.5 + 1.5*\Cdelta,0 + 1.5*\Cdelta) -- (-1. + 1.5*\Cdelta, 0 + 1.5*\Cdelta);
    \draw[lightorange(ryb), line width = 0.65 pt] (0 + 1.5*\Cdelta,1.5 + 1.5*\Cdelta) -- (0 + 1.5*\Cdelta,-1. + \Cdelta);
    \draw[lightorange(ryb), line width = 0.65 pt] (0 + 1.5*\Cdelta, 0 + 1.5*\Cdelta) -- (1.5 + 1.5*\Cdelta, 0 + 1.5*\Cdelta) -- (1.5 + 1.5*\Cdelta, 1.5 + 1.5*\Cdelta) -- (0 + 1.5*\Cdelta, 1.5 + 1.5*\Cdelta);
    \draw[lightcrimson, line width = 0.65 pt] (1.5 + 1.2*\aAdeltax,0 + 2.5*\aAdeltay) -- (-1. + 1.2*\aAdeltax, 0 + 2.5*\aAdeltay);
    \draw[lightcrimson, line width = 0.65 pt] (0 + 1.2*\aAdeltax,1.5 + 2.5*\aAdeltay) -- (0 + 1.2*\aAdeltax,-1. + \aAdeltay);
    \node at (0.5 + 1.2*\aAdeltax, 2.5*\aAdeltay) {\small $\AR{X}$};
    \node at (-0.5 + 1.2*\aAdeltax, 2.5*\aAdeltay) {\small $\AR{X}$};
    \node at (1.2*\aAdeltax, 0.5 + 2.5*\aAdeltay) {\small $\AR{X}$};
    \node at (1.2*\aAdeltax, -0.5 + 2.5*\aAdeltay) {\small $\AR{X}$};
    \draw[color = black, line width = 0.75 pt](0, 0.8) -- (0.75 + 1.5*\Cdelta, 1.5 + 1.5*\Cdelta);
    \filldraw[color = dodgerblue] (0, 0.8) circle (2 pt);
    \filldraw[color = orange(ryb)] ((0.75 + 1.5*\Cdelta, 1.5 + 1.5*\Cdelta) circle (2 pt);
    \draw[color = lightgray, line width = 0.75 pt](0.8, 0) -- (1.5 + 1.5*\Cdelta, 0.75 + 1.5*\Cdelta);
    \filldraw[color = dodgerblue] (0.8, 0) circle (2 pt);
    \filldraw[color = orange(ryb)] (1.5 + 1.5*\Cdelta, 0.75 + 1.5*\Cdelta) circle (2 pt);
    \end{tikzpicture} \qquad \begin{tikzpicture}[scale = 0.75, baseline = {([yshift=-.5ex]current bounding box.center)}]
    \draw[lightdodgerblue, line width = 0.65 pt] (1.,0) -- (-1., 0);
    \draw[lightdodgerblue, line width = 0.65 pt] (0,1.5) -- (0,-1.);
    \draw[lightdodgerblue, line width = 0.65 pt] (0, 0) -- (1.5, 0) -- (1.5, 1.5) -- (0, 1.5);
    \draw[lightorange(ryb), line width = 0.65 pt] (1.5 + 1.5*\Cdelta,0 + 1.5*\Cdelta) -- (-1. + 1.5*\Cdelta, 0 + 1.5*\Cdelta);
    \draw[lightorange(ryb), line width = 0.65 pt] (0 + 1.5*\Cdelta,1.5 + 1.5*\Cdelta) -- (0 + 1.5*\Cdelta,-1. + \Cdelta);
    \draw[lightorange(ryb), line width = 0.65 pt] (0 + 1.5*\Cdelta, 0 + 1.5*\Cdelta) -- (1.5 + 1.5*\Cdelta, 0 + 1.5*\Cdelta) -- (1.5 + 1.5*\Cdelta, 1.5 + 1.5*\Cdelta) -- (0 + 1.5*\Cdelta, 1.5 + 1.5*\Cdelta);
    \draw[lightcrimson, line width = 0.65 pt] (1.5 + 1.2*\aAdeltax,0 + 2.5*\aAdeltay) -- (-1.5 + 1.2*\aAdeltax, 0 + 2.5*\aAdeltay);
    \draw[lightcrimson, line width = 0.65 pt] (0 + 1.2*\aAdeltax,1.5 + 2.5*\aAdeltay) -- (0 + 1.2*\aAdeltax,-1.5 + 2.5*\aAdeltay);
    \node at (0.5 + 0, 0) {\small $\bB{X}$};
    \node at (-0.5 + 0, 0) {\small $\bB{X}$};
    \node at (0, 0.5 + 0) {\small $\bB{X}$};
    \node at (0, -0.5 + 0) {\small $\bB{X}$};
    \draw[color = lightgray, line width = 0.75 pt](-0.9 + 1.2*\aAdeltax, 2.5 *\aAdeltay) -- (1.5*\Cdelta, 0.75 + 1.5*\Cdelta);
    \draw[color = black, line width = 0.75 pt](0 + 1.2*\aAdeltax, -0.9 + 2.5 *\aAdeltay) -- (0.75 + 1.5*\Cdelta, 0 + 1.5*\Cdelta);
    \filldraw[color = orange(ryb)] (1.5*\Cdelta, 0.75 + 1.5*\Cdelta) circle (2 pt);
    \filldraw[color = orange(ryb)](0.75 + 1.5*\Cdelta, 0 + 1.5*\Cdelta) circle (2 pt);
    \filldraw[color = crimson] (-0.9 + 1.2*\aAdeltax, 2.5 *\aAdeltay)  circle (2 pt);
    \filldraw[color = crimson] (1.2*\aAdeltax, -0.9 + 2.5 *\aAdeltay)  circle (2 pt);
    \end{tikzpicture}\qquad \begin{tikzpicture}[scale = 0.75, baseline = {([yshift=-.5ex]current bounding box.center)}]
    \draw[lightdodgerblue, line width = 0.65 pt] (1.,0) -- (-1.5, 0);
    \draw[lightdodgerblue, line width = 0.65 pt] (0,1.) -- (0,-1.5);
    \draw[lightdodgerblue, line width = 0.65 pt] (0, 0) -- (-1.5, 0) -- (-1.5, -1.5) -- (0, -1.5);
    \draw[lightorange(ryb), line width = 0.65 pt] (1. + 1.5*\Cdelta,0 + 1.5*\Cdelta) -- (-1.5 + 1.5*\Cdelta, 0 + 1.5*\Cdelta);
    \draw[lightorange(ryb), line width = 0.65 pt] (0 + 1.5*\Cdelta,1. + 1.5*\Cdelta) -- (0 + 1.5*\Cdelta,-1. + \Cdelta);
    \draw[lightcrimson, line width = 0.65 pt] (1. + 1.2*\aAdeltax,0 + 2.5*\aAdeltay) -- (-1. + 1.2*\aAdeltax, 0 + 2.5*\aAdeltay);
    \draw[lightcrimson, line width = 0.65 pt] (0 + 1.2*\aAdeltax,1. + 2.5*\aAdeltay) -- (0 + 1.2*\aAdeltax,-1. + \aAdeltay);
    \draw[lightcrimson, line width = 0.65 pt] (0 + 1.2*\aAdeltax, 0 + 2.5*\aAdeltay) -- (0 + 1.2*\aAdeltax, -1.5 + 2.5*\aAdeltay)  -- (-1.5 + 1.2*\aAdeltax, -1.5 + 2.5*\aAdeltay) -- (-1.5 + 1.2*\aAdeltax, 0 + 2.5*\aAdeltay)--cycle;
    \node at (0.5 + 1.5*\Cdelta, 1.5*\Cdelta) {\small $\oC{X}$};
    \node at (-0.5 + 1.5*\Cdelta, 1.5*\Cdelta) {\small $\oC{X}$};
    \node at (1.5*\Cdelta, 0.5 + 1.5*\Cdelta) {\small $\oC{X}$};
    \node at (1.5*\Cdelta, -0.5 + 1.5*\Cdelta) {\small $\oC{X}$};
    \draw[color = lightgray, line width = 0.75 pt] (-0.75 + 1.2*\aAdeltax, 2.5 *\aAdeltay) -- (-0.75, 0);
    \draw[color = black, line width = 0.75 pt] (1.2*\aAdeltax, -0.75 + 2.5 *\aAdeltay) -- (0, -0.75);
    \filldraw[color = dodgerblue] (-0.75, 0) circle (2 pt);
    \filldraw[color = dodgerblue] (0, -0.75) circle (2 pt);
    \filldraw[color = crimson] (1.2*\aAdeltax, -0.75 + 2.5 *\aAdeltay)  circle (2 pt);
    \filldraw[color = crimson] (-0.75 + 1.2*\aAdeltax, 2.5 *\aAdeltay)  circle (2 pt);
    \end{tikzpicture} 
\end{aligned},
\end{equation}
as well as diagonal plaquette stabilizers on the $\AR{A}$, $\bB{B}$, and $\oC{C}$ sublattices. 
As an added remark, we note that the ``$X$-stabilizers'' above do not commute with one another.
Instead, they commute up to a products of other $Z$ stabilizers and consequently will commute in the ground state subspace.

Since we will follow the same sequence of steps in our non-Abelian qLDPC constructions, we summarize the three essential steps of the gauging procedure:
\begin{enumerate}
    \item First, introduce two sublattices of ancilla qubits labeled by vertices $\textcolor{goodpurple}{a}$ and edges $\textcolor{red}{A}$, which are prepared in a cluster SPT state, which has a global $0$-form symmetry $G_{\aP{a}}$.
    
    \item Subsequently, apply a symmetry enrichment operator $\Omega$ satisfying $\Omega G_{\aP{a}} \Omega^{\dagger} = G_{\aP{a}} U$ such that the 0-form symmetry of the cluster state is tied to the global symmetry action $U$.
    
    \item Finally, ``measure out" the vertex ancilla qubits on $\textcolor{goodpurple}{a}$, yielding the final gauged code.
\end{enumerate}

\subsection{Preparation Procedure from Measurement and Feedback}

As a concluding remark, we mention that the gauging procedure used here operates as both a method to derive the non-Abelian $\mathcal{D}(D_4)$ code as well as a protocol to prepare the code with measurement and feedforward.
Indeed, the $\mathcal{D}(D_4)$ topological order can be prepared with a finite-depth \textit{adaptive} circuit where measurement implements exactly this gauging procedure.
In this measurement procedure, the $\ket{+}$ state projectors will be replaced with projectors associated with the non-deterministic measurement outcomes.
Nevertheless, these outcomes can be corrected back to the $\ket{+}$ state!
This is because the random measurement outcomes only produce $e_{\mathsf{A}}$ anyons---which are Abelian.
Consequently, they can be paired up via a finite-depth circuit (serving as a decoder) to return to the $\mathcal{D}(D_4)$ ground state \cite{verresen_efficiently_2022}.
In fact, using a different gauging procedure, the $\mathcal{D}(D_4)$ state can be minimally prepared with just one round of measurement and feedforward \cite{tantivasadakarn_shortest_2023}.

\section{Gauging Construction of Non-Abelian qLDPC Codes}\label{sec:gauging-theory}

Given the construction of non-Abelian topological codes in the previous section, a natural question is how to extend the gauging construction to the qLDPC setting. 
Surprisingly, we find that two qualitatively distinct approaches to gauging---which reduce to one another in the manifold context---naturally emerge in this setting.
The essential difference between these approaches can be distilled down to a difference in the structure of the auxiliary gauge fields introduced in each gauging procedure. As  we will see, different structures of ancilla gauge fields enable the gauging of different symmetries.

The first, which we term \textbf{homological gauging}, applies to quantum codes that have a certain algebraic structure analogous to Poincar\'e duality in manifolds.
In this approach, the redundancy structure of a qLDPC code's stabilizers is tied to the structure of its logical Clifford gates. 
We will see that this correspondence enables gauging of the symmetries associated with these gates using gauge fields that are structurally similar to those of the original qLDPC codes.
In contrast, the second approach, which we term \textbf{graph gauging}, applies to more general qLDPC codes but relies on case-dependent considerations of how the structure of the code can be embedded into an underlying graph geometry.
In what follows, we describe these two procedures in detail and use them to develop distinct constructions of non-Abelian qLDPC codes.
Before introducing our full formalism, we will first illustrate these two distinct approaches to gauging in the context of a familiar example. We then briefly review the essential concepts from homology underlying our constructions before giving the explicit constructions themselves.

\subsection{Toy Model for the Two Approaches to Gauging}\label{sec:toy}
The differences between our two gauging constructions lies in the structure of the ancilla qubits introduced to gauge a symmetry. We illustrate this distinction with a toy example showing how the same input code may be gauged with two structurally different choices of ancilla gauge fields. Specifically, we consider two stacks of $N$ decoupled $2D$ toric codes, which we denote $\bB{B}$ and $\oC{C}$:
\newcommand{\toricstack}[2]{
\begin{scope}[xshift=#1]
\coordinate (eone) at (2,0);
\coordinate (etwo) at (0.8,0.6);
\coordinate (etwotwo) at (1.2,0.9);
\coordinate (ez)   at (0,2.5);

\coordinate (eoneone) at (3,0);
\def\Nx{2}
\def\Ny{2}
\def\Nz{2}

\foreach \k in {0,...,\Nz} {

\coordinate (A0) at ($(0,0) + \k*(ez)$);
\coordinate (B0) at ($\Ny*(etwotwo) + \k*(ez)$);
\coordinate (C0) at ($\Nx*(eoneone) + \Ny*(etwotwo) + \k*(ez)$);
\coordinate (D0) at ($\Nx*(eoneone) + \k*(ez)$);

\fill[#2, opacity=0.2] (A0)--(B0)--(C0)--(D0)--cycle;

  \foreach \i in {0,...,\Nx} {
    \foreach \j in {0,...,\Ny} {

      \coordinate (A) at ($\i*(eone)+\j*(etwo)+\k*(ez)$);
      \coordinate (B) at ($(A)+(etwo)$);
      \coordinate (C) at ($(A)+(eone)+(etwo)$);
      \coordinate (D) at ($(A)+(eone)$);

      \draw[#2] (A)--(B)--(C)--(D)--cycle;

      \coordinate (Eh) at ($(A)!0.5!(D)$);
      \coordinate (Ev) at ($(A)!0.5!(B)$);

      \filldraw[#2] (Eh) circle (2.3pt);
      \filldraw[#2] (Ev) circle (2.3pt);
    }
  }

  \foreach \j in {0,...,\Ny} {
    \coordinate (A) at ($\Nx*(eone)+\j*(etwo)+\k*(ez)$);
    \coordinate (B) at ($(A)+(etwo)$);
    \coordinate (Ev) at ($(A)!0.5!(B)$);
    \filldraw[#2] (Ev) circle (2.3pt);
  }

  \foreach \i in {0,...,\Nx} {
    \coordinate (A) at ($\i*(eone)+\Ny*(etwo)+\k*(ez)$);
    \coordinate (D) at ($(A)+(eone)$);
    \coordinate (Eh) at ($(A)!0.5!(D)$);
    \filldraw[#2] (Eh) circle (2.3pt);

    \draw[dotted] (0,0) -- ($(0,0)+2*(ez)$);
    \draw[dotted] ($3*(eone)$) -- ($3*(eone)+2*(ez)$);
    \draw[dotted] ($3*(etwo)$) -- ($3*(etwo)+2*(ez)$);
    \draw[dotted] ($3*(etwo)+3*(eone)$) -- ($3*(etwo)+3*(eone)+2*(ez)$);
  }
}
\end{scope}
}
\begin{equation}
\quad
\begin{tikzpicture}[
    scale=0.35,
    baseline={([yshift=-.5ex]current bounding box.center)}
]

\toricstack{0cm}{dodgerblue}
\toricstack{10cm}{orange(ryb)}

\end{tikzpicture}
\end{equation}

Our first approach to gauging the input \BandC codes introduces a stack of $N$ Lieb lattices of ancilla qubits, such that each layer realizing a 2D cluster state:
%
\newcommand{\toricstackredgoodpurple}[1]{
\begin{scope}[xshift=#1]
\coordinate (eone) at (2,0);
\coordinate (etwo) at (0.8,0.6);
\coordinate (ez)   at (0,2.5);
\def\Nx{2}
\def\Ny{2}
\def\Nz{2}
%
\foreach \k in {0,...,\Nz} {
%
  \foreach \i in {0,...,\Nx} {
    \foreach \j in {0,...,\Ny} {

      \coordinate (A) at ($\i*(eone)+\j*(etwo)+\k*(ez)$);
      \coordinate (B) at ($(A)+(etwo)$);
      \coordinate (C) at ($(A)+(eone)+(etwo)$);
      \coordinate (D) at ($(A)+(eone)$);

      \draw[red, thin] (A)--(B)--(C)--(D)--cycle;

      \filldraw[goodpurple] (A) circle (3pt);
      \filldraw[goodpurple] (B) circle (3pt);
      \filldraw[goodpurple] (C) circle (3pt);
      \filldraw[goodpurple] (D) circle (3pt);

      \coordinate (Eh1) at ($(A)!0.5!(B)$); 
      \coordinate (Eh2) at ($(B)!0.5!(C)$); 
      \coordinate (Eh3) at ($(C)!0.5!(D)$); 
      \coordinate (Eh4) at ($(D)!0.5!(A)$); 

      \filldraw[red] (Eh1) circle (2.5pt);
      \filldraw[red] (Eh2) circle (2.5pt);
      \filldraw[red] (Eh3) circle (2.5pt);
      \filldraw[red] (Eh4) circle (2.5pt);

    \draw[dotted] (0,0) -- ($(0,0)+2*(ez)$);
    \draw[dotted] ($3*(eone)$) -- ($3*(eone)+2*(ez)$);
    \draw[dotted] ($3*(etwo)$) -- ($3*(etwo)+2*(ez)$);
    \draw[dotted] ($3*(etwo)+3*(eone)$) -- ($3*(etwo)+3*(eone)+2*(ez)$);
    }
  }

}
\end{scope}
}
\begin{equation}
\begin{tikzpicture}[
    scale=0.35,
    baseline={([yshift=-.5ex]current bounding box.center)}
]
\toricstackredgoodpurple{0cm}
\end{tikzpicture}
\end{equation}

For this choice of ancilla lattice, each layer of ancilla qubits has an independent $\aP{\mathbb{Z}_2^{(0)}}$ 0-form symmetry. This structure allows us to independently gauge each $\mathsf{CZ}$ symmetry which acts between a single layer of \BandC toric codes. This can be achieved by using a symmetry enrichment operator which ties the $\aP{\mathbb{Z}_2^{(0)}}$ 0-form symmetry of each layer of the cluster state to the corresponding $\mathsf{CZ}$ symmetry in each layer of the \BandC codes. Conceptually, this approach mirrors the $2$D construction of Sec.~\ref{sec:prelim}: the edges of the ancilla lattice are copies of the original \BandC code edges, and the same gauging procedure as in Sec.~\ref{sec:prelim} is repeated independently for each layer in the stacks.

However, the same \BandC stacks of toric codes can also be gauged by introducing ancilla qubits with an underlying structure that is distinct from the original codes \BandC. For instance, we could have instead introduced an ancilla cluster state formed from a 3D cubic lattice of ancilla qubits:

\begin{center}
\begin{tikzpicture}[
    lattice edge/.style={red, thin},
    vertex dot/.style={fill=goodpurple, circle, inner sep=1.3pt},
    midpoint dot/.style={fill=red, circle, inner sep=1pt},
    x={(1cm,0cm)}, 
    y={(0cm,1cm)}, 
    z={(-0.4cm,-0.3cm)}
]

    \def\N{2} 

    \foreach \y in {0,...,\N} {
        \foreach \z in {0,...,\N} {
            \draw[lattice edge] (0,\y,\z) -- (\N,\y,\z);
        }
    }
    \foreach \x in {0,...,\N} {
        \foreach \z in {0,...,\N} {
            \draw[lattice edge] (\x,0,\z) -- (\x,\N,\z);
        }
    }
    \foreach \x in {0,...,\N} {
        \foreach \y in {0,...,\N} {
            \draw[lattice edge] (\x,\y,0) -- (\x,\y,\N);
        }
    }

    \pgfmathsetmacro{\M}{\N-1}
    
    \foreach \x in {0,...,\M} {
        \foreach \y in {0,...,\N} {
            \foreach \z in {0,...,\N} {
                \node[midpoint dot] at (\x+0.5, \y, \z) {};
            }
        }
    }
    
    \foreach \x in {0,...,\N} {
        \foreach \y in {0,...,\M} {
            \foreach \z in {0,...,\N} {
                \node[midpoint dot] at (\x, \y+0.5, \z) {};
            }
        }
    }
    
    \foreach \x in {0,...,\N} {
        \foreach \y in {0,...,\N} {
            \foreach \z in {0,...,\M} {
                \node[midpoint dot] at (\x, \y, \z+0.5) {};
            }
        }
    }

    \foreach \x in {0,...,\N} {
        \foreach \y in {0,...,\N} {
            \foreach \z in {0,...,\N} {
                \node[vertex dot] at (\x,\y,\z) {};
            }
        }
    }
\end{tikzpicture}
\end{center}

In this case, the ancilla lattice has a single $\aP{\mathbb{Z}_2^{(0)}}$ 0-form symmetry. Per the symmetry enrichment condition of the previous section, this ancilla cluster state can be used to gauge a single global symmetry on codes \BandC -- namely the single transversal $\mathsf{CZ}$ which acts simultaneously between the \BandC toric codes in each layer. This choice meets our ``symmetry matching" requirement between the 0-form symmetry of ancillas and global symmetry of the input code.

However, the outcome of gauging this single global symmetry is qualitatively different than the first approach; the resulting theory has 3D gauge fields which have non-Abelian mutual statistics with 2D gauge fields of restricted mobility. Despite this added complexity, this second approach is also more general, since it does not rely on the existence independent $\mathsf{CZ}$ symmetries acting within the two stacks, and applies even in cases where the underlying codes do not have independently addressable symmetries inherited from the stacked layer construction used in this example.

We emphasize the above example demonstrates two consistent approaches to gauging the input \BandC codes. The key difference lies in the choice of ancilla cluster state used in the gauging procedure.
For our homological gauging approach, the ancilla cluster state inherits the same structure as the input codes, whereas for our graph gauging approach, the ancilla cluster state is on a graph distinct from the underlying input codes.
As we will see, the difference in these two choices in ancilla gauge fields will be the key distinction in our approaches to gauging general qLDPC codes.

\subsection{Generalizing Gauging to qLDPC Codes}\label{sec:homological}
Having developed some intuition for the conceptual differences between our two gauging procedures, we describe in detail the process of gauging a qLDPC code.
In particular, we start by first introducing the homological formalism of quantum error-correction~\cite{Bombin_2007}.
This formalism applies to any  Calderbank–Shor–Steane (CSS) code \cite{CalderBankShor,Steane}---i.e. codes whose stabilizers are either entirely products of $X$ or entirely products of $Z$---and will be the language we need to extend the $2$D non-Abelian construction of Section~\ref{sec:prelim} to an enormous class of qLDPC codes with arbitrary connectivity. 
Using this language, we describe in detail both the symmetries and the protocol for gauging a qLDPC code.
We conclude by giving a qualitative overview on the two approaches to gauging a qLDPC code that we will explore in later parts of this section.

\subsubsection{Homological Formalism of qLDPC Codes}

As previously stated, CSS codes are specified by two sets of commuting $X$ and $Z$ stabilizers called \textit{parity checks}.
For instance, the toric code is a CSS code whose $X$-checks are identified with the vertices while $Z$-checks are identified with the plaquettes of the 2D square lattice.
The sets of $X$- and $Z$-type checks of CSS codes and their mutual commutativity is naturally captured by the algebraic structure of a \textit{chain complex}.

In particular, a chain complex $(D_\bullet,\partial_\bullet)$ is a sequence of vector spaces $D_n$ over $\mathbb{Z}_2$\footnote{Here, the restriction to $\mathbb{Z}_2$ is because we restrict our attention to codes defined for qubits.
For the case of qudit codes, $\mathbb{Z}_2$ is replaced with $\mathbb{Z}_n$.},
together with a set of linear boundary maps $\partial_n$:
 \begin{equation}
    0\longrightarrow \substack{\displaystyle D_2\\ \text{Z-checks}}\xrightarrow{\partial_2}\substack{\displaystyle D_1\\ \text{qubits}}\xrightarrow{\partial_1}\substack{\displaystyle D_0\\ \text{X-checks}}\rightarrow0
\end{equation}
which satisfy the characteristic condition that $\partial_n:D_n\rightarrow D_{n-1}$ such that $\partial_{n-1}\circ \partial_{n}=0$.
We henceforth refer to the elements of $a \in D_n$ as $n$-chains.
As we will see shortly, in a quantum code, the vector spaces $D_2$,  $D_1$, and $D_0$ are identified with the ``space'' of $Z$-checks, qubits, and $X$-checks respectively.
As a concrete example, in the toric code, $D_2$ is a vector space whose basis elements are individual plaquettes on the square lattice (or equivalently the location of the $Z$-stabilizers), $D_1$ is the space of edges (the locations of the qubits), and $D_0$ is the space of vertices (the location of the $X$-stabilizers).
The boundary maps in this case are the usual boundary maps; for example, a square plaquette is mapped to the four edges at its boundary.

Associated to any chain complex is a dual cochain complex which can be represented by the sequence:
\begin{equation} \label{eq-cochaincomplex}
    0\leftarrow D^2\xleftarrow{\exd_{1}}D^1\xleftarrow{\exd_{0}}D^0\leftarrow0
\end{equation}
where $D^n=\mathsf{Hom}_{\mathbb{Z}_2}(D_n,\mathbb{Z}_2)$ is the space of linear maps from $D_n$ to $\mathbb{Z}_2$. 
One can also think of $D^n$ as the space of row vectors associated with the column vectors in $D_n$, which map elements in $D_n$ to $\mathbb{Z}_2$ via the dot product.
We refer to elements $\mathbf{a} \in D^n$ as \textit{n-cochains} and denote them with boldface letters.
Moreover, their action on, or dot product with, a chain $b \in D_n$ is denoted as $\mathbf{a}(b)$.
Finally, the ``co-boundary maps'' above are defined as $\exd_n = \partial_{n + 1}^{\mathsf{T}}$ and naturally satisfy $\exd_{n + 1} \circ \exd_n = 0$.
As a concrete example of a co-boundary map, in the toric code, the co-boundary of a vertex is the four edges adjacent to this vertex.

Given a chain complex and its associated co-chain complex , we can specify a CSS code.
In particular, we associate $Z$ operators with $1$-chains in $D_1$ and $X$ operators with $1$-cochains in $D^1$ as follows.
More precisely, given a $Z_b$ operator associated with a $1$-chain $b \in D_1$, for every basis element ``edge'' $e$ that appears in $b$, we place a Pauli $Z$ operator on the lattice.
Similarly,given a $X_{\mathbf{a}}$ operator associated with a $1$-cochain $\mathbf{a} \in D^1$, for every basis element edge indicator function ``$\mathbf{e}$'' that appears in the $1$-cochain $\mathbf{a}$, we place a Pauli $X$ operator on the lattice.
Written out mathematically, we can identify products on $Z$ and $X$ operators on the lattice with these $1$-chains and co-chains as:
\begin{equation} \label{eq-ZXchainscochains}
    Z_{b \in D_1} \equiv \prod_{\mathbf{e}} Z_e^{\mathbf{e}(b)}
    \qquad
    X_{\mathbf{a} \in D^1} \equiv \prod_{e} X_{e}^{\mathbf{a}(e)}
\end{equation} 
where the product is performed over basis elements $\mathbf{e}$ of $D^1$ and basis elements $e$ of $D_1$.

The commutation relations between the operators above can be expressed succinctly as
\begin{equation}
Z_{b} X_{\mathbf{a}} = (-1)^{\mathbf{a}(b)} X_{\mathbf{a}} Z_{b}.    
\end{equation}
A key feature of Eq.~\eqref{eq-ZXchainscochains} is that \textit{sums} of (co-)chains become \textit{products} of $Z$ ($X$) operators.
Moreover, in this language, as stated before, the $Z$- and $X$- stabilizers are associated with \textit{boundaries} of $2$-chains, $\mathsf{Im}(\partial_2)$, and \textit{co-boundaries} of $0$-cochains, $\mathsf{Im}(\exd_0)$, respectively.
Consequently, the primitive stabilizers of a qLDPC code can be expressed as: 
\begin{equation} \label{eq-qLDPCstab}
    Z_{\partial_2 p}, \qquad X_{\exd_0 \mathbf{v}}
\end{equation}
where $p$ is a basis element in $D_2$ and $\mathbf{v}$ is a basis element of $D^0$.
We highlight first that the ``LDPC'' condition for the stabilizers becomes a statement that the maps $\partial_n$ (and consequently their transposes) are sparse matrices.
Moreover, the commutation of the stabilizers has a homological origin.
In particular: $Z_{\partial_2 p}X_{\exd_0 \mathbf{v}}Z_{\partial_2 p} X_{\exd_0 \mathbf{v}}  = (-1)^{\exd_0 \mathbf{v}(\partial_2 p)} = (-1)^{\mathbf{v}(\partial_1 \circ \partial_2 (p))} = 1$, with the second step followed from $\exd_0 = \partial_1^{\mathsf{T}}$\footnote{We can see that the transpose relation between the boundary and co-boundary map is a discrete form of generalized Stokes' theorem.}, and the last following from $\partial_1 \circ \partial_2 = 0$.

We also note that the logical operators of the qLDPC code can be easily understood in the homological formalism.
Indeed, recall that in the toric code, logical operators correspond to ``non-contractible'' loops on the lattice and its dual: loops with no (co-)boundary [dubbed  \textit{(co-)cycles}] that are not the (co-)boundary of anything.
Similarly, for a qLDPC code, the $Z$ ($X$) logicals of a code are (co-)cycle representatives in the elements of the (co-)homology groups $H_1$ ($H^1$) defined as: 
\begin{equation} H_{\alpha} = \frac{\mathsf{ker}(\partial_{\alpha})}{\mathsf{Im}(\partial_{\alpha + 1})}, \qquad H^{\alpha} = \frac{\mathsf{ker}(\exd_{\alpha})}{\mathsf{Im}(\exd_{\alpha-1})}
\end{equation}
With this, we conclude with a summary of the identification of objects in the chain complex and their interpretation in the quantum code in Table~\ref{table:dictionary}.

\begin{table}[h]
\centering
\setlength{\tabcolsep}{3pt}
\begin{tabular}{c c}
\toprule
\textbf{Chain Complex} & \textbf{Quantum Code} \\ 
\midrule
\text{1-chain} $e \in D_1$ & $Z$-\text{operator} $Z_{b} = \prod_{e} Z_{e}^{\mathbf{e}(b)}$ \\ \vspace{1 mm}
$\text{1-boundary } \partial_2 (p \in D_2)$ & $Z$-stabilizers \\ \vspace{1 mm}
$1$-cycle $e \in H_1$  & $Z$-logical \\
$2$-cycle $\mathcal{M} \in H_2$ & global $Z$-redundancy $Z_{\partial \mathcal{M}} = 1$\\
\midrule
\vspace{1 mm}
\text{1-cochain} $\mathbf{e} \in D^1$ & $X$-\text{operator} $X_{\mathbf{e}} = \prod_{e} X_{e}^{\mathbf{a}(e)}$ \\ \vspace{1 mm}
$\text{1-co-boundary }\exd_{0}(\mathbf{v} \in D_0)$ & $X$-stabilizers  \\ \vspace{1 mm}
 $1$-cocycle $\mathbf{e} \in H^1$ & $X$-logical\\ 
 $0$-cocycle $\boldsymbol{\mu} \in H^0$ & global $X$-redundancy $X_{\exd\boldsymbol{\mu}} = 1$\\
\bottomrule
\end{tabular}
\caption{Dictionary mapping between homological and quantum code description of CSS code.}\label{table:dictionary}
\end{table}
We note that in Table~\ref{table:dictionary}, we also provide interpretations for the homology group $H_2$ and $H^0$ in this table, which will play a special role in the construction of non-Abelian qLDPC codes.
We will discuss more about these groups in the coming subsections.

As a final remark, throughout this work, we will be predominantly interested in gauging level-$(2 + 1)$ chain complexes, where the highest $n$ for which $D_n$ is nonzero is $2$---though much of our formalism carries through exactly for higher chain complexes.
Physically, this means that there are no \textit{local redundancies} of the $X$ and $Z$ stabilizers such that products of $\mathcal{O}(1)$ $X$ or $Z$ stabilizers that equal $1$.
In Refs.~\cite{rakovszky2023physicsgoodldpccodes, rakovszky2024physicsgoodldpccodes, homologicalorder}, it is shown that many concepts which appear in $2+1$D topological codes can be generalized to level-$(2+1)$ chain complexes.
For instance, one can show that all excitations of codes described by level-$(2+1)$ complexes are point-like, similar to the point-like anyonic excitations of $2+1$D topological orders.

\subsubsection{Symmetries and Gauging in qLDPC  Codes}\label{sec:gen}

With this homological perspective on quantum codes in hand, we now turn to explicit constructions of gauged non-Abelian qLDPC codes. Motivated by the two dimensional examples of Sec.~\ref{sec:prelim} and our discussion of the two different possible types of gauging explored in Sec.~\ref{sec:toy}, we return to two key questions motivating our construction: 
\begin{enumerate}
    \item[1.] Given a qLDPC code, what symmetry, or equivalently what transversal gate, do we wish to gauge?

    \item[2.] What structure is required of the ancilla gauge fields in order to gauge a given symmetry?
\end{enumerate}

The answers to these questions are closely intertwined and will differ for our homological and graph gauging procedures.
For both questions, let us take as our starting setup two copies of a qLDPC code we again label as \BandC, whose structure is captured by a common chain complex $D_\bullet$:
 \begin{equation}
    0\longrightarrow D_2\xrightarrow{\partial_2}D_1\xrightarrow{\partial_1}D_0\rightarrow 0
\end{equation}
and an associated cochain complex of the form of Eq.~\eqref{eq-cochaincomplex}. 

Code in hand, we begin by discussing which of its symmetries we would like to gauge.
Let us start by noting that, in general, a qLDPC code can have a wide array of non-trivial symmetries.
First, every qLDPC code enjoys non-trivial \textbf{1-cycle symmetries} $Z_{\eta}$ and \textbf{1-cocycle symmetries} $X_{\boldsymbol{\gamma}}$, corresponding to the logical $Z$ and $X$ operators of the code.
Here, $\eta$ is a representative in an element of $H_1$ and $\boldsymbol{\gamma}$ is a representative in an element of $H^1$.
These symmetries generalize the $1$-form symmetries of the toric code's logical operators to the qLDPC case.
However, crucially, unlike conventional $1$-form symmetries, they need not be supported on $(D - 1)$-dimensional surfaces.
These are crucially \textit{not} the symmetries that we would like to gauge to produce a qLDPC code.

Instead, our work aims to gauge transversal Clifford gates of a qLDPC code.
These transversal Clifford gates need not leave the Hamiltonian invariant.
Rather, they act to send stabilizers to products of other stabilizers.
In this way, they can be thought of as generators of emergent symmetries of the code space.
Throughout, when we reference gauging our input \BandC codes, it is this class of transversal Clifford gates we are gauging.
Unlike the 1-co(cycle) symmetries discussed above, the generators of these transversal Clifford gates are rigid in that they have fixed support and may act globally (i.e. have a nontrivial action on every qubit in the input code) or may act on only a subset or \textit{subsystem} of the code qubits.

Symmetries in hand, our construction follows Ref.~\cite{Hierarchy} as reviewed in Sec.~\ref{sec:prelim}, adapted to the LDPC setting.
In this context, the second question above amounts to a choice of which cluster state of ancilla qubits and which symmetry enrichment operator $\Omega_{\aP{a},\textcolor{dodgerblue}{B}\textcolor{orange(ryb)}{C}}$ should be used to gauge a given symmetry. 
In both of our qLDPC constructions, the ancilla lattice of Sec.~\ref{sec:lattice2D} is replaced by an ancilla chain  complex $A_\bullet$:
 \begin{equation} \label{eq-Acomplex}
    0\longrightarrow A_2\xrightarrow{\partial^a_2}A_1\xrightarrow{\partial^a_1}A_0\rightarrow0
\end{equation}
and its associated cochain complex.
We identify the vertex ancilla qubits $\aP{a}$ with $A_0$ and the edge ancilla qubits $\textcolor{red}{A}$ with $A_1$. For the example of gauging a 2D toric code, the code and ancilla chain complexes were identical, i.e. $A_\bullet\sim D_\bullet$. However, as demonstrated by the example Sec.~\ref{sec:toy}, this is not strictly a requirement. Rather, it is the choice of which symmetry we are gauging on codes \BandC that determines the ancilla complex $A_\bullet$. 

The first step of the gauging procedure introduces a cluster state on the $\aP{a}$ and $\textcolor{red}{A}$ ancilla qubits, whose stabilizers can be specified in a homological language analogous to Eq.~\eqref{eq-qLDPCstab}.
In particular, the stabilizers are:
\begin{equation}\label{eq:LDPCcluster}
    \textcolor{goodpurple}{X^a_{\mathbf{v}}}\textcolor{red}{X^A_{\exd^a\mathbf{v}}},\qquad \textcolor{red}{Z^A_{e}}\textcolor{goodpurple}{Z^a_{\partial^a e}}
\end{equation}
The resulting cluster state will have $\textcolor{red}{\mathbb{Z}_2^{(1)}}$ ``1-cycle symmetries'' acting on edge qubits $\textcolor{red}{A}$ and $\textcolor{goodpurple}{\mathbb{Z}_2^{(0)}}$ ``0-cocycle subsystem symmetries'' acting on the vertex qubits $\textcolor{goodpurple}{a}$.
The 1-cycle symmetries $G_{\AR{A}}$ are the $Z$ operators along cycles: $G_{\AR{A}}=\AR{Z_\eta}=\prod_{\mathbf{e}}\AR{Z_e}^{\mathbf{e}(\eta)}$ where $\eta\in\mathsf{ker}(\partial^a_1)$ is a non-trivial generator of $H_1$. 
Likewise, the 0-cocycle symmetries $G_{\aP{a}}$ are the $X$ operators along cocycles: $G_{\aP{a}}=\aP{X_{\bm{\mu}}}=\prod_{v}\aP{X_{\mathbf{v}}}^{\boldsymbol{\mu}(v)}$ where $\bm{\mu}\in\mathsf{ker}(\exd^a_0)$ is a nontrivial representative in the generator of $H^0$ (see Table~\ref{table:dictionary}). 
If we project $\aP{X_{\mathbf{v}}^a} = +1$, the stabilizers of Eq.~\eqref{eq:LDPCcluster} become stabilizers of a code defined only on the $\AR{A}$ sublattice. 
In this case, the product of the first stabilizer in Eq.~\eqref{eq:LDPCcluster} over any nontrivial element $\bm{\mu}$ will result in $\textcolor{red}{X^A_{\exd^a\bm{\mu}}}=1$. 
Hence, we can also think of $\bm{\mu}$ as labeling redundancies on the resulting $\AR{A}$ stabilizers.

As a more concrete example, in the 2D square lattice toric code, there is a single generator $\bm{\mu}$ corresponding to the product over all vertex stabilizers. For our stacked toric code example of Sec.~\ref{sec:toy}, the product over all stabilizers in each layer of toric codes is the identity, i.e., each layer is associated with a distinct generator $\bm{\mu}$.
This motivates the naming of $\aP{X_{\bm \mu}}$  as $0$-cocycle \textit{subsystem} symmetry since elements of $\bm{\mu}$ specify subsets of stabilizers on the $\AR{A}$ sublattice which will product to 1 after qubits in the $\aP{a}$ sublattice is measured out.

These redundancies play a key role in defining the gauging map. In the general LDPC case, the symmetry enrichment condition that the 0-cocycle symmetries $\aP{X_{\bm{\mu}}}$ are tied to a global symmetry action on the code becomes:
\begin{equation}
    \Omega_{\aP{a},\textcolor{dodgerblue}{B}\textcolor{orange(ryb)}{C}} \aP{X_{\bm{\mu}} }\Omega^{\dagger}_{\aP{a},\textcolor{dodgerblue}{B}\textcolor{orange(ryb)}{C}} = \aP{X_{\bm{\mu}}} U_{\boldsymbol{\mu}}
\end{equation}
where $U_{\boldsymbol{\mu}}$ denotes a transversal Clifford gate between the \BandC code that we wish to gauge. Note that this requirement must be met for \textit{every} element $\bm{\mu}$.
Further note that $U_{\boldsymbol{\mu}}$ may or may not be distinct for each $\boldsymbol{\mu}$.
However, for the gauging procedure to be non-trivial, at least one $U_{\boldsymbol{\mu}}$ must be a non-trivial transversal Clifford gate as we will later see.

We can distill the symmetry enrichment condition into a requirement on the allowed structure of our ancilla complex; given a specific symmetry $U$ of our input \BandC codes that we wish to gauge, we require that the cluster state on the ancilla chain complex satisfies the above symmetry enrichment constraint. However, we emphasize this does $\textit{not}$ require that the ancilla complex be a copy of the original chain complex (though this may be a natural choice). In fact, given a transversal gate (or set of gates) $U$, we can gauge this gate with \textit{any} ancilla chain complex with subsystem symmetries $\bm{\mu}\in\mathsf{ker}(\exd_0^a)$ such that the symmetry enrichment condition is satisified for all $\bm{\mu}$.
After satisfying the symmetry-enrichment condition, the gauging procedure concludes by projecting each of the $\aP{a}$ qubits into the $\ket{+}$ state.
In Section~\ref{subsec:measurement_prep}, we demonstrate that this projection can be performed \textit{deterministically} using measurement and a classically conditioned feedforward operation.
Subsequently, we will present gauging procedures for two natural (but not exhaustive) choices of ancilla complex.

\subsubsection{Two Approaches to Gauging qLDPC Codes}
Our homological gauging procedure requires that the chain complex underlying codes \BandC have a notion of Poincar\'e duality, i.e. a local isomorphism between $H^2$ and $H_0$.
With this method of gauging, the ancilla complex is chosen to be a copy of the original chain complex of the input codes, i.e $D_\bullet\simeq A_\bullet$.
We will find that the various subsystem symmetries on the ancilla complex labeled by $\bm{\mu}\in\mathsf{ker}(\exd_0^a)$ can be associated to collections of transversal gates acting on the input codes.

In our graph gauging procedure, we will always choose the ancilla complex $D_\bullet$ to be a graph, such that it has a single global 0-cocycle symmetry acting on its vertices.
This procedure will be applicable whenever there is a single transversal gate we wish to gauge (for instance, the transversal $\mathsf{SWAP}$ gate that is always present between two copies \BandC of the same code).
Unlike our homological gauging procedure, there is no need for Poincar\'e duality.
The choice of graph will be dependent on the underlying chain complex in this case.

\subsection{Non-Abelian Codes from Poincar\'e Duality}\label{sec:addressable}

In this section, we detail the homological approach to gauging alluded to earlier. 
In particular, this approach to gauging applies to CSS codes whose chain complexes have an algebraic structure analogous to Poincar\'e duality in manifolds.
We will find that this duality ensures that the structure of the codes' global redundancies $\boldsymbol{\mu} \in H^0$ can be identified with transversal $\mathsf{CZ}$ gates that exist between two copies of the code.
Consequently, the cluster state used for gauging can be described by a chain complex that is identical to the chain complex for the two qLDPC codes.

In what follows, we first review Poincar\'e duality and the associated mathematical concept of a cup product in its usual setting (i.e. chain complexes associated with cellulations of manifolds).
Subsequently, we discuss how these concepts have been extended to the general qLDPC setting and furthermore, how they imply the existence of transversal $\mathsf{CZ}$ gates between two copies of qLDPC codes.
We conclude by showing how to gauge these symmetries.
The result is a simple homological form for a non-Abelian qLDPC code.

\subsubsection{Poincar\'e Duality in the Toric Code}

Prior to discussing Poincar\'e duality for general qLDPC chain complexes, we first discuss it for chain complexes associated with cellulations of 2D manifolds, specifically the toric code~\cite{Chen_2023, hatcher2002algebraic}.
Recall that, in the toric code, for every $X$ logical there is a $Z$ logical that can be chosen to ``run parallel'' to it.
Specifically, note that for every $X$ logical operator (element of the cohomology group $H^1$) wrapping a around handle of the torus, there is a $Z$ logical operator (element of the homology group $H_1$) that wraps around that same handle, e.g.
\begin{equation} \label{eq-logicalpairing}
    \begin{tikzpicture}[scale = 0.8, baseline = {([yshift=-.5ex]current bounding box.center)}]
        \foreach \i in {0, 1, 2}{
            \draw[color = lightgray] (\i, -0.5) -- (\i, 1.5);}
        \foreach \i in {0, 1}{
            \draw[color = lightgray] (-0.5, \i) -- (2.5, \i);}
        \draw[color = gray, line width = 1 pt, dashed] (-0.5, 0.5) -- (2.5, 0.5);
        \draw[color = gray, line width = 1 pt, dashed] (-0.5, 0.) -- (2.5, 0.);
        \foreach \i in {0, 1, 2}{
            \node at (\i, 0.5) {${X}$};
            \node at (\i - 0.5, 0) {${Z}$};
        }
    \end{tikzpicture}
    \qquad     \begin{tikzpicture}[scale = 0.8, baseline = {([yshift=-.5ex]current bounding box.center)}]
        \foreach \i in {0, 1, 2}{
            \draw[color = lightgray] (\i, -0.5) -- (\i, 1.5);
        }
        \foreach \i in {0, 1}{
            \draw[color = lightgray] (-0.5, \i) -- (2.5, \i); 
        }
        \draw[color = gray, line width = 1pt, dashed] (0, -0.5) --  (0, 1.5);
        \draw[color = gray, line width = 1pt, dashed] (0.5, -0.5) --  (0.5, 1.5);
        \foreach \i in {0, 1}{
            \draw[color = lightgray] (-0.5, \i) -- (2.5, \i);
            \node at (0.5, \i) {${X}$};
            \node at (0, \i - 0.5) {${Z}$};
        }
    \end{tikzpicture}
\end{equation}
This association between $X$ and $Z$ logical operators can be described as an isomorphism between logical operators known as \textit{Poincar\'e duality}.

To be precise, Poincar\'e duality for a level $(2 + 1)$-chain complex is an isomorphism between the cohomology group $H^\alpha$ and homology group $H_{2 - \alpha}$ built from two ingredients: the \textit{cup product} and the \textit{fundamental class}.
The cup product is an LDPC local,  bilinear cochain product $\cupp: C^p \times C^q \to C^{p + q}$.
For the square lattice, it is operationally defined using the following rules~\cite{Chen_2023}:
\vspace{-3 mm}
\begin{equation}
    \vspace{-3 mm}
    \begin{tikzpicture}[scale = 0.85, baseline = {([yshift=-.5ex]current bounding box.center)}]
    \foreach \i in {1.5}{
    \filldraw[fill = gray, draw = none, opacity = 0.1] (0 + \i,0) -- (0.75 + \i,0) -- (0.75 + \i, -0.75) -- (0 + \i, -0.75) -- cycle;
    \draw[color = black, line width = 0.8pt](0 + \i, -0.75) -- (0 + \i,0);
    \draw[color = black, line width = 0.8pt](0 + \i, 0) -- (0.75 + \i,0);
    \node at (-0.15 + \i, -0.375) {${\mathbf{e}}$};
    \node at (0.375 + \i, 0.18) {${\mathbf{e}'}$};
    \node at (0.375 + \i, -0.75 - 0.19) {$\textcolor{white}{\mathbf{e}'}$};
    \node at (0.375 + \i, -0.375) {\textcolor{red}{$\mathbf{p}$}};
    }
    \foreach \i in {-0.4}{
    \filldraw[fill = gray, draw = none, opacity = 0.1] (0 + \i,0) -- (0.75 + \i,0) -- (0.75 + \i, -0.75) -- (0 + \i, -0.75) -- cycle;
    \draw[color= black, line width = 0.8pt](0.75 + \i, -0.75) -- (0.75 + \i,0);
    \draw[color = black, line width = 0.8pt](0 + \i ,-0.75) -- (0.75 + \i,-0.75);
    \node at (0.94 + \i, -0.375) {${\mathbf{e}'}$};
    \node at (0.375 + \i, -0.75 - 0.19) {${\mathbf{e}}$};
    \node at (0.375 + \i, -0.375) {\textcolor{red}{$\mathbf{p}$}};
    \node at (0.375 + \i, -0.375) {\textcolor{red}{$\mathbf{p}$}};
    \node at (0.375 + \i, 0.19) {$\textcolor{white}{\mathbf{e}'}$};
    };

    \node at (1, 0.65) {\small $\underline{{\mathbf{e}} \cupp {\mathbf{e}'} = \textcolor{red}{\mathbf{p}}}$};
    \foreach \i in {3.5}{
    \filldraw[fill = gray, draw = none, opacity = 0.1] (0 + \i,0) -- (0.75 + \i,0) -- (0.75 + \i, -0.75) -- (0 + \i, -0.75) -- cycle;
    \filldraw[fill = black] (0 + \i, -0.75) circle (1.5pt);
    \node at (-0.2 + \i, -0.65) {$\mathbf{v}$};
    \node at (0.375 + \i, -0.375) {\textcolor{red}{$\mathbf{p}$}};
    \node at (0.325 + \i, 0.64) {\small $\underline{\mathbf{v} \cupp \textcolor{red}{\mathbf{p}} = \textcolor{red}{\mathbf{p}}}$};
    };

    \foreach \i in {6}{
    \filldraw[fill = gray, draw = none, opacity = 0.1] (0 + \i,0) -- (0.75 + \i,0) -- (0.75 + \i, -0.75) -- (0 + \i, -0.75) -- cycle;
    \filldraw[fill = black] (0.75 + \i,0) circle (1.5pt);
    \node at (1.0 + \i, 0) {$\mathbf{v}$};
    \node at (0.375 + \i, -0.375) {\textcolor{red}{$\mathbf{p}$}};
    \node at (0.325 + \i, 0.64) {\small $\underline{\textcolor{red}{\mathbf{p}} \cupp \mathbf{v} = \textcolor{red}{\mathbf{p}}}$};
    };
    \end{tikzpicture}
    \label{eq-cupproduct}
\end{equation}
 Moreover,  for two vertices $\mathbf{v} $ and $\mathbf{w},$ $\mathbf{v} \cupp \mathbf{w} = \mathbf{v} \delta_{\mathbf{v}, \mathbf{w}}$. For a vertex and an edge, we have $\mathbf{v} \cupp \mathbf{e} = \mathbf{e}$ if the edge is adjacent to $\mathbf{v}$ and to the right or above it and $\mathbf{e} \cupp \mathbf{v} = \mathbf{e}$ if the edge is adjacent to $\mathbf{v}$ and to the left or below it. 
Outside of these cases, the cup product is zero.
Note that the edges that are paired by the cup product reflect the edges of logicals that are associated to each other in Eq.~\eqref{eq-logicalpairing}.

The cup product is a \textit{cohomology} operation in the sense that if $\mathbf{a}_p$ and $\mathbf{b}_q$ are cocycle representatives of some elements of the cohomology groups $H^p$ and $H^q$ then $\mathbf{a}_p \cupp \mathbf{b}_q$ will be a cocycle representative of $H^{p + q}$.
This follows from the fact that the cup product satisfies the \textit{Leibniz rule}: 
\begin{equation}
    \exd ( \mathbf{a}_p \cupp \mathbf{b}_q) = \exd \mathbf{a}_p \cupp \mathbf{b}_q  + \mathbf{a}_p \cupp \exd \mathbf{b}_q
    \label{eq-Leibniz}
\end{equation}
where $\mathbf{a}_p \in C^p$ and $\mathbf{b}_q \in C^q$ are $\mathbb{Z}_2$-valued cochains.

Cup product in hand, our goal is to use this product to map cocycles to cycles---i.e. $X$ to $Z$ logicals.
To do so, we require an element of homology group $H_2\simeq \mathsf{ker}(\partial_2)/\mathsf{Im}(\partial_3)$ called the \textit{fundamental class} $\mathcal{M}$.
For chain complexes with coefficients in $\mathbb{F}_2$ describing cellulations of 2D orientable manifolds, the fundamental class is the unique generator of $H_2 \simeq \mathbb{Z}_2$.
In a level-$(2 + 1)$ complex, $\partial_3 = 0$ and hence $H_2 \simeq \mathsf{ker}(\partial_2)$---i.e. the set of plaquettes whose boundary is zero.
Equivalently, $H_2$ describes the sets of $Z$-stabilizers whose product is the identity.
Consequently, the fundamental class is identified with the \textit{global redundancies} of the $Z$-stabilizers [c.f. Table~\ref{table:dictionary}].
This is to be contrasted with global redundancies of the $X$-stabilizers, which we say in Section~\ref{sec:gen} are identified with elements $\boldsymbol{\mu}$ of the cohomology group $H^0$~\footnote{Though, under Poincar\'e duality, these redundancies map into one another.}.

In the toric code, the only global $Z$ redundancy is the product of all stabilizers is the identity.
Consequently, $\mathcal{M}$ is the sum of all plaquettes on the lattice:
\begin{equation} \label{eq-Mtoriccode}
    \mathcal{M} = \left(\sum_{p} p \right) \in \mathsf{ker}(\partial_2) \simeq H_2
\end{equation}
where the sum over $p$ is a sum over the basis for $D_2$.
Using the cup product and the fundamental class, the Poincar\'e dual of a cochain $\mathbf{a}_p$ is then defined via an operation called the \textit{cap product} \cite{hatcher2002algebraic}: 
\begin{equation}
    \mathbf{a}_p \capp \mathcal{M} \equiv \sum_{x_{2 - p} } \left( \int_{\mathcal{M}} \mathbf{x}_{2 - p} \cupp \mathbf{a}_p \right) x_{2 -p}
    \label{eq-capproduct}
\end{equation}
where the sum over $x_{2-p}$ is implicitly a sum over the basis for $D_{2 - p}$ and $\mathbf{x}_{2 - p}$ is the co-chain associated to $x_{2 - p}$ by transposition.
Moreover, $\int_{\mathcal{M}} \mathbf{p}$ means evaluate the cochain $\mathbf{p}$ on $\mathcal{M}$ and $\mathcal{M} \capp \mathbf{a}_p$ is defined similarly to the above but with the integrand changed to $\mathbf{a}_p \cupp \mathbf{x}$.
In words, given a cochain $\mathbf{a}_p$, the cap product builds a chain $\mathbf{a}_p \capp \mathcal{M}$ by summing over indicator cochains $\mathbf{x}_{2 - p}$ and checking if $\mathbf{x}_{2-p} \cupp \mathbf{a}_p$ has support on $\mathcal{M}$; if so, $x_{2 - p}$ is included in the chain $\mathbf{a}_{p} \capp \mathcal{M}$. 
Crucially, from the Leibniz rule:
\begin{equation}
 \partial \left(\mathbf{a}_p \capp \mathcal{M}\right) = \left( \exd \mathbf{a}_p \capp \mathcal{M} \right)  
\end{equation}
and consequently if $\mathbf{a}_p$ is a cocycle, $\mathbf{a}_p \capp \mathcal{M}$ is a cycle, as expected.

\subsubsection{Poincar\'e Duality in qLDPC Codes}\label{Sec:PD_LDPC}

In the qLDPC setting, per our discussion in the last section, we know that the concept of a ``fundamental class'' readily generalizes to the global redundancies of a qLDPC code's $Z$ stabilizers, described by $H_2$---though, unlike the toric code, $H_2$ need not have a single generator.
However, it is not generically possible to define a cup product for arbitrary qLDPC codes in which case there is no appropriate notion of Poincar\'e duality.
Nevertheless, a flurry of recent work has shown that for a large variety of qLDPC codes---including ``good'' qLDPC codes and several codes of interest---generalizations of these notions can still be defined \cite{breuckmann2025cupsgatesicohomology, lin2024transversalnoncliffordgatesquantum,zhu2025topological, menon2025magic, li_poincare_2025}. 
In particular, the cup products present in literature typically take one of two forms.
The first form follows from generalizing the cup product on manifolds to the mathematical structure of a sheaf.
In doing so, one is able to define a cup product on so-called ``sheaf codes'' \cite{first2024good2querylocallytestable, panteleev2024maximallyextendablesheafcodes}, which encompasses all recent constructions of good qLDPC codes \cite{lin2024transversalnoncliffordgatesquantum, li_poincare_2025}.
The second form (i.e. the ``cups and gates'' formalism) defines a cup product by mimicking the local rules of Eq.~\eqref{eq-cupproduct}.
In doing so, the cup product no longer necessarily satisfies the Leibniz rule of Eq.~\eqref{eq-Leibniz}.
Instead, these cup products satisfy the so-called ``$\Lambda$-fold integrated Leibniz rule'': 
\begin{equation}
    \sum_{k = 1}^{\Lambda} \int_{\mathcal{M}} \mathbf{a}_1 \cupp\left( \cdots \cupp (\exd \mathbf{a}_k \cupp ( \cdots  \mathbf{a}_{\Lambda} ))\right) = 0 
\end{equation}
for a fixed $\mathcal{M} \in H_2$---which in Refs.~\cite{breuckmann2025cupsgatesicohomology, menon2025magic} was assumed to be a uniform sum over all basis elements of $D_2$----and all $p_k$-cochains $\mathbf{a}_k$ such that $\sum_k p_k = 2$.
Such a cup product has been shown to apply to a wide variety of quantum codes including  bivariate-bicycle codes \cite{Bravyi_2024}, Sipser-Spielman codes \cite{SipserSpiel}, and fracton models.

We further comment that an additional general form of the cup product will appear in future work \cite{addressablegates}, an example of which will be shown in Section~\ref{subsec:addressable-gauging}.
For our purposes, any of these constructions are sufficient, with $\Lambda \geq 3$ required from the second for our non-Abelian codes.
As an added technical remark, if a cup product satisfies the $\Lambda \geq 3$ integrated Leibniz rule for a fundamental class $\mathcal{M}$, then it is easy to see that the cup product at least satisfies the $\Lambda -1$ integrated Leibniz rule for any $\mathcal{M}(\boldsymbol{\mu})$ of the following form: 
\begin{equation}
    \mathcal{M}(\boldsymbol{\mu}) = \mathcal{M} \capp \boldsymbol{\mu}
\end{equation}
where $\boldsymbol{\mu} \in H^0$.
This simply follows from the fact that $\exd \boldsymbol{\mu} = 0$ and the fact that $\int_{\mathcal{M}} \boldsymbol{\mu} \cupp \mathbf{A} = \int_{\mathcal{M} \capp \boldsymbol{\mu}} \mathbf{A}$ for any $2$-cochain $\mathbf{A}$.

Given a qLDPC code with a cup product satisfying the Leibniz rule (or $\Lambda \geq 2$ integrated Leibniz rule), one can naturally define a transversal $\mathsf{CZ}$ gate for such a code.
In particular, the transversal $\mathsf{CZ}$ gates take the form: 
\begin{equation} \label{eq-cupCZsymmetry}
        U_{\mathsf{CZ}}[\mathcal{M}] = \prod_{e,e'} \left(\mathsf{CZ}_{\bB{e}, \oC{e'}}\right)^{\int_{\mathcal{M}} \mathbf{e} \cupp \mathbf{e}'}
\end{equation}
Indeed, for toric code, given the cup product and fundamental class of Eqs.~\eqref{eq-cupproduct}~and~\eqref{eq-Mtoriccode}, we recover the $\mathsf{CZ}$ gate shown in Eq.~\eqref{eq:CZ} of Section~\ref{sec:prelim}.
While the fundamental class is unique in the toric code---and consequently it has one logical $\mathsf{CZ}$ gate---for a generic qLDPC code, it need not be.
This leads to several independent transversal $\mathsf{CZ}$ gate each labeled by $\mathcal{M}$.

\subsubsection{Homological Expression for non-Abelian $D_4$ Codes}\label{subsubsec:d4-homological}

Equipped with the cup product, we now describe our first construction of non-Abelian qLDPC codes.
In particular, let us consider two copies of some qLDPC codes labeled by $\bB{B}$ and $\oC{C}$ that are described by a cochain complex: 
\begin{equation} \label{eq-3B3chaincomplex}
    0\leftarrow D^2\xleftarrow{\exd_{1}}D^1\xleftarrow{\exd_{0}}D^0\leftarrow0
\end{equation}
and whose stabilizers are given by:
\begin{equation}
    \bB{X^B_{\exd \mathbf{v}}},  \quad  \bB{Z^B_{\partial p}},\quad \oC{X^C_{\exd \mathbf{v}}}, \quad \oC{Z^C_{\partial p}}.  
\end{equation}
Let us suppose further that the qLDPC code is equipped with a cup product $\cupp$ that satisfies the Leibniz rule.

Our goal now is to gauge the transversal $\mathsf{CZ}$ gate of Eq.~\eqref{eq-cupCZsymmetry}.
Per our previous discussion, doing so requires (1) a choice of ancillary ``cluster'' state and (2) the appropriate symmetry enrichment operator.
We now show that, for codes equipped with a cup product, there is a natural choice for both of these.
In particular, we will choose our ancillary cluster state to be the one associated with the chain complex of Eq.~\eqref{eq-3B3chaincomplex}.
In particular, it will have $\aP{a}$ qubits living on basis elements of $D_0$ and $\AR{A}$ qubits living on basis elements of $D_1$.
Moreover, its stabilizers will be given by: 
\begin{equation}
    \aP{X^a_{\mathbf{v}}} \AR{X^A_{\exd \mathbf{v}}}, \qquad \AR{Z^A_{e}} \aP{Z^a_{\partial e}}
\end{equation}
Such a state is invariant under a ``0-cocycle symmetry'' given by $H^0$, generated by $\aP{X^a_{\boldsymbol{\mu}}}$ for a $0$-cocycle $\boldsymbol{\mu} \in H^0$, and a $1$-cycle symmetry $H_1$, generated by $\AR{Z^A_{\gamma}}$ for $\gamma \in H_1$ [c.f. Section~\ref{sec:gen} for a discussion of $0$-cocycle and $1$-cycle symmetries].

With this choice of cluster state, we now must choose a symmetry enrichment operator satisfying the requirement that each 0-cocycle symmetry of our cluster state is tied to a symmetry of the $\bB{B}$ and $\oC{C}$ codes that we are gauging.
For codes with a cup product, a natural choice of this symmetry enrichment operator is given by: 
\begin{equation} \label{eq-CCZSEop}
\Omega_{\aP{a},\textcolor{dodgerblue}{B}\textcolor{orange(ryb)}{C}}=\prod_{v_0,e,e'} (\mathsf{CCZ}_{\aP{v}, \bB{e}, \oC{e'}})^{\int_{\mathcal{M}}\mathbf{v}\cupp (\mathbf{e}\cupp\mathbf{e'})}.
\end{equation}
To see the above is a valid symmetry enrichment operator, we can check that $\aP{X^a_{\boldsymbol{\mu}}}\Omega_{\aP{a}, \bB{B}\oC{C}} \aP{X^a_{\boldsymbol{\mu}}}\Omega_{\aP{a}, \bB{B}\oC{C}}^{\dagger}$ is a symmetry of the $\bB{B}$ and $\oC{C}$ codes.
We can compute explicitly that:
\begin{equation}
\aP{X^a_{\boldsymbol{\mu}}}\Omega_{\aP{a}, \bB{B}\oC{C}} \aP{X^a_{\boldsymbol{\mu}}}\Omega_{\aP{a}, \bB{B}\oC{C}}^{\dagger} = \prod_{e,e'} (\mathsf{CZ}_{\bB{e}, \oC{e'}})^{\int_{\mathcal{M}}\boldsymbol{\mu}\cupp (\mathbf{e}\cupp\mathbf{e'})}
\end{equation}
To see that the right hand side above is indeed a symmetry, let us note that: 
\begin{equation}
    \int_{\mathcal{M}} \boldsymbol{\mu} \cupp ( \mathbf{e} \cupp \mathbf{e}') = \int_{\mathcal{M} \capp \boldsymbol{\mu}}  \mathbf{e} \cupp \mathbf{e}' = \int_{\mathcal{M}'} \mathbf{e} \cupp \mathbf{e}'
\end{equation}
where in the first equality, we used the definition of the cap product [Eq.~\eqref{eq-capproduct}] and in the second equality we used  the fact that, since the cap product implements Poincar\'e duality, for $\boldsymbol{\mu} \in H^0$, $\mathcal{M} \capp \boldsymbol{\mu} \in H_2$.
We can therefore re-write: 
\begin{equation}
\aP{X_{\boldsymbol{\mu}}}\Omega_{\aP{a}, \bB{B}\oC{C}} \aP{X_{\boldsymbol{\mu}}}\Omega_{\aP{a}, \bB{B}\oC{C}}^{\dagger} = U_{\mathsf{CZ}}[\mathcal{M} \capp \boldsymbol{\mu}]
\end{equation}
which is indeed a symmetry of the $\bB{B}$ and $\oC{C}$ qLDPC codes, cementing $\Omega$ as a valid symmetry enrichment operator.
The symmetry enrichment operator further pre-empts that the symmetries of the $\bB{B}$ and $\oC{C}$ codes that are gauged by this procedure are precisely $U_{\mathsf{CZ}}[\mathcal{M} \capp \boldsymbol{\mu}]$ for all $\boldsymbol{\mu} \in H^0$.

Given both the cluster state and the symmetry enrichment operator, we can freely carry out the gauging procedure to arrive at the final stabilizers of the non-Abelian qLDPC code (see Appendix~\ref{D4-cup-derivation} for detailed derivation).
We find that the off-diagonal stabilizers take the following form:
\begin{align}
&\AR{\mathcal{A}^{A}_{\mathbf{v}}}=\AR{X_{\exd\mathbf{v}}^A}\prod_{e,e'} 
\left(\mathsf{CZ}_{\bB{e},\oC{e'}}\right)^{\int_{\mathcal{M}}\mathbf{v}\cupp (\mathbf{e}\cupp\mathbf{e'})}
\label{eq-homologicalA}\\
&\bB{\mathcal{A}^{B}_{\mathbf{v}}}=\bB{X_{\exd\mathbf{v}}^B}\prod_{e,e'} 
\left(\mathsf{CZ}_{\AR{e},\oC{e'}}\right)^{\int_{\mathcal{M}}\mathbf{e}\cupp (\mathbf{v}\cupp\mathbf{e'})} 
\label{eq-homologicalB}\\
&\oC{\mathcal{A}^{C}_{\mathbf{v}}}=\oC{X_{\exd\mathbf{v}}^C}\prod_{e,e'} 
\left(\mathsf{CZ}_{\bB{e},\AR{e'}}\right)^{\int_{\mathcal{M}}\mathbf{e'}\cupp (\mathbf{e}\cupp\mathbf{v})}
\label{eq-homologicalC}
\end{align}
and the diagonal stabilizers of the system are: 
\begin{equation}
     \AR{Z_{\partial p}^A}, \qquad \bB{Z_{\partial p}^{B}}, \qquad \oC{Z_{\partial p}^C}
\end{equation}
as expected.

At this point, a few remarks are in order.
First, we remark that, using the cup product for the toric code Eq.~\eqref{eq-cupproduct}, one can show the above formulas reproduce the stabilizers of the $\mathcal{D}(D_4)$ topological order in Section~\ref{sec:prelim}.
Indeed, the above homological form above closely mirrors the form of the ``magic stabilizer codes'' derived for Euclidean manifolds in Ref.~\cite{hsin_non-abelian_2025}, with the differences being that in the manifold case, the fundamental class is unique, while here it can be chosen arbitrarily in $H_2$.
As a broader point, since the construction of these codes depend only on a single chain complex, these codes represent examples of homological qLDPC codes beyond CSS codes.
Second, the gauging procedure of we outline is only capable of simultaneously gauging all symmetries of the form $U_{\mathsf{CZ}}[\mathcal{M} \capp \boldsymbol{\mu}]$.
Finally, we note that the homological form of the stabilizers above closely resembles the partition function of the $(2 + 1)$D $\mathbb{Z}_2^3$ (``type-III'') twisted quantum double, whose action is $\propto \int \mathbf{a}_1 \cupp (\mathbf{b}_1 \cupp \mathbf{c}_1)$ for $1$-co-chain gauge fields $\mathbf{a}_1, \mathbf{b}_1, \mathbf{c}_1$.
Indeed, when restricting this spacetime action to a single spatial slice, one would naturally expect expressions of the form $\int \mathbf{a}_0 \cupp (\mathbf{b}_1 \cupp \mathbf{c}_1), \int \mathbf{a}_1 \cupp (\mathbf{b}_0 \cupp \mathbf{c}_1),\text{ and } \int \mathbf{a}_1 \cupp (\mathbf{b}_1 \cupp \mathbf{c}_0)$ where now $\mathbf{a}_0, \mathbf{b}_0, \mathbf{c}_0$ are $0$-cochains.
These are precisely the cup product expressions that appear in our non-Abelian $X$-stabilizers.

\subsubsection{Homological Gauging of transversal $\mathsf{SWAP}$ Gates}

Having established the homological gauging procedure for the example of gauging transversal $\mathsf{CZ}$ gate, we will perform an analogous procedure for the case of gauging transversal $\mathsf{SWAP}$ gate. Similar to the $\mathsf{CZ}$ gate case, we will see that each homologically distinct $\bm{\mu}\in\text{ker}(\text{d}_0)$ is associated to a symmetry of the codespace. Gauging these $\mathsf{SWAP}$ gates will yield a non-Abelian qLDPC code. 

Our starting point is again two qLDPC codes \BandC along with ancilla codes $\AR{A}$ and $\aP{a}$ with the same structure as codes \BandC (i.e. $A_\bullet\sim D_\bullet$). We assume the chain complex underlying codes \BandC has a cup product and Poincar\'e duality as defined in Sec.~\ref{Sec:PD_LDPC}.
We further make the stronger assumption that there exists a $\mathcal{M}_0 \in H_2$ such that the function:
\begin{equation}
    \mathsf{PD}_{\mathcal{M}_0}: D^p \to D_{2 - p} \qquad \mathsf{PD}_{\mathcal{M}_0}(\mathbf{a}) \equiv \mathbf{a} \capp \mathcal{M}_0
\end{equation}
is LDPC and a bijective function.
We claim the following family of unitary operators defined for any $\bm{\mu}\in\text{ker}(\exd^a_0)$ acts as a symmetry on the combined codespace of the \BandC codes:
\begin{equation}
        U_{\mathsf{SWAP}}=\prod_{e,e'}\left(\textcolor{dodgerblue}{X^{B}_{\mathbf{e}}}\textcolor{orange(ryb)}{X^{C}_{\mathbf{e}}}\right)^{\left(\textcolor{dodgerblue}{n^{B}_{e'}}+\textcolor{orange(ryb)}{n^{C}_{e'}}\right)\int_{\mathcal{M}}\bm{\mu}\cupp( \mathsf{PD}_{\mathcal{M}_0}^{-1}(e)\cupp\mathbf{e'})}
\end{equation}
for $\mathcal{M}_0 \in H_2$.
To see this is true, let us assume the conditions specified above hold and take $\bm{\mu}\in\text{ker}(\exd_0)$. $U_{\mathsf{SWAP}}$ will leave the ground state subspace invariant if and only if $U_{\mathsf{SWAP}}$ commutes with all stabilizers of the \textcolor{dodgerblue}{B} and \textcolor{orange(ryb)}{C} codes (up to products of additional stabilizers which do not take us out of the ground state manifold). Therefore, we can check $U_{\mathsf{SWAP}}$ is a symmetry by computing the group commutators (defined $\left[V,U\right]=V^{-1}U^{-1}VU$):
\begin{align}\label{eq:com1}
\left[\textcolor{dodgerblue}{X^B_{d\mathbf{v}}},U_{\mathsf{SWAP}}\right]&=\prod_{e}\left(\textcolor{dodgerblue}{X^{B}_{\mathbf{e}}}\textcolor{orange(ryb)}{X^{C}_{\mathbf{e}}}\right)^{\int_{\mathcal{M}}\bm{\mu}\cupp (\mathsf{PD}^{-1}_{\mathcal{M}_0}(e)\cupp d\mathbf{v})}  \\
\left[\textcolor{dodgerblue}{Z^B_{\partial p}},U_{\mathsf{SWAP}}\right]&=\prod_{e}\left(\textcolor{dodgerblue}{Z^{B}_{e}}\textcolor{orange(ryb)}{Z^{C}_{e}}\right)^{\int_{\mathcal{M}}\bm{\mu}\cupp (\mathsf{PD}^{-1}_{\mathcal{M}_0}(\partial p)\cupp \mathbf{e})} \label{eq:com2}
\end{align}
The resulting commutators commute exactly with all stabilizers of the original \textcolor{dodgerblue}{B} and \textcolor{orange(ryb)}{C} codes:
    \begin{equation}
    \begin{split}
       & \left[\prod_{e}\left(\textcolor{dodgerblue}{Z^{B}_{e}}\textcolor{orange(ryb)}{Z^{C}_{e}}\right)^{\int_{\mathcal{M}}\bm{\mu}\cupp (\mathsf{PD}^{-1}_{\mathcal{M}_0}(\partial p)\cupp \mathbf{e})},\textcolor{dodgerblue}{X^B_{d\mathbf{v}}}\right]\\&=\left[\prod_{e}\left(\textcolor{dodgerblue}{X^{B}_{\mathbf{e}}}\textcolor{orange(ryb)}{X^{C}_{\mathbf{e}}}\right)^{\int_{\mathcal{M}}\bm{\mu}\cupp (\mathsf{PD}^{-1}_{\mathcal{M}_0}(e)\cupp d\mathbf{v})},\textcolor{dodgerblue}{Z^B_{\partial p}}\right]\\&=(-1)^{\int_{\mathcal{M}}\bm{\mu}\cupp (\mathsf{PD}^{-1}_{\mathcal{M}_0}(\partial p)\cupp \exd\mathbf{v})}=1
    \end{split}
    \end{equation}
    This means the original commutators of Eq.~\eqref{eq:com1} and Eq.\eqref{eq:com2} are stabilizers or logical operators which preserve the ground state subspace. If the cup product and function $\mathsf{PD}^{-1}_{\mathcal{M}_0}$ are both LDPC, the commutators cannot be logical operators. We therefore conclude that for any $\bm{\mu}\in\text{ker}(\exd_0)$, $U_{\mathsf{SWAP}}$ is either the identity or a symmetry of the ground state codespace.

Having established the existence of the symmetry $U_{\mathsf{SWAP}}$ for every element $\bm{\mu}\in\text{ker}(\exd_0)$, we now turn to the procedure for gauging these symmetries. The protocol is the same as the $\mathsf{CZ}$ case, but using the following symmetry enrichment operator:
\begin{equation}
    \Omega_{a,\textcolor{dodgerblue}{B}\textcolor{orange(ryb)}{C}}=\prod_{v,e,e'}\left(\textcolor{dodgerblue}{X^{B}_{\mathbf{e}}}\textcolor{orange(ryb)}{X^{C}_{\mathbf{e}}}\right)^{\textcolor{goodpurple}{n^a_v}\left(\textcolor{dodgerblue}{n^{B}_{e'}}+\textcolor{orange(ryb)}{n^{C}_{e'}}\right)\int_{\mathcal{M}}\mathbf{v}\cupp ( \mathsf{PD}^{-1}_{\mathcal{M}_0}(e)\cupp\mathbf{e'})}
\end{equation}
After following the same gauging procedure as in the previous section, we obtain the following final vertex stabilizers:
\begin{align}
    &\bB{X^B_{\exd \mathbf{v}}}\prod_{e}\left(\AR{Z_{(\mathsf{PD}^{-1}_{\mathcal{M}_0}(e)\cupp\mathbf{v} )\capp\mathcal{M}}^A} \right)^{\bB{m_{\mathbf{e}}^B} + \oC{m_{\mathbf{e}}^C}} \\& \oC{X^C_{\exd \mathbf{v}}}\prod_{e}\left(\AR{Z_{(\mathsf{PD}^{-1}_{\mathcal{M}_0}(e)\cupp\mathbf{v} )\capp\mathcal{M}}^A} \right)^{\bB{m_{\mathbf{e}}^B} + \oC{m_{\mathbf{e}}^C}}\\&\textcolor{red}{X^A_{\exd\mathbf{v}}}\prod_{e,e'}\left(\textcolor{dodgerblue}{X^{B}_{\mathbf{e}}}\textcolor{orange(ryb)}{X^{C}_{\mathbf{e}}}\right)^{\left(\textcolor{dodgerblue}{n^{B}_{e'}}+\textcolor{orange(ryb)}{n^{C}_{e'}}\right)\int_{\mathcal{M}}\mathbf{v}\cupp ( \mathsf{PD}^{-1}_{\mathcal{M}_0}(e)\cupp\mathbf{e'})}
\end{align}
where $m_{\mathbf{e}} = (1 - X_{\mathbf{e}})/2$.
Furthermore, the plaquette stabilizers are given by: 
\begin{align}
  & \bB{Z^B_{\partial p}}\prod_{e'}\left( \AR{Z_{(\mathbf{v}_p\cupp\mathbf{e}' )\capp\mathcal{M}}^A}\right)^{\bB{n_{e'}^B} + \oC{n_{e'}^C}} \\
  &\oC{Z^C_{\partial p}}\prod_{e'}\left( \AR{Z_{(\mathbf{v}_p\cupp\mathbf{e}' )\capp\mathcal{M}}^A}\right)^{\bB{n_{e'}^B} + \oC{n_{e'}^C}}
\end{align}
along with $\AR{Z_{\partial p}^{A}}$.
In the above, we have defined $\mathbf{v}_p$ via the relation $\exd \mathbf{v}_p=\mathsf{PD}^{-1}_{\mathcal{M}_0}(\partial p)$.

\subsection{Non-Abelian Codes from Gauging with Graph States}\label{sec:gauging-graphs}

We now turn to our graph gauging construction. In the previous section, we saw that when a chain complex has certain additional structures (namely Poincar\'e duality and a cup product), we could identify a set of transversal gates acting on codes \BandC that can be matched with the subsystem symmetries of the cluster state built from this complex.
However, this approach doesn't apply to codes whose underlying chain complex does not obey Poincar\'e duality.
%
Nevertheless, \BandC can still have transversal gates that we may wish to gauge.
An example is the global $\mathsf{SWAP}$ acting between the two copies of a code \BandC.

 Here we will describe how to gauge the transversal $\mathsf{SWAP}$ gate (or any other transversal gate acting uniformly on the code) using an ancilla complex $A_\bullet$ built from a graph. 
 This procedure shares some features with \cite{williamson_tanner}, which used a graph construction to measure logical operators of a qLDPC code; the key difference in our approach is that we measure a transversal gate on the logical operators rather than the logical operators themselves.

Crucially, this graph-based gauging approach can be applied to any qLDPC code, without additional structural assumptions on the underlying chain complex.
In contrast to the homological gauging procedure, where the ancilla chain complex was inherited from the input codes \BandC, here the ancilla complex $A_\bullet$ is generally distinct from the original code chain complex $D_\bullet$ and must be associated with a graph.
Constructing the cluster state on this graph yields stabilizers of the form $\textcolor{goodpurple}{Z_{v}} \textcolor{red}{Z_{\langle v, w \rangle}} \textcolor{goodpurple}{Z_{w}}$, where $\langle v, w \rangle$ denotes an edge on the graph connecting two endpoint vertices $v, w$.
Provided the ancilla graph is connected, the resulting cluster state has a single $0$-cocycle symmetry $H^0 \simeq \mathbb{Z}_2$ generated by $\aP{X_{\boldsymbol{\mu}}}$ where $\boldsymbol{\mu}$ is the sum over of all vertices of the graph. 
This construction offers a more general gauging procedure capable of gauging any $\mathbb{Z}_2$ transversal gate of a qLDPC code---at the expense of selecting a bespoke graph that introduces additional homological structure to the code as we will soon see.

\subsubsection{Graph requirements}
The choice of graph and symmetry enrichment operator must satisfy a few general constraints.
First, we require the graph be connected such that its $0$-cocycle symmetry is given by $H^0 \simeq \mathbb{Z}_2$.
Second, we require the symmetry enrichment operator $\Omega$ must be chosen such that the gauging procedure will tie the action of the transversal gate $U$ we are gauging to the 0-cocycle symmetry $\aP{X_{\boldsymbol{\mu}}}$ of the ancilla cluster state; in other words, we assert $\aP{X_{\boldsymbol{\mu}}}\Omega_{\aP{a},\textcolor{dodgerblue}{B}\textcolor{orange(ryb)}{C}} \aP{X_{\boldsymbol{\mu}}}\Omega_{\aP{a},\textcolor{dodgerblue}{B}\textcolor{orange(ryb)}{C}}^{\dagger} = U$.

Finally, the graph and symmetry enrichment operators must be selected to ensure the stabilizers after gauging remain LDPC; the final stabilizers resulting from the gauging procedure should each involve a finite number of qubits, with each qubit involved in a finite number of stabilizers.
In what follows, we will run through the steps of the gauging procedure with the graph ancilla complex and comment on some general candidates for an ancilla graph.

\subsubsection{Example: Gauging transversal $\mathsf{SWAP}$ gate}\label{sec:graphygraph}
We now demonstrate our graph gauging procedure, gauging the transversal $\mathsf{SWAP}$ gate which exchanges $\textcolor{dodgerblue}{X_{\mathbf{e}}}\leftrightarrow\textcolor{orange(ryb)}{X_{\mathbf{e}}}$ and $\textcolor{dodgerblue}{Z_e}\leftrightarrow\textcolor{orange(ryb)}{Z_e}$ as a representative example. As before, our starting point consists of two copies \BandC of a qLDPC code. We assume an ancilla graph with qubits at the vertices $\textcolor{goodpurple}{a}$ and edges $\textcolor{red}{A}$, but make no further assumptions on the graph beyond its connectivity, as mentioned above. In the graph gauging procedure, the symmetry enrichment operator takes the form:
\begin{equation}
    \Omega_{\aP{a},\textcolor{dodgerblue}{B}\textcolor{orange(ryb)}{C}}=\prod_{v, e}\left(\mathsf{SWAP}_{\bB{e}, \oC{e}} \right)^{\aP{n^a_{v}}\int_{\phii(\mathbf{e})} \mathbf{v}}
\end{equation}
where the linear function $\phii: D^1 \oplus D_1 \rightarrow D_0$ determines which ancilla will control the $\mathsf{SWAP}$ action associated with edge $e$ (treated as either a chain or co-chain). 
For simplicity, we assume that $\phii$ is local on the chain complex; for instance we assume $\phii(e)$ is within a bounded hypergraph distance of $e$, or stronger $\phii(e)\in \partial e$.

Our requirement that $\Omega_{\aP{a},\textcolor{dodgerblue}{B}\textcolor{orange(ryb)}{C}} \aP{X_{\boldsymbol{\mu}}}\Omega_{\aP{a},\textcolor{dodgerblue}{B}\textcolor{orange(ryb)}{C}}^{\dagger} = U$ for the symmetry enrichment operator translates into requirements on the function $\phii$. In particular, we require that $\int_{\phii(\mathbf{e})} \boldsymbol{\mu} = 1$ for all  $e$---i.e. for all $\mathbf{e}$,  $\phii(\mathbf{e})$ maps to an odd number of vertices.
For simplicity, we henceforth assume that for all $e$, $\phii(e)$ is a single vertex $v$ (i.e. an element of the basis of $D_0$).
Moreover, to preserve the LDPC property of the code, we require that the matrix representing $\phii$ is sparse.

With these conditions, we can proceed with the same gauging procedure as in the previous sections. 
In doing so, we will naturally see what constraints we must place on the edges of the graph to ensure the resulting non-Abelian stabilizers are LDPC.
After preparing a cluster state on the ancilla graph and applying the symmetry enrichment operator, the vertex stabilizers are transformed to:
\begin{align}
    \textcolor{dodgerblue}{X^B_{\exd\mathbf{v}}}\prod_{\mathbf{e}\in\exd \mathbf{v}}&\left(\textcolor{dodgerblue}{X^B_{\mathbf{e}}} \textcolor{orange(ryb)}{X^C_{\mathbf{e}}}\right)^{\aP{n^{a}_{\phii(e)}}} \quad \textcolor{orange(ryb)}{X^C_{\exd\mathbf{v}}}\prod_{\mathbf{e}\in\exd \mathbf{v}}\left(\textcolor{dodgerblue}{X^B_{\mathbf{e}}} \textcolor{orange(ryb)}{X^C_{\mathbf{e}}}\right)^{\aP{n^{a}_{\phii(e)}}} \\&\qquad \aP{X^a_{\mathbf{v}}}\textcolor{red}{X^A_{\exd\mathbf{v}}}\mathsf{SWAP}_{\phii^{\mathsf{T}}(v)} \nonumber
\end{align} 
and the plaquette stabilizers are transformed to:
\begin{equation}
\begin{split}
    \textcolor{dodgerblue}{Z^B_{\partial p}}\prod_{e\in\partial p}&\left(\textcolor{dodgerblue}{Z^B_{e}} \textcolor{orange(ryb)}{Z^C_{e}}\right)^{\aP{n^{a}_{\phii(e)}}} \qquad \textcolor{orange(ryb)}{Z^C_{\partial p}}\prod_{e\in\partial p}\left(\textcolor{dodgerblue}{Z^B_{e}} \textcolor{orange(ryb)}{Z^C_{e}}\right)^{\aP{n^{a}_{\phii(e)}}} \\&\qquad\qquad \aP{Z^a_{\partial e}}\textcolor{red}{Z^A_{e}}
\end{split}
\end{equation} 
Next, the $\aP{a}$ qubits are measured in the $\ket{+}$ state.
After measurement, the gauged vertex stabilizer associated with the $\AR{A}$ qubits is: 
\begin{equation}
    \textcolor{red}{X^A_{\exd\mathbf{v}}}\mathsf{SWAP}_{\phii^{\mathsf{T}}(v)} \qquad 
\end{equation}
We then use the fact that the vertex and plaquette stabilizers can safely be set to $+1$ prior to measurement to re-write the stabilizers in a form independent of $\aP{a}$.
Starting with the vertex stabilizers, if we multiply both the \BandC stabilizers with $\left(\bB{X^B_{\exd\mathbf{v}}} \oC{X^C_{\exd\mathbf{v}}}\right)^{\aP{n^a_v}}$, we find: 
\begin{equation}
     \textcolor{dodgerblue}{X^B_{\exd\mathbf{v}}}\prod_{\mathbf{e}\in\exd \mathbf{v}}\left(\textcolor{dodgerblue}{X^B_{\mathbf{e}}} \textcolor{orange(ryb)}{X^C_{\mathbf{e}}}\right)^{\aP{n^{a}_{\phii(e)}} + \aP{n^{a}_{v}}} = \bB{X_{\exd \mathbf{v}}} \prod_{\mathbf{e}\in\exd \mathbf{v}} (\aP{Z_{\phii(e)} Z_v})^{\bB{m_{\mathbf{e}}^B}  + \oC{m_{\mathbf{e}}^C} }
\end{equation}
where $m_{\mathbf{e}} = (1 - X_{\mathbf{e}})/2$.
Similarly, for every plaquette $p$ on the \BandC codes, we can define a function $v(p)$ such that $v(p)$ is an arbitrary vertex in the \textbf{closure} of $p$ (i.e. vertices in the boundary of any edge contained in $\partial p$).
Now, multiply by:
$\left(\bB{Z^B_{\exd\mathbf{v}}} \oC{Z^C_{\exd\mathbf{v}}}\right)^{\aP{n^a_{v(p)}}}$.
Then, we have for the vertex stabilizers:
\begin{equation}
     \textcolor{dodgerblue}{Z^B_{\exd\mathbf{v}}}\prod_{e\in\partial p}\left(\textcolor{dodgerblue}{Z^B_{\mathbf{e}}} \textcolor{orange(ryb)}{Z^C_{\mathbf{e}}}\right)^{\aP{n^{a}_{\phii(e)}} + \aP{n^{a}_{v(p)}}} = \bB{Z^B_{\exd \mathbf{v}}} \prod_{e\in\partial p} (\aP{Z_{\phii(e)} Z_{v(p)}})^{\bB{n_{e}^B}  + \oC{n_{e}^C} }
\end{equation}
At this stage, since the ancilla vertex qubits live on a connected graph, we are guaranteed a connected path between $\phii(e)$ and $v$ (and $\phii(e)$ and $v(p)$). Therefore, we can rewrite the final stabilizers in terms of the $\textcolor{red}{A}$ qubits. The vertex stabilizers are:
\begin{equation}
\begin{split}
    &\bB{X^B_{\exd \mathbf{v}}}\prod_{\mathbf{e}\in\exd\mathbf{v}}\left(\AR{Z_{\langle \phii(e),v\rangle}^A} \right)^{\bB{m_e^B} + \oC{m_e^C}} \quad \oC{X^C_{\exd \mathbf{v}}}\prod_{\mathbf{e}\in\exd\mathbf{v}}\left(\AR{Z_{\langle \phii(e),v\rangle}^A} \right)^{\bB{m_e^B} + \oC{m_e^C}}\\&\qquad \qquad \qquad\qquad \textcolor{red}{X^A_{\exd\mathbf{v}}}\mathsf{SWAP}_{\phii^{\mathsf{T}}(v)} \qquad 
\end{split}
\end{equation}
and the plaquette stabilizers are: 
\begin{equation}
\begin{split}
   & \bB{Z^B_{\partial p}}\prod_{e\in\partial p}\left( \AR{Z_{\langle \phii(e),v(p)\rangle}^A}\right)^{\bB{n_e^B} + \oC{n_e^C}}\quad \oC{Z^C_{\partial p}}\prod_{e\in\partial p}\left( \AR{Z_{\langle \phii(e),v(p)\rangle}^A}\right)^{\bB{n_e^B} + \oC{n_e^C}}\\&\qquad \qquad \qquad\qquad\qquad\qquad \textcolor{red}{Z^A_{\partial p}}
\end{split}
\end{equation}
With these final stabilizers, we return to our requirement that the resulting code remain LDPC. The potentially problematic terms are those involving edges $\langle v(p),\phii(e)\rangle$ which for a generic graph with vertices $\textcolor{goodpurple}{a}$ are strings which are not guaranteed to be local on the graph.

This observation highlights that the choice of graph is crucial. In general, the appropriate graph will depend both on the underlying chain complex and on the desired properties of the resulting code. Nevertheless, we give a few comments on strategies for explicit constructions.

One approach is as follows: given a symmetry enrichment operator and associated function $\phii$, we construct a graph by introducing an edge $\langle \phii(e), v \rangle$ for all $e\in\exd\mathbf{v}$ and for all $e \in \partial p$ define an edge $\langle \phii(e), v(p) \rangle$.
We can define a boundary map:
\begin{equation}
    \partial_{a} \langle \phii(e), v \rangle = \phii(e) + v \qquad \partial_{a}\langle \phii(e), v(p) \rangle = \phii(e) + v(p)
\end{equation}
We can also define a co-boundary map: 
\begin{equation}
   \exd_A \mathbf{v} = \sum_{e} \langle v, \phii(e) \rangle \exd \mathbf{v}(e)
\end{equation}

However, one must verify the resulting graph will have closed loops (and other favorable properties like distance and encoding rate). 

Another option could be a similar choice to the explicit construction given in \cite{williamson_tanner}. More choices may be considered which respect certain underlying properties of the hypergraph like a clique expansion (replacing each hyperedge of the original chain complex with a complete graph of its boundary) or a modification thereof. Generally, we leave the question of which graph to use open; indeed, it is likely that particular codes and target properties will require particular constructions tailored to the task at hand.

\subsubsection{Gauging Other Symmetries}
Finally, we remark on the gauging of transversal gates beyond those discussed in our examples so far (namely $\mathsf{CZ}$ and $\mathsf{SWAP}$). 
Our procedure in the graph gauging case is quite general
and can be readily adapted to gauge other transversal gate (both acted globally or on a substeym) of a $\mathbb{Z}_N$ qLDPC code. 
For example, we could have chosen the input qLDPC code to be a $\mathbb{Z}_3$ qLDPC code, which allows for a transversal gate that is a $\mathbb{Z}_2$ charge conjugation symmetry that maps between the two nontrivial generators of $\mathbb{Z}_3$. We can use our graph gauging procedure to gauge this transversal gate and obtain a non-Abelian $S_3$ qLDPC code, where $S_3$ is the symmetry group on three elements. We provide an example of such $S_3$ qLDPC code in Appendix~\ref{app:s3-ALP}. Alternatively, we can choose a graph state which only spans the support of a subsystem symmetry present in an input qLDPC code, which allows for gauging transversal gate on the subsystem to create a non-Abelian ``slice" within the input Abelian qLDPC code.

\subsection{Measurement Preparation of Non-Abelian qLDPC Codes} \label{subsec:measurement_prep}

Throughout our discussion, we have assumed in the gauging procedure that it is possible to project the qubits on the $\aP{a}$ sublattice into the $\ket{+}$ state.
In this section, we show that this projection can be implemented \textit{deterministically} using a measurement and non-locally conditioned classical feedforward protocol.

Let us recall that, in the preparation of the $2$D non-Abelian $\mathcal{D}(D_4)$ topological order reviewed in Section~\ref{sec:prelim}, the projection could be implemented deterministically because the anyons created by the randomness of the measurement outcomes were all Abelian.
Consequently, the same decoder used to pair up anyons in preparing the Abelian toric code could be used to pair up the Abelian anyons in the non-Abelian $\mathcal{D}(D_4)$ code.
In the qLDPC case, we will find an analogous statement applies.

In particular, let us suppose that there exists a decoder for the qLDPC code associated with the ancilla complex we use for gauging.
An important consequence of this is that the \textit{Abelian} qLDPC code of the complex can be prepared deterministically.
To see this, let us note that one route to preparing the Abelian qLDPC code associated with the ancilla complex on the $\AR{A}$ sublattice is to start with the $\ket{0}^{\otimes N}$ state on this sublattice and directly measure the $\AR{X_{\exd^a \mathbf{v}}}$ stabilizers \footnote{Equivalently, we could prepare the cluster state on the $\aP{a}\AR{A}$ sublattice, with stabilizers $\aP{X_{\mathbf{v}}} \AR{X_{\exd_a \mathbf{v}}}$ and $\aP{Z_{\partial_a e}}\AR{Z_e}$, and measure $\aP{X_{\mathbf{v}}}$.}.
Then the stabilizers of the $\AR{A}$ sublattice immediately after measurement are given by:
\begin{equation}
    x_{\mathbf{v}} \AR{X_{\exd^a \mathbf{v}}^A \qquad Z_{\partial p}^A}
\end{equation}
where $\{x_{\mathbf{v}} = \pm 1\}$ are the set of random measurement outcomes.
The existence of a decoder for the ancilla complex implies that there exists a classical algorithm that reads in these measurement outcomes $\{x_{\mathbf{v}}\}$ and determines a product of $\AR{Z_e}$ that can be applied to the measured state to return it to the code space.

Decoder in hand, let us return to the non-Abelian qLDPC context.
For the sake of simplicity, let us consider the (Clifford) stabilizers of our code after measurement in the homological gauging approach.
These stabilizers take the following form:
\begin{align}
&\AR{\mathcal{A}^{A}_{\mathbf{v}}}[x_\mathbf{v}]=x_{\mathbf{v}} \AR{X_{\exd\mathbf{v}}^A}\prod_{e,e'} 
\left(\mathsf{CZ}_{\bB{e},\oC{e'}}\right)^{\int_{\mathcal{M}}\mathbf{v}\cupp (\mathbf{e}\cupp\mathbf{e'})}
\\
&\bB{\mathcal{A}^{B}_{\mathbf{v}}}=\bB{X_{\exd\mathbf{v}}^B}\prod_{e,e'} 
\left(\mathsf{CZ}_{\AR{e},\oC{e'}}\right)^{\int_{\mathcal{M}}\mathbf{e}\cupp (\mathbf{v}\cupp\mathbf{e'})} 
\\
&\oC{\mathcal{A}^{C}_{\mathbf{v}}}=\oC{X_{\exd\mathbf{v}}^C}\prod_{e,e'} 
\left(\mathsf{CZ}_{\bB{e},\AR{e'}}\right)^{\int_{\mathcal{M}}\mathbf{e'}\cupp (\mathbf{e}\cupp\mathbf{v})}
\end{align}
Along with the plaquette stabilizers $Z_{\partial p}$ on the $\AR{A}$, $\bB{B}$, and $\oC{C}$ sublattices.
Notice the appearance of the random measurement outcomes $\{x_{\mathbf{v}} = \pm 1\}$ in the $\AR{\mathcal{A}_{\mathbf{v}}^A}$.
Now, let us make the observation that the operator $\AR{Z_{e}^A}$ commutes with all stabilizers except for the stabilizers $\AR{\mathcal{A}_{\mathbf{v}}^A}[x_{\mathbf{v}}]$. 
As a consequence, the same classical decoding algorithm that could be used to prepare the Abelian qLDPC code associated with the ancilla complex, can be used to deterministically prepare the non-Abelian code!
In particular, the output of the decoding algorithm yields a certain product of Pauli $\AR{Z_e}$ operators that can be applied to systematically correct each $\AR{\mathcal{A}_{\mathbf{v}}^A}[x_{\mathbf{v}}=-1]$ back to $\AR{\mathcal{A}_{\mathbf{v}}^A}[1]$ as desired, without violating additional stabilizers since the Pauli $\AR{Z}$ operator commutes with all other stabilizers.
This demonstrates that any non-Abelian codes obtained from sequential gauging of transversal gates can be prepared deterministically with finite-depth adaptive circuits.
Note that the same argument applies in the graph gauging case and will apply for any symmetry enrichment operator of the ``controlled-symmetry'' form discussed in Section~\ref{sec:prelim}.
As a final remark, we note that, because the correction step can be done with a depth-$1$ Pauli circuit, stabilizer measurements of the non-Abelian code can be corrected in software without needing to physically perform the gate operations on the code. 

\section{Examples of Non-Abelian qLDPC Codes}\label{sec:gauging-examples}

Having described the general procedure for obtaining our non-Abelian codes, we now use this procedure to write down explicit models for non-Abelian qLDPC codes.
We first focus on a particular model of fractons \cite{Shirley_2019,You_2018,Burnell_2022,Vijay_2016,Haah_2011,Gorantla_2023} and concretely illustrate both approaches to gauging in this setting.
In particular, we will start by using the ``graph method'' to gauge a transversal $\mathsf{CZ}$ gate present between two copies of this code.
Subsequently, we will show that two copies of this code in fact have an \textit{addressable} family of $\mathsf{CZ}$ gates, which can be understood via a cup product construction\footnote{The method used to find these gates and construct the cup product for this code will be the subject of a forthcoming work \cite{addressablegates}}.
We then use our homological formalism for gauging to determine the model that results from gauging these addressable symmetries.
We remark that numerous past works have studied various approaches to obtaining non-Abelian versions of fractons \cite{Bulmash_2019,Prem_2019,Wang_2020,Tu_2021,Williamson_2023}.
One interesting feature that we discover through our homological gauging, is a non-Abelian fractonic system where only a portion of the full code has a non-Abelian character, with the remainder being fully Abelian.

We conclude this section by using the graph formalism to arrive at non-Abelian bi-variate bicycle codes from gauging a non-local (but nevertheless qLDPC) transversal $\mathsf{CZ}$ gate present in these codes\footnote{Our construction of non-Abelian bivariate bicycle code via gauging is distinct from what is meant by ``non-Abelian bicycle code'' in Ref.~\cite{guo_toward_2026}.
There, they construct an \textit{Abelian} qLDPC code based on a group algebra construction where the group in question is non-Abelian.}.

\subsection{Non-Abelian Fractons by Gauging transversal $\mathsf{CZ}$ Gates}\label{subsec:graph-gauging-fracton}

Let us start by considering how to use the ``graph method'' to concretely obtain a non-Abelian qLDPC code.
In particular, our initial code of choice is the so-called anisotropic lineon-planeon (ALP) model~\cite{xu2004strong, Vijay_2016, Shirley_2019}, which is defined on a cubic lattice and is associated with a level-$(2 + 1)$ chain complex (with the usual associated cochain complex): 
\begin{equation} \label{eq-ALPchaincomplex}
        0\rightarrow D_2\xrightarrow{\partial_2}D_1\xrightarrow{\partial_1}D_0\rightarrow0
\end{equation}
where geometrically $D_2$ is associated with the space of cubes on the lattice (associated with $Z$-checks), $D_1$ is space of plaquettes in the $xy$ planes and edges in the $z$ direction (where our qubits live), and $D_0$ is the space of vertices (associated with $X$-checks).
This is illustrated concretely via the stabilizers of the model: 
\begin{equation}\label{eq:qubit-ALP-stabilizers}
    H = - \sum_{p \in \mathsf{B}_2}\ 
    \underbrace{\begin{tikzpicture}[scale = 0.4, baseline={([yshift=-.5ex]current bounding box.center)}]
  \coordinate (A) at (0, 0);
  \coordinate (B) at (2, 0);
  \coordinate (C) at (2, 2);
  \coordinate (D) at (0, 2);
  \coordinate (E) at (0.7, 0.7);
  \coordinate (F) at (2.7, 0.7);
  \coordinate (G) at (2.7, 2.7);
  \coordinate (H) at (0.7, 2.7);
  \draw[color = gray] (E) -- (F) -- (G) -- (H) -- cycle;
  \draw[color = gray] (A) -- (B) -- (C) -- (D) -- cycle;
  \draw[color = gray] (A) -- (E);
  \draw[color = gray] (B) -- (F);
  \draw[color = gray] (C) -- (G);
  \draw[color = gray] (D) -- (H);
  \node at (0, 1) {\scriptsize $Z$};
  \node at (0.7, 1.5) {\scriptsize $Z$};
  \node at (2, 1) {\scriptsize $Z$};
  \node at (2.7, 1.5) {\scriptsize $Z$};
  \node at (1.3, 0.35) {\scriptsize $Z$};
  \node at (1.5, 0.35 + 2) {\scriptsize $Z$};
\end{tikzpicture}}_{Z_{\partial p}} 
- \sum_{\mathbf{v} \in \mathsf{B}^0}
\underbrace{\begin{tikzpicture}[scale = 0.4, baseline={([yshift=-.5ex]current bounding box.center)}]
    \coordinate (A) at (0, 0);
    \coordinate (Azp) at (0, 1.5); 
    \coordinate (Azm) at (0, -1.5); 
    \coordinate (B) at (0.7, 0.7);
    \coordinate (C) at (2.7, 0.7);
    \coordinate (D) at (2, 0);
    \coordinate (E) at (-0.7, -0.7);
    \coordinate (F) at (-2.7, -0.7);
    \coordinate (G) at (-2, 0);
    \coordinate (H) at (-1.3, 0.7);
    \coordinate (I) at (1.3, -0.7);
    \draw[color = gray] (A) -- (B) -- (C) -- (D)--cycle;
    \draw[color = gray] (A) -- (E) -- (F) -- (G)--cycle;
    \draw[color = gray] (A) -- (B) -- (H) -- (G)--cycle;
    \draw[color = gray] (A) -- (E) -- (I) -- (D)--cycle;
    \draw[color = gray] (Azm) -- (Azp);

    \node at (-2.6/2, -0.7/2) {\scriptsize $X$}; 
    \node at (2.6/2, 0.7/2) {\scriptsize $X$};
    \node at (-1.3/2, 0.7/2) {\scriptsize $X$};
    \node at (1.3/2, -0.7/2) {\scriptsize $X$};
    \node at (0, 1.) {\scriptsize $X$};
    \node at (0, -1.) {\scriptsize $X$};
\end{tikzpicture}}_{X_{\exd \mathbf{v}}}
\end{equation}
We remark that for a $L_x \times L_y \times L_z$ cubic lattice with period boundary conditions, the ALP code encodes $2(L_x + L_y -1)$ logical qubits and has a code distance of $\sim \mathsf{min}(L_x, L_y, L_z)$.
We will now take two copies of this code, defined on the $\bB{B}$ and $\oC{C}$ sublattices and use the graph state method to gauge the following transversal $\mathsf{CZ}$ gate:
\begin{equation}\label{eq:cubic-CZ-sym}
    U_{\mathsf{CZ}} = 
    \begin{tikzpicture}[scale = 0.5, baseline={([yshift=-.5ex]current bounding box.center)}]
    \coordinate (A) at (0, 0);
    \coordinate (Azp) at (0, 2);
    \coordinate (Azm) at (0, -1.6); 
    \coordinate (B) at (0.7, 0.7);
    \coordinate (Bzp) at (0.7, 0.7 + 2);
    \coordinate (C) at (2.7, 0.7);
    \coordinate (Czp) at (2.7, 2.7);
    \coordinate (D) at (2, 0);
    \coordinate (Dzp) at (2, 2);
    \coordinate (E) at (-0.7, -0.7);
    \coordinate (F) at (-2.7, -0.7);
    \coordinate (G) at (-2, 0);
    \coordinate (H) at (-1.3, 0.7);
    \coordinate (I) at (1.3, -0.7);
    \draw[color = gray] (A) -- (B) -- (C) -- (D)--cycle;
    \draw[color = gray] (Azp) -- (Bzp) -- (Czp) -- (Dzp)--cycle;
    \draw[color = gray] (B) -- (Bzp);
    \draw[color = gray] (C) -- (Czp);
    \draw[color = gray] (D) -- (Dzp);
    \draw[color = gray] (A) -- (Azp);

    \draw[color = black, line width = 1pt]  (1+0.35,0.35) -- (2+0.7,1+0.7);
    \draw[color = black, line width = 1pt]  (0,1) -- (1+0.35,2+0.35);

    \filldraw[color = orange(ryb)] (0,1) circle (3 pt);
    \filldraw[color = orange(ryb)] (1+0.35,0.35) circle (3 pt);
    \filldraw[color = dodgerblue] (1+0.35,2+0.35) circle (3 pt);
    \filldraw[color = dodgerblue] (2+0.7,1+0.7) circle (3 pt);

\end{tikzpicture}
\end{equation}
where the black lines denote $\mathsf{CZ}$ gates between the \BandC sublattices; this pattern of $\mathsf{CZ}$ gates is repeated for each unit cell of the cubic lattice.

To gauge the transversal $\mathsf{CZ}$ gate, we need to use an appropriate graph cluster state.
To that end, we introduce ancilla degrees of freedom on the cubic lattice, where qubits at the vertices and edges are in the $\aP{a}$ and $\AR{A}$ sublattices respectively.
We initialize the ancilla in the $\ket{+}$ product state. Then, we apply $\mathsf{CZ}$ between the nearest neighbor qubits in the $\aP{a}$ and $\AR{A}$ sublattices, followed by a Hadamard gate $H$ on the $\AR{A}$ sublattice, to obtain a 3D cluster state, with the stabilizers given by:
\begin{equation}\label{eq:cubic-graph-cluster-state}
    \begin{tikzpicture}[scale = 0.4, baseline={([yshift=-.5ex]current bounding box.center)}]
    \coordinate (A) at (0, 0);
    \coordinate (Azp) at (0, 2);
    \coordinate (Azm) at (0, -1.6); 
    \coordinate (B) at (0.7, 0.7);
    \coordinate (Bzp) at (0.7, 0.7 + 2);
    \coordinate (C) at (2.7, 0.7);
    \coordinate (Czp) at (2.7, 2.7);
    \coordinate (D) at (2, 0);
    \coordinate (Dzp) at (2, 2);
    \coordinate (E) at (-0.7, -0.7);
    \coordinate (F) at (-2.7, -0.7);
    \coordinate (G) at (-2, 0);
    \coordinate (H) at (-1.3, 0.7);
    \coordinate (I) at (1.3, -0.7);
    \draw[color = gray] (A) -- (B) -- (C) -- (D)--cycle;
    \draw[color = gray] (A) -- (E) -- (F) -- (G)--cycle;
    \draw[color = gray] (A) -- (B) -- (H) -- (G)--cycle;
    \draw[color = gray] (A) -- (E) -- (I) -- (D)--cycle;
    \draw[color = gray] (Azm) -- (Azp);


    \node at (A) {\scriptsize $\aP{X}$};
    \node at (1,0) {\scriptsize $\AR{X}$};
    \node at (-1,0) {\scriptsize $\AR{X}$};
    \node at (0,1) {\scriptsize $\AR{X}$};
    \node at (0,-1) {\scriptsize $\AR{X}$};
    \node at (0.35,0.35) {\scriptsize $\AR{X}$};
    \node at (-0.35,-0.35) {\scriptsize $\AR{X}$};

\end{tikzpicture},
\quad 
\begin{tikzpicture}[scale = 0.55, baseline = {([yshift=-.5ex]current bounding box.center)}]
    \draw[lightcrimson] (0,0) -- (1.5, 0);
    \node at (0.75, 0) {\normalsize $\AR{Z}$};
    \node at (1.5, 0) {\normalsize $\aP{Z}$};
    \node at (0,0) {\normalsize $\aP{Z}$};
\end{tikzpicture}, 
\quad 
\begin{tikzpicture}[scale = 0.55, baseline = {([yshift=-.5ex]current bounding box.center)}]
    \draw[lightcrimson] (0,0) -- (0, 1.5);
    \node at (0, 0.75) {\normalsize $\AR{Z}$};
    \node at (0, 1.5) {\normalsize $\aP{Z}$};
    \node at (0,0) {\normalsize $\aP{Z}$};
\end{tikzpicture},
\quad
\begin{tikzpicture}[scale = 0.75, baseline = {([yshift=-.5ex]current bounding box.center)}]
    \draw[lightcrimson] (0,0) -- (0.9, 0.7);
    \node at (0.45, 0.35) {\normalsize $\AR{Z}$};
    \node at (0.9, 0.7) {\normalsize $\aP{Z}$};
    \node at (0,0) {\normalsize $\aP{Z}$};
\end{tikzpicture}.
\end{equation}
Next, we perform the symmetry enrichment to tie together the $\aP{\mathbb{Z}^{(0)}_2}$ 0-cycle symmetry in the ancilla graph cluster state, given by $\prod_{\aP{a}} \aP{X}$, with the transversal $\mathsf{CZ}$ gate in the two copies of the ALP model given in Eq.~\eqref{eq:cubic-CZ-sym}. We act a $\mathsf{CCZ}$ gate, controlled on the qubits in the $\aP{a}$ sublattice, and target on the qubits on the \BandC sublattices given by $\phii^{\mathsf{T}}(\aP{\mathbf{v}})$ as follows:
\begin{equation}\label{eq:cubic-symm-enrich}
    \begin{tikzpicture}[scale = 0.5, baseline={([yshift=-.5ex]current bounding box.center)}]
    \coordinate (A) at (0, 0);
    \coordinate (Azp) at (0, 2);
    \coordinate (Azm) at (0, -1.6); 
    \coordinate (B) at (0.7, 0.7);
    \coordinate (Bzp) at (0.7, 0.7 + 2);
    \coordinate (C) at (2.7, 0.7);
    \coordinate (Czp) at (2.7, 2.7);
    \coordinate (D) at (2, 0);
    \coordinate (Dzp) at (2, 2);
    \coordinate (E) at (-0.7, -0.7);
    \coordinate (F) at (-2.7, -0.7);
    \coordinate (G) at (-2, 0);
    \coordinate (H) at (-1.3, 0.7);
    \coordinate (I) at (1.3, -0.7);
    \draw[color = gray] (A) -- (B) -- (C) -- (D)--cycle;
    \draw[color = gray] (Azp) -- (Bzp) -- (Czp) -- (Dzp)--cycle;
    \draw[color = gray] (B) -- (Bzp);
    \draw[color = gray] (C) -- (Czp);
    \draw[color = gray] (D) -- (Dzp);
    \draw[color = gray] (A) -- (Azp);

    \draw[color = cz1color, line width = 1pt]  (A) -- (1+0.35,0.35) -- (2+0.7,1+0.7);
    \draw[color = cz2color, line width = 1pt]  (A) -- (0,1) -- (1+0.35,2+0.35);

    \filldraw[color = goodpurple] (A) circle (3 pt);
    \filldraw[color = orange(ryb)] (0,1) circle (3 pt);
    \filldraw[color = orange(ryb)] (1+0.35,0.35) circle (3 pt);
    \filldraw[color = dodgerblue] (1+0.35,2+0.35) circle (3 pt);
    \filldraw[color = dodgerblue] (2+0.7,1+0.7) circle (3 pt);

\end{tikzpicture},
\end{equation}
where the two different shades of green lines denote two distinct $\mathsf{CCZ}$ gates; this symmetry enrichment is acted for each $\aP{\mathbf{v}}$ in the $\aP{a}$ sublattice. This specifies the $\phii$ function as for each vertex $\aP{\mathbf{v}}$ in the $\aP{a}$ sublattice, $\phii^{\mathsf{T}}(\aP{v})$ gives pairs of edges on the \BandC sublattices; and above implementation of $\phii$ function is surjective and local. 

Using the cluster state and symmetry-enrichment operator above, we can directly obtain the stabilizers of the non-Abelian LDPC code:

\begin{equation}\label{eq:fracton-stabilizers-graph-gauging}
\begin{aligned}
    \begin{tikzpicture}[scale = 0.4, baseline={([yshift=-.5ex]current bounding box.center)}]
    \coordinate (A) at (0, 0);
    \coordinate (Azp) at (0, 2);
    \coordinate (Azm) at (0, -1.6); 
    \coordinate (B) at (0.7, 0.7);
    \coordinate (Bzp) at (0.7, 0.7 + 2);
    \coordinate (C) at (2.7, 0.7);
    \coordinate (Czp) at (2.7, 2.7);
    \coordinate (D) at (2, 0);
    \coordinate (Dzp) at (2, 2);
    \coordinate (E) at (-0.7, -0.7);
    \coordinate (Ezp) at (-0.7, -0.7+2);
    \coordinate (F) at (-2.7, -0.7);
    \coordinate (Fzp) at (-2.7, -0.7+2);
    \coordinate (G) at (-2, 0);
    \coordinate (Gzp) at (-2, 0+2);
    \coordinate (H) at (-1.3, 0.7);
    \coordinate (Hzp) at (-1.3, 0.7+2);
    \coordinate (I) at (1.3, -0.7);
    \coordinate (Izp) at (1.3, -0.7+2);
    \coordinate (tNE) at (1+0.35,2+0.35);
    \coordinate (tNW) at (0.35-1,2+0.35);
    \coordinate (tSE) at (0.55,2-0.35);
    \coordinate (tSW) at (0.7-2,2-0.35);
    \coordinate (mNE) at (2+0.7,1+0.5);
    \coordinate (mNW) at (0.7,1+0.5);
    \coordinate (mSE) at (0,0.85);
    \coordinate (mSW) at (2,0.85);
    \coordinate (uNE) at (0, -0.8-0.1);
    \coordinate (uNW) at (-2, -0.8-0.3);
    \coordinate (uSE) at (-0.7, -0.7-0.8-0.25);
    \coordinate (uSW) at (-2-0.7, -0.7-0.8+0.1);
    \coordinate (bNE) at (0, -1.6);
    \coordinate (bNW) at (-2, -1.6);
    \coordinate (bSE) at (-0.7, -0.7-1.6);
    \coordinate (bSW) at (-2-0.7, -0.7-1.6);
    
    \coordinate (bN) at (-1, -1.6);
    \coordinate (bE) at (-0.35, -0.35-1.6);
    \coordinate (bS) at (-1.7, -0.7-1.6);
    \coordinate (bW) at (-2-0.35, -0.35-1.6);

    \draw[color = gray] (A) -- (B) -- (C) -- (D)--cycle;
    \draw[color = gray] (A) -- (E) -- (F) -- (G)--cycle;
    \draw[color = gray] (A) -- (B) -- (H) -- (G)--cycle;
    \draw[color = gray] (A) -- (E) -- (I) -- (D)--cycle;
    \draw[color = gray] (Azm) -- (Azp);    
    \draw[color = gray] (F) -- (bSW);    
    \draw[color = gray] (G) -- (bNW);    
    \draw[color = gray] (E) -- (bSE);    
    \draw[color = gray] (bNE) -- (bNW) -- (bSW) -- (bSE) -- cycle;

    \draw[color = cz1color, line width = 1pt]  (uSW) -- (-2.6/2-0.2, -0.7/2);
    \draw[color = cz1color, line width = 1pt]  (bN) -- (uNE);
    \draw[color = cz1color, line width = 1pt]  (bS) -- (uSE);

    \draw[color = cz2color, line width = 1pt]  (uNW) -- (bW) -- (uNE);
    \filldraw[color = crimson] (uSW) circle (3 pt);
    \filldraw[color = crimson] (bN) circle (3 pt);
    \filldraw[color = crimson] (bS) circle (3 pt);
    \filldraw[color = crimson] (bW) circle (3 pt);

    \filldraw[color = orange(ryb)] (-2.6/2-0.2, -0.7/2) circle (3 pt);
    \filldraw[color = orange(ryb)] (uNE) circle (3 pt);
    \filldraw[color = orange(ryb)] (uSE) circle (3 pt);
    \filldraw[color = orange(ryb)] (uNW) circle (3 pt);

    \node at (-2.6/2, -0.7/2) {\scriptsize $\bB{X}$}; 
    \node at (2.6/2, 0.7/2) {\scriptsize $\bB{X}$};
    \node at (-1.3/2, 0.7/2) {\scriptsize $\bB{X}$};
    \node at (1.3/2, -0.7/2) {\scriptsize $\bB{X}$};
    \node at (0, 1) {\scriptsize $\bB{X}$};
    \node at (0, -1.) {\scriptsize $\bB{X}$};
    
    \end{tikzpicture}, 
    \quad&
    \begin{tikzpicture}[scale = 0.4, baseline={([yshift=-.5ex]current bounding box.center)}]
    \coordinate (A) at (0, 0);
    \coordinate (Azp) at (0, 2);
    \coordinate (Azm) at (0, -1.6); 
    \coordinate (B) at (0.7, 0.7);
    \coordinate (Bzp) at (0.7, 0.7 + 2);
    \coordinate (C) at (2.7, 0.7);
    \coordinate (Czp) at (2.7, 2.7);
    \coordinate (D) at (2, 0);
    \coordinate (Dzp) at (2, 2);
    \coordinate (E) at (-0.7, -0.7);
    \coordinate (Ezp) at (-0.7, -0.7+2);
    \coordinate (F) at (-2.7, -0.7);
    \coordinate (Fzp) at (-2.7, -0.7+2);
    \coordinate (G) at (-2, 0);
    \coordinate (Gzp) at (-2, 0+2);
    \coordinate (H) at (-1.3, 0.7);
    \coordinate (Hzp) at (-1.3, 0.7+2);
    \coordinate (I) at (1.3, -0.7);
    \coordinate (Izp) at (1.3, -0.7+2);
    \coordinate (tNE) at (1+0.35,2+0.35);
    \coordinate (bNE) at (1+0.35,0.35);
    \coordinate (tNW) at (0.35-1,2+0.35);
    \coordinate (tSE) at (0.55,2-0.35);
    \coordinate (tSW) at (0.7-2,2-0.35);
    \coordinate (mNE) at (2+0.7,1+0.5+0.2);
    \coordinate (mNW) at (0.7,1+0.5-0.1);
    \coordinate (mSE) at (2,0.85+0.2);
    \coordinate (mSW) at (0,0.85+0.4);

    \draw[color = gray] (A) -- (B) -- (C) -- (D)--cycle;
    \draw[color = gray] (Azp) -- (Bzp) -- (Czp) -- (Dzp)--cycle;
    \draw[color = gray] (B) -- (Bzp);
    \draw[color = gray] (C) -- (Czp);
    \draw[color = gray] (D) -- (Dzp);
    \draw[color = gray] (A) -- (E) -- (F) -- (G)--cycle;
    \draw[color = gray] (A) -- (B) -- (H) -- (G)--cycle;
    \draw[color = gray] (A) -- (E) -- (I) -- (D)--cycle;
    \draw[color = gray] (Azm) -- (Azp);    


    \draw[color = cz1color, line width = 1pt]  (-1,0) -- (mNW);
    \draw[color = cz2color, line width = 1pt]  (-0.35-2, -0.35) -- (mSW);
    \draw[color = cz2color, line width = 1pt]  (-1, 0) -- (mSW);
    \draw[color = cz2color, line width = 1pt]  (-0.35, -0.35) -- (mSE);
    
    \draw[color = cz1color, line width = 1pt]  (0, -1.+0.2) -- (bNE);

    \filldraw[color = dodgerblue] (bNE) circle (3 pt);

    \filldraw[color = dodgerblue] (mNW) circle (3 pt);
    \filldraw[color = dodgerblue] (mSE) circle (3 pt);
    \filldraw[color = dodgerblue] (mSW) circle (3 pt);

    \filldraw[color = crimson] (0, -1.+0.2) circle (3 pt);
    \filldraw[color = crimson] (-1, 0) circle (3 pt);
    \filldraw[color = crimson] (-0.35, -0.35) circle (3 pt);
    \filldraw[color = crimson] (-0.35-2, -0.35) circle (3 pt);

    \node at (-2.6/2, -0.7/2) {\scriptsize $\oC{X}$}; 
    \node at (2.6/2, 0.7/2) {\scriptsize $\oC{X}$};
    \node at (-1.3/2, 0.7/2) {\scriptsize $\oC{X}$};
    \node at (1.3/2, -0.7/2) {\scriptsize $\oC{X}$};
    \node at (0, 1) {\scriptsize $\oC{X}$};
    \node at (0, -1.) {\scriptsize $\oC{X}$};
    
    \end{tikzpicture},
    \quad
    \begin{tikzpicture}[scale = 0.4, baseline={([yshift=-.5ex]current bounding box.center)}]
    \coordinate (A) at (0, 0);
    \coordinate (Azp) at (0, 2);
    \coordinate (Azm) at (0, -1.6); 
    \coordinate (B) at (0.7, 0.7);
    \coordinate (Bzp) at (0.7, 0.7 + 2);
    \coordinate (C) at (2.7, 0.7);
    \coordinate (Czp) at (2.7, 2.7);
    \coordinate (D) at (2, 0);
    \coordinate (Dzp) at (2, 2);
    \coordinate (E) at (-0.7, -0.7);
    \coordinate (F) at (-2.7, -0.7);
    \coordinate (G) at (-2, 0);
    \coordinate (H) at (-1.3, 0.7);
    \coordinate (I) at (1.3, -0.7);
    
    \draw[color = gray] (A) -- (B) -- (C) -- (D)--cycle;
    \draw[color = gray] (Azp) -- (Bzp) -- (Czp) -- (Dzp)--cycle;
    \draw[color = gray] (B) -- (Bzp);
    \draw[color = gray] (C) -- (Czp);
    \draw[color = gray] (D) -- (Dzp);
    \draw[color = gray] (A) -- (E) -- (F) -- (G)--cycle;
    \draw[color = gray] (A) -- (B) -- (H) -- (G)--cycle;
    \draw[color = gray] (A) -- (E) -- (I) -- (D)--cycle;
    \draw[color = gray] (Azm) -- (Azp);    

    \draw[color = cz1color, line width = 1pt]  (1+0.35,0.35) -- (2+0.7,1+0.7);
    \draw[color = cz2color, line width = 1pt]  (0,1) -- (1+0.35,2+0.35);

    \filldraw[color = orange(ryb)] (0,1) circle (3 pt);
    \filldraw[color = orange(ryb)] (1+0.35,0.35) circle (3 pt);
    \filldraw[color = dodgerblue] (1+0.35,2+0.35) circle (3 pt);
    \filldraw[color = dodgerblue] (2+0.7,1+0.7) circle (3 pt);

    \node at (1,0) {\scriptsize $\AR{X}$};
    \node at (-1,0) {\scriptsize $\AR{X}$};
    \node at (0,1+0.3) {\scriptsize $\AR{X}$};
    \node at (0,-1) {\scriptsize $\AR{X}$};
    \node at (0.35,0.35) {\scriptsize $\AR{X}$};
    \node at (-0.35,-0.35) {\scriptsize $\AR{X}$};

    \end{tikzpicture}, \\
    \begin{tikzpicture}[scale = 0.55, baseline = {([yshift=-.5ex]current bounding box.center)}]
        \draw[lightcrimson] (0,0) -- (1.5, 0) -- (1.5,1.5) -- (0,1.5) -- cycle;
        \node at (0.75, 0) {\normalsize $\AR{Z}$};
        \node at (0.75, 1.5) {\normalsize $\AR{Z}$};
        \node at (0, 0.75) {\normalsize $\AR{Z}$};
        \node at (1.5, 0.75) {\normalsize $\AR{Z}$};
    \end{tikzpicture}, 
    \quad &
    \begin{tikzpicture}[scale = 0.55, baseline = {([yshift=-.5ex]current bounding box.center)}]
        \draw[lightcrimson] (0,0) -- (1, 1) -- (1.5+1,1) -- (1.5,0) -- cycle;
        \node at (0.75, 0) {\scriptsize $\AR{Z}$};
        \node at (0.5, 0.5) {\scriptsize $\AR{Z}$};
        \node at (1+0.75, 1) {\scriptsize $\AR{Z}$};
        \node at (2, 0.5) {\scriptsize $\AR{Z}$};
    \end{tikzpicture},
    \quad
    \begin{tikzpicture}[scale = 0.55, baseline = {([yshift=-.5ex]current bounding box.center)}]
        \draw[lightcrimson] (0,0) -- (1, 1) -- (1,1+1.5) -- (0,1.5) -- cycle;
        \node at (0.5, 0.5) {\scriptsize $\AR{Z}$};
        \node at (0, 0.85) {\scriptsize $\AR{Z}$};
        \node at (1, 1+0.65) {\scriptsize $\AR{Z}$};
        \node at (0.5, 2) {\scriptsize $\AR{Z}$};
    \end{tikzpicture},
\end{aligned}
\end{equation}
where the green lines (shaded differently for easier visualization) all represent $\mathsf{CZ}$ gates.
A detailed derivation of these stabilizers is deferred to Appendix ~\ref{append-sub:fracton-from-graph-gauging}. We note that above stabilizers only commute up to stabilizers; they need to be symmetrized with respect to the $U_{\mathsf{CZ}}$ symmetry to form commuting projectors.

\subsection{Non-Abelian Fractons by Gauging Addressable Gates}\label{subsec:addressable-gauging}

Having obtained a non-Abelian ALP code from the graph gauging method of Section~\ref{sec:gauging-graphs}, we now show that the ALP code can alternatively be gauged using the homological formalism of Section~\ref{sec:addressable}.
To do so, we first point out that, while we so far have discussed a global transversal $\mathsf{CZ}$ gate for two copies of the ALP code of Eq.~\eqref{eq:qubit-ALP-stabilizers}, two ALP codes in fact have \textit{addressable} $\mathsf{CZ}$ gates between them, which take the following form:

\begin{equation} \label{eq-addressableCZ}
    \mathcal{U}_{\mathsf{CZ}}[\boldsymbol{\mu}_y] = \prod_{\mathbf{v} \in \boldsymbol{\mu}_y}\,  \begin{tikzpicture}[scale = 0.5, baseline={([yshift=-.5ex]current bounding box.center)}]
    \coordinate (A) at (0, 0);
    \coordinate (Azp) at (0, 2); 
    \coordinate (Azm) at (0, -2); 
    \coordinate (B) at (0.7, 0.7);
    \coordinate (Bzp) at (0.7, 2.7);
    \coordinate (Bzm) at (0.7, -1.3);
    \coordinate (C) at (2.7, 0.7);
    \coordinate (D) at (2, 0);
    \coordinate (E) at (-0.7, -0.7);
    \coordinate (F) at (-2.7, -0.7);
    \coordinate (G) at (-2, 0);
    \coordinate (H) at (-1.3, 0.7);
    \coordinate (I) at (1.3, -0.7);
    \draw[color = black, line width = 0.7pt]  (0.7, -0.65) -- (1.35, 0.35);
    \draw[color = black, line width = 0.7pt]  (0, 0.9) -- (-0.65, 0.35);
    \draw[color = black, line width = 0.7pt]  (0.7, 2.) -- (-0.65, 0.35);
    \draw[color = gray] (Azm) -- (Azp);
    \draw[color = gray] (Bzm) -- (Bzp);%
    \draw[color = black, line width = 0.7pt]  (0, -1) -- (1.35, 0.35);
    \draw[color = gray] (A) -- (B) -- (C) -- (D)--cycle;
    \draw[color = gray] (A) -- (B) -- (H) -- (G)--cycle;
    \filldraw[color = dodgerblue] (-0.65, 0.35) circle (3 pt);
    \filldraw[color = orange(ryb)] (0.7, 2.) circle (3 pt);
    \filldraw[color = orange(ryb)] (0, 0.9) circle (3 pt);
    \filldraw[color = dodgerblue] (0, -1) circle (3 pt);
    \filldraw[color = dodgerblue] (0.7, -0.65) circle (3 pt);
    \filldraw[color = orange(ryb)]  (1.35, 0.35) circle (3 pt);
    \filldraw[color = black]  (B) circle (2 pt);
    \node at (0.7, 1) {\scriptsize $\mathbf{v}$};
    %
\end{tikzpicture}
\end{equation}
where $\boldsymbol{\mu}_y$ is the collection of vertices along one of the $xz$ planes of the ALP model at a fixed $y$ (naturally, an identical gate exists for $yz$ planes at fixed $x$).
Crucially, this means that the above transversal gate only acts on a single $xz$ plane of qubits.
Nevertheless, one can readily check that the above gates leave the stabilizer group of the ALP model invariant and implements a logical $\mathsf{CZ}$ gate on an $\mathcal{O}(1)$ number of logical qubits of the ALP code (and are hence addressable).
The construction of these gates and the cup product we will associate with them in the next subsubsection is the subject of a forthcoming work \cite{addressablegates}.

\subsubsection{A Cup Product for the ALP Code}

We will now show that these addressable $\mathsf{CZ}$'s can be gauged using the homological formalism of Section~\ref{sec:addressable}.
To do so, we must associate the gates with a cup product that satisfies the Leibniz rule.
To define this cup product, we first note that the ALP code can be expressed as a \textit{hypergraph product} between two classical codes~\cite{Tillich_2014}---namely, the 1D classical Ising model and the 2D plaquette Ising model: 
\begin{equation}
    H_{\text{1D}} = -\sum_{x = 1}^{N} Z_{x} Z_{x + 1} \qquad H_{\text{2D}} = -\sum_{p}\,  \begin{tikzpicture}[scale = 0.5, baseline={([yshift=-.5ex]current bounding box.center)}]
    \draw[color = gray] (0, 0) -- (1.5, 0) -- (1.5, 1.5) -- (0, 1.5)--cycle;
    \node at (0,0) {\small $Z$};
    \node at (1.5,0) {\small $Z$};
    \node at (0,1.5) {\small $Z$};
    \node at (1.5,1.5) {\small $Z$};
    \end{tikzpicture}
\end{equation}
where in the first sum, we implicitly assume periodic boundary conditions, and the second sum is performed over plaquettes on the 2D square lattice with the global topology of a torus.

We can also understand the 1D classical Ising model and 2D plaquette Ising model in the homological code language via a chain complex.
In particular, let $I_1 \to I_0$ be the chain complex of the Ising model, which consists of the space of edges and vertices of the $1$D line with the usual boundary map between them;  $Z$-checks associated with $I_1$ and its bits associated with $I_0$.
Similarly, let $P_1 \to P_0$ be the chain complex of the plaquette Ising model with $P_1$ and $P_0$ being the space of plaquettes and vertices on the 2D square lattice respectively.
The ALP chain complex of Eq.~\eqref{eq-ALPchaincomplex} can be expressed as a tensor product of the $(1 + 1)$D chain complexes associated with Ising and Plaquette Ising models:
\begin{equation}
        I^1 \otimes P^1 \xleftarrow{\exd_{1} }(I^0 \otimes P^1) \oplus (I^1 \otimes P^0)\xleftarrow{\exd_{0}} I^0 \otimes P^0,
\end{equation}
cementing it as a hypergraph product of the two codes.

Given the hypergraph construction of the ALP code, we can express a cup product for these codes in terms of cup products of their constituent codes.
In particular, the cup product between two cochains $\mathbf{a} = (\mathbf{a}_I, \mathbf{a}_P)$ and $\mathbf{b} = (\mathbf{b}_I, \mathbf{b}_P)$ will be $\mathbf{a} \cupp \mathbf{b} = (\mathbf{a}_I \cupp \mathbf{b}_I, \mathbf{a}_P \cupp \mathbf{b}_P)$.
The cup product for the Ising model is well known and is given by the following rules: $\mathbf{v} \cupp \mathbf{w} = \delta_{\mathbf{v}, \mathbf{w}} \mathbf{v}$ and $\mathbf{v} \cupp \mathbf{e} = \mathbf{e}$ ($\mathbf{e} \cupp \mathbf{v} = \mathbf{e}$) if $\mathbf{e}$ is directly to the right (left) of $\mathbf{v}$.
For the plaquette Ising model, we will operationally define the cup product with the following rules\footnote{Strictly speaking, this cup product is only sufficient to generate the addressable $\mathsf{CZ}$ gates in the $xz$ plane. 
A more complete cup product will appear in Ref.~\cite{addressablegates}.
}:
\begin{equation} \label{eq-cupproductrules}
    \begin{tikzpicture}[scale = 0.4, baseline={([yshift=-.5ex]current bounding box.center)}]
    \foreach \i in {0, ..., 4}{
        \foreach \j in {0, ..., 5}{
            \filldraw[color = black] (\i, \j) circle (2 pt);
        }
    }
    \foreach \i in {0, ..., 4}{
    \foreach \j in {0, ..., 5}{
            \draw[color = black] (\i, \j) -- (\i + 1, \j);
            \draw[color = black] (\i, 6) -- (\i + 1, 6);
            \draw[color = black] (\i, \j) -- (\i, \j + 1);
            \draw[color = black] (5, \j) -- (5, \j + 1);
        }
    }
    \filldraw[color = red] (1 + 1, 1) circle (3 pt);
    \filldraw[color = red] (1 + 1, 2) circle (3 pt);
    \filldraw[color = red] (1 + 1, 3) circle (3 pt);
    \node at (0.5 + 1, 1) {\scriptsize $\textcolor{red}{u}$};
    \node at (0.5 + 1, 2) {\scriptsize $\textcolor{red}{v}$};
    \node at (0.5 + 1, 3) {\scriptsize $\textcolor{red}{w}$};
    \node at (0.5 + 1, 1.5) {\scriptsize $A$};
    \node at (0.5 + 1, 2.5) {\scriptsize$C$};
    \node at (1.5 + 1, 2.5) 
    {\scriptsize$D$};
    \node at (1.5 + 1, 1.5) 
    {\scriptsize$B$};
    \node at (0.5 + 1, 3.5) {\scriptsize$E$};
    \node at (1.5 + 1, 3.5) {\scriptsize$F$};
    \node at (7, 4) {\scriptsize $\mathbf{v} \cupp \mathbf{w} = \mathbf{w}$};
    \node at (7.5, 3) {\scriptsize $\mathbf{v} \cupp \mathbf{v} = \mathbf{v} + \mathbf{w}$};
    \node at (6.9, 2) {\scriptsize $\mathbf{v} \cupp \mathbf{u} = \mathbf{v}$};
    \node at (14, 4) {\scriptsize $\mathbf{v} \cupp \mathbf{D} = \mathbf{D} + \mathbf{F}$};
    \node at (14, 3) {\scriptsize $\mathbf{v} \cupp \mathbf{B} = \mathbf{B} + \mathbf{D}$};
    \node at (14, 2) {\scriptsize $\mathbf{C} \cupp \mathbf{v} = \mathbf{C} + \mathbf{E}$};
    \node at (14, 1) {\scriptsize $\mathbf{A} \cupp \mathbf{v} = \mathbf{A} + \mathbf{C}$};
\end{tikzpicture}
\end{equation}
which we take to be translationally invariant across the lattice; one can check that this cup product satisfies the Leibniz rule of Eq.~\eqref{eq-Leibniz}.

Cup product in hand, the only remaining ingredient that we need from our homological formalism is a fundamental class $\mathcal{M}_{y'}$, which is some chosen element of the second homology group $H_2$.
For hypergraph product codes, the K\"unneth formula dictates that $H_2 = (H_1)_I \otimes (H_1)_P$ \cite{homologicalorder}.
Let us recall that for $(1 + 1)$D codes, $H_1 \simeq \mathsf{ker}(\partial)$, corresponding to sets of classical checks that product to one.
In the Ising model, the only such product is the global product over all checks, implying that $(H_1)_I = \mathbb{Z}_2$ and is generated by $\mathcal{M}^I = \sum_{e} e$.
In the Plaquette Ising model, any product of plaquettes around a ``strip'' of the square lattice products to the identity; these strips form a basis for $(H_1)_{P}$.

For our purposes, let us arbitrarily pick the fundamental class of the plaquette Ising model to be of the form:
\begin{equation} \label{eq-globalredundancy}
        \mathcal{M}^P_{y'} = \begin{tikzpicture}[scale = 0.4, baseline={([yshift=-.5ex]current bounding box.center)}]
        \filldraw[color = red, opacity = 0.2] (0, 3) -- (0, 6) -- (6, 6) -- (6, 3);
    \foreach \i in {0, ..., 5}{
    \foreach \j in {0, ..., 5}{
            \draw[color = gray] (\i, \j) -- (\i + 1, \j);
            \draw[color = gray] (\i, 6) -- (\i + 1, 6);
            \draw[color = gray] (\i, \j) -- (\i, \j + 1);
            \draw[color = gray] (6, \j) -- (6, \j + 1);
        }
    }
    \foreach \i in {0, ..., 5}{
        \foreach \j in {0, ..., 5}{
            \filldraw[color = black] (\i, \j) circle (2 pt);
        }
    }
    \node at (-0.5, 3) {\scriptsize $y'$};
    \draw[color = black, line width = 1 pt] (0,3) -- (6, 3);
    \end{tikzpicture} = \sum_{x} \sum_{y \geq y'} p_{x, y} \in H_1^P
\end{equation}
where $p_{x, y} \in P_1$ denotes a plaquette chain on the square lattice, whose lower left corner has coordinate $(x, y)$ and the coordinate $y'$ is chosen arbitrarily.
This implies that the fundamental class we will use for the ALP code is $\mathcal{M}_{y'} = \mathcal{M}_{y'}^P \otimes \mathcal{M}^I$, which graphically is the set of all cubes whose $y$ coordinate is greater than $y'$.
With $\mathcal{M}_{y'}$, one can easily show through explicit computation that our addressable gate in Eq.~\eqref{eq-addressableCZ} can be expressed in a cup product form as: 
\begin{equation}
    U_{\mathsf{CZ}}[\boldsymbol{\mu}_{y'}] = \prod_{e,e'} \mathsf{CZ}_{\bB{e}, \oC{e'}}^{\int_{\mathcal{M}_{y'}} \mathbf{e} \cupp \mathbf{e}'}
\end{equation}
as desired.

\subsubsection{Gauging the Addressable $\mathsf{CZ}$ Gates}

With the cup product and fundamental class from the previous section, our goal now is to gauge the addressable gates of the previous section.
To do so, we start by noting that the cluster state we utilize for gauging will be the cluster state associated with the chain complex of the ALP code, whose stabilizers are: 
\begin{equation}
    \begin{tikzpicture}[scale = 0.5, baseline={([yshift=-.5ex]current bounding box.center)}]
    \coordinate (A) at (0, 0);
    \coordinate (Azp) at (0, 1.5); 
    \coordinate (Azm) at (0, -1.5); 
    \coordinate (B) at (0.7, 0.7);
    \coordinate (C) at (2.7, 0.7);
    \coordinate (D) at (2, 0);
    \coordinate (E) at (-0.7, -0.7);
    \coordinate (F) at (-2.7, -0.7);
    \coordinate (G) at (-2, 0);
    \coordinate (H) at (-1.3, 0.7);
    \coordinate (I) at (1.3, -0.7);
    \draw[color = gray] (A) -- (B) -- (C) -- (D)--cycle;
    \draw[color = gray] (A) -- (E) -- (F) -- (G)--cycle;
    \draw[color = gray] (A) -- (B) -- (H) -- (G)--cycle;
    \draw[color = gray] (A) -- (E) -- (I) -- (D)--cycle;
    \draw[color = gray] (Azm) -- (Azp);

    \node at (-2.6/2, -0.7/2) {\small $\textcolor{red}{X}$}; 
    \node at (2.6/2, 0.7/2) {\small $\textcolor{red}{X}$};
    \node at (-1.3/2, 0.7/2) {\small $\textcolor{red}{X}$};
    \node at (1.3/2, -0.7/2) {\small $\textcolor{red}{X}$};
    \node at (0, 1.) {\small $\textcolor{red}{X}$};
    \node at (0, -1.) {\small $\textcolor{red}{X}$};
    \node at (0,0) {\small $\aP{X}$};
\end{tikzpicture} \qquad    \begin{tikzpicture}[scale = 0.5, baseline={([yshift=-.5ex]current bounding box.center)}]
    \coordinate (A) at (0, 0);
    \coordinate (Azp) at (0, 1.5); 
    \coordinate (Azm) at (0, -1.5); 
    \coordinate (B) at (0.7, 0.7);
    \coordinate (C) at (2.7, 0.7);
    \coordinate (D) at (2, 0);
    \coordinate (E) at (-0.7, -0.7);
    \coordinate (F) at (-2.7, -0.7);
    \coordinate (G) at (-2, 0);
    \coordinate (H) at (-1.3, 0.7);
    \coordinate (I) at (1.3, -0.7);
    \draw[color = gray] (A) -- (B) -- (C) -- (D)--cycle;
    \node at (A) {\small $\aP{Z}$};
    \node at (B) {\small $\aP{Z}$};
    \node at (C) {\small $\aP{Z}$};
    \node at (D) {\small $\aP{Z}$};
    \node at (1.35, 0.35) {\small $\AR{Z}$};
\end{tikzpicture} \qquad 
\begin{tikzpicture}[scale = 0.5, baseline={([yshift=-.5ex]current bounding box.center)}]
    \draw[color = gray] (0,0) -- (0, 1.5);
    \node at (0, 1.5) {\small $\aP{Z}$};
    \node at (0, 0) {\small $\aP{Z}$};
    \node at (0, 0.75) {\small $\AR{Z}$};
\end{tikzpicture}
\end{equation}
Moreover, in our homological formalism, we were able to explicitly write down the symmetry enrichment operator given the fundamental class and cup product: 
\begin{align}
\Omega_{\aP{a}, \bB{B}\oC{C}} &= \prod_{v_0,e,e'} (\mathsf{CCZ}_{\aP{v}, \bB{e}, \oC{e'}})^{\int_{\mathcal{M}_{y'}}\mathbf{v}\cupp (\mathbf{e}\cupp\mathbf{e'})}\\
&= \prod_{\aP{\mathbf{v}} \in \boldsymbol{\mu}_{y'}}  
\begin{tikzpicture}[scale = 0.52, baseline={([yshift=-.5ex]current bounding box.center)}]
    \coordinate (A) at (0, 0);
    \coordinate (Azp) at (0, 2);
    \coordinate (Azm) at (0, -2); 
    \coordinate (B) at (0.7, 0.7);
    \coordinate (Bzm) at (0.7, 0.7 -2);
    \coordinate (Bzp) at (0.7, 0.7 + 2);
    \coordinate (C) at (2.7, 0.7);
    \coordinate (Czm) at (2.7, 0.7 -2);
    \coordinate (Czp) at (2.7, 2.7);
    \coordinate (D) at (2, 0);
    \coordinate (Dzm) at (2, -2);
    \coordinate (Dzp) at (2, 2);
    \coordinate (E) at (-0.7, -0.7);
    \coordinate (Ezp) at (-0.7, -0.7 + 2);
    \coordinate (Ezm) at (-0.7, -0.7 - 2);
    \coordinate (F) at (-2.7, -0.7);
    \coordinate (Fzm) at (-2.7, -0.7-2);
    \coordinate (Fzp) at (-2.7, -0.7+2);
    \coordinate (G) at (-2, 0);
    \coordinate (Gzm) at (-2, -2);
    \coordinate (Gzp) at (-2, +2);
    \coordinate (H) at (-1.3, 0.7);
    \coordinate (Hzm) at (-1.3, 0.7 - 2);
    \coordinate (Hzp) at (-1.3, 0.7 + 2);
    \coordinate (I) at (1.3, -0.7);
    \coordinate (Izm) at (1.3, -0.7 -2);
    \coordinate (Izp) at (1.3, -0.7 + 2);
    \draw[color = gray] (A) -- (B) -- (C) -- (D)--cycle;
    \draw[color = gray] (A) -- (E) -- (I) -- (D)--cycle; 
    \draw[color = gray] (Azm) -- (A);   
    \draw[color = gray] (Bzm) -- (B);
    \draw[color = gray] (C) -- (Czp);
    \draw[color = gray] (D) -- (Dzp);
    \draw[color = gray] (Ezm) -- (E);
    \draw[color = gray] (I) -- (Izp);
    \draw[color = black, line width = 1.25 pt] (B) -- (C);
    \node at (B) {\scriptsize \aP{$\mathsf{C}$}};
    \draw[color = cz1color, line width = \CZlwtwo pt] (0, -1.3) -- (1.35 - 0.45, 0.35);
    \draw[color = cz1color, line width = \CZlwtwo pt] (0, -1.3) -- (0.65 - 0.4, -  0.35);
    \draw[color = cz1color, line width = \CZlwtwo pt] (-0.7, -1.7) -- (0.65 - 0.4, -  0.35);
    \draw[color = cz1color, line width = \CZlwtwo pt] (0.7, -0.9) -- (1.35 - 0.45, 0.35);
    \draw[color = cz2color, line width = \CZlwtwo pt] (1.35 + 0.3, 0.35) --  (2.7, 2.3);
    \draw[color = cz2color, line width = \CZlwtwo pt] (1.35 + 0.3, 0.35) --  (2, 1.6);
    \draw[color = cz2color, line width = \CZlwtwo pt] (0.65 + 0.4, -0.35) --  (2, 1.6);
    \draw[color = cz2color, line width = \CZlwtwo pt] (0.65 + 0.4, -0.35) -- (1.3, 0.9);
    \filldraw[color = dodgerblue] (0, -1.3) circle (3pt);
    \filldraw[color = dodgerblue] (-0.7, -1.7) circle (3pt);
    \filldraw[color = dodgerblue] (0.7, -0.9) circle (3pt);
    \filldraw[color = dodgerblue] (1.35 + 0.3, 0.35) circle (3pt);
    \filldraw[color = dodgerblue] (0.65 + 0.4, -0.35) circle (3pt);
    \filldraw[color = orange(ryb)] (1.35 - 0.45, 0.35) circle (3pt);
    \filldraw[color = orange(ryb)] (0.65 - 0.4, -0.35) circle (3pt);
    \filldraw[color = orange(ryb)] (2, 1.6) circle (3pt);
    \filldraw[color = orange(ryb)] (1.3, 0.9) circle (3pt);
    \filldraw[color = orange(ryb)] (2.7, 2.3) circle (3pt);
\end{tikzpicture} \times 
\prod_{\aP{\mathbf{v}} \in \boldsymbol{\mu}_{y'-1}} \begin{tikzpicture}[scale = 0.52, baseline={([yshift=-.5ex]current bounding box.center)}]
    \coordinate (A) at (0, 0);
    \coordinate (Azp) at (0, 2);
    \coordinate (Azm) at (0, -2); 
    \coordinate (B) at (0.7, 0.7);
    \coordinate (Bzm) at (0.7, 0.7 -2);
    \coordinate (Bzp) at (0.7, 0.7 + 2);
    \coordinate (C) at (2.7, 0.7);
    \coordinate (Czm) at (2.7, 0.7 -2);
    \coordinate (Czp) at (2.7, 2.7);
    \coordinate (D) at (2, 0);
    \coordinate (Dzm) at (2, -2);
    \coordinate (Dzp) at (2, 2);
    \coordinate (E) at (-0.7, -0.7);
    \coordinate (Ezp) at (-0.7, -0.7 + 2);
    \coordinate (Ezm) at (-0.7, -0.7 - 2);
    \coordinate (F) at (-2.7, -0.7);
    \coordinate (Fzm) at (-2.7, -0.7-2);
    \coordinate (Fzp) at (-2.7, -0.7+2);
    \coordinate (G) at (-2, 0);
    \coordinate (Gzm) at (-2, -2);
    \coordinate (Gzp) at (-2, +2);
    \coordinate (H) at (-1.3, 0.7);
    \coordinate (Hzm) at (-1.3, 0.7 - 2);
    \coordinate (Hzp) at (-1.3, 0.7 + 2);
    \coordinate (I) at (1.3, -0.7);
    \coordinate (Izm) at (1.3, -0.7 -2);
    \coordinate (Izp) at (1.3, -0.7 + 2);
    \draw[color = gray] (A) -- (B) -- (C) -- (D)--cycle;
    \draw[color = gray] (A) -- (E) -- (I) -- (D)--cycle; 
    \draw[color = gray] (Azm) -- (A);   
    \draw[color = gray] (Bzm) -- (B);
    \draw[color = gray] (C) -- (Czp);
    \draw[color = gray] (D) -- (Dzp);
    \draw[color = gray] (Ezm) -- (E);
    \draw[color = gray] (I) -- (Izp);
    \draw[color = black, line width = 1.25 pt] (B) -- (C);
    \node at (A) {\scriptsize \aP{$\mathsf{C}$}};
    \draw[color = cz1color, line width = \CZlwtwo pt] (0, -1.3) -- (1.35 - 0.45, 0.35);
    \draw[color = cz1color, line width = \CZlwtwo pt] (0, -1.3) -- (0.65 - 0.4, -  0.35);
    \draw[color = cz1color, line width = \CZlwtwo pt] (-0.7, -1.7) -- (0.65 - 0.4, -  0.35);
    \draw[color = cz1color, line width = \CZlwtwo pt] (0.7, -0.9) -- (1.35 - 0.45, 0.35);
    \draw[color = cz2color, line width = \CZlwtwo pt] (1.35 + 0.3, 0.35) --  (2.7, 2.3);
    \draw[color = cz2color, line width = \CZlwtwo pt] (1.35 + 0.3, 0.35) --  (2, 1.6);
    \draw[color = cz2color, line width = \CZlwtwo pt] (0.65 + 0.4, -0.35) --  (2, 1.6);
    \draw[color = cz2color, line width = \CZlwtwo pt] (0.65 + 0.4, -0.35) -- (1.3, 0.9);
    \filldraw[color = dodgerblue] (0, -1.3) circle (3pt);
    \filldraw[color = dodgerblue] (-0.7, -1.7) circle (3pt);
    \filldraw[color = dodgerblue] (0.7, -0.9) circle (3pt);
    \filldraw[color = dodgerblue] (1.35 + 0.3, 0.35) circle (3pt);
    \filldraw[color = dodgerblue] (0.65 + 0.4, -0.35) circle (3pt);
    \filldraw[color = orange(ryb)] (1.35 - 0.45, 0.35) circle (3pt);
    \filldraw[color = orange(ryb)] (0.65 - 0.4, -0.35) circle (3pt);
    \filldraw[color = orange(ryb)] (2, 1.6) circle (3pt);
    \filldraw[color = orange(ryb)] (1.3, 0.9) circle (3pt);
    \filldraw[color = orange(ryb)] (2.7, 2.3) circle (3pt);
\end{tikzpicture}
\end{align}
where in the above the $\aP{\mathsf{C}}$ represents the first (vertex) control of the $\mathsf{CCZ}$ and the solid lines indicate the $\mathsf{CZ}$'s controlled by this vertex.
Moreover, the thick black line above is meant to represent the $xz$ plane at the edge of the $\mathcal{M}_{y}$'s support [c.f. Eq.~\eqref{eq-globalredundancy}].
Moreover, as with the addressable $\mathsf{CZ}$ gate of Eq.~\eqref{eq-addressableCZ},  $\boldsymbol{\mu}_{y'}$ and $\boldsymbol{\mu}_{y'-1}$ are the collection of vertices along one of the $xz$ planes of the ALP model at a fixed $y'$ and $y' - 1$ respectively.
Notice that similar to the addressable $\mathsf{CZ}$ gates, the above symmetry enrichment operator acts on only part of the pair of ALP codes.
Consequently, we can anticipate that the result will only gauge a few of the addressable $\mathsf{CZ}$'s (in fact two) and the result will be an ordinary Abelian qLDPC code with a non-Abelian slab living in its $xz$ plane!

We can see this intuition borne out by explicitly writing down the gauged stabilizers.
In particular, recall that they take the form of Eqs.~\eqref{eq-homologicalA}~-~\eqref{eq-homologicalC}.
With some care, we can compute these explicitly using the cup product for the ALP model.
In particular, the $Z$-stabilizers of the code are given by $Z_{\partial p}$ on each sublattice as expected.
Moreover, the $\AR{A}$ $X$-stabilizers are mostly left unchanged, except in the vicinity of the $xz$ plane at $y'$.
In particular, for vertices that are adjacent to this $xz$ plane:
\begin{equation}
    \textcolor{red}{A_{\mathbf{v}}} = \textcolor{red}{X_{\exd \mathbf{v}}} \cdot 

\end{equation}
where the location of $\AR{\mathbf{v}}$ indicates the location of the $X$ stabilizer's vertex relative to the $xz$ plane at $y'$.
Similarly, we can compute the $\bB{B}$ stabilizers and we find them to be similarly largely unchanged except near the edge of $\mathcal{M}_{y'}$:
\begin{align}
    \textcolor{dodgerblue}{A_{\mathbf{v}}} = &\textcolor{dodgerblue}{X_{\exd \mathbf{v}}} \cdot     %
\end{align}
We conclude by reporting the results for the $\oC{C}$ stabilizers, which proceed similarly:
\begin{align}
    \textcolor{orange(ryb)}{A_{\mathbf{v}}} = &\textcolor{orange(ryb)}{X_{\exd \mathbf{v}}} \cdot  
    %
\end{align}

\subsection{Non-Abelian Bivariate Bicycle Codes from Gauging Non-Local $\mathsf{CZ}$s}

Here, we take the input qLDPC code to be a bivariate bicycle (BB) code \cite{Bravyi_2024}. These codes have recently garnered significant attention for surpassing the encoding rate and distance scaling of the surface code while requiring only a small number of additional long-range connections, making them attractive candidates for practical implementation \cite{xu2025batchedhighratelogicaloperations,yoder2025tourgrossmodularquantum,eberhardt2024logicaloperatorsfoldtransversalgates,cross2025improvedqldpcsurgerylogical}.
To construct a non-Abelian BB code, we will gauge the transversal $\mathsf{CZ}$ gate acting between the two copies of a BB code, using the ancilla graph construction of Sec.~\ref{sec:gauging-graphs}.

The qubits of the bicycle code reside on a $L_x \times L_y$ square lattice with periodic boundary conditions. 
It is convenient to label the qubits on horizontal edges as $H$ and those on vertical edges as $V$. 
Similar to the square lattice toric code, every $H$ qubit and $V$ qubit can be associated to a single vertex (and plaquette) at position $(i,j)$.

The vertex and plaquette checks of a BB code at position $(i,j)$ each contain three $H$ and three $V$ qubits. Which qubits participate in each check are specified by a pair of polynomials in the commuting shift matrices $x=S_{L_x}\otimes I_{L_y}$ and $y=I_{L_x}\otimes S_{L_y}$ where $S_{n}$ denotes the cyclic $n\times n$ shift matrix---i.e. in some basis $\{\ket{k}\}_{k = 1,...,n}$, $S_{n} = \sum_{k = 1}^{n} \ket{k + 1} \bra{k}$ where $\ket{n+1} \equiv \ket{1}$---for some $n$ \cite{Haah_2013}.
For the sake of simplicity, and to connect our discussion to the familiar example of the toric code, we restrict ourselves to considering polynomials of the form $f(x,y)=1+x+x^\alpha y^\beta$ and $g(x,y)=1+y+x^a y^b$, though general bivariate polynomials are allowed for both $f$ and $g$. We define the notational shorthand $\bar{x} = x^{-1}$ and $\bar{y} = y^{-1}$.

For a code with $2L_{x}L_y$ total qubits, a vertex parity check (or equivalently an $X$ stabilizer) is specified by associating each monomial in $f(x,y)$ with an $H$ qubit of the check and each monomial in $g(x,y)$ with a $V$ qubit of the check. This definition for the $X$ stabilizers can be succinctly encoded by the parity check matrix $H_X=\left(f(x,y)|g(x,y) \right)$ where the rows of $H_X$ are the individual vertex parity checks, the first $L_xL_y$ columns correspond to $H$ qubits and the last $L_xL_y$ columns correspond to $V$ qubits. For example, the vertex parity at position $(i,j)$ given the polynomials $f(x,y)=1+x+\bar{x}y$ and $g(x,y)=1+y+\bar{x}^2\bar{y}$ will include $H$ qubits at positions $(i,j)$, $(i+1,j)$, $(i-1,j+1)$ and $V$ qubits at positions $(i,j)$, $(i,j+1)$, $(i-2,j-1)$:
\begin{equation}\label{eq:BBstabX}
H_X = \begin{tikzpicture}[scale = 0.5, baseline = {([yshift=-.5ex]current bounding box.center)}]
        \foreach \i in {0, 1, 2, 3, 4}{
            \draw[color = lightgray] (\i, -0.5) -- (\i, 4.5);}
        \foreach \i in {0, 1, 2, 3, 4}{
            \draw[color = lightgray] (-0.5, \i) -- (4.5, \i);}  
        \node at (1.5, 2) {\small $X$};
        \node at (2.5, 2) {\small $X$};
        \node at (2, 2.5) {\small $X$};
        \node at (2, 1.5) {\small $X$};
        \node at (0.5, 3) {\small $X$};
        \node at (0, 0.5) {\small $X$};
\end{tikzpicture}
\end{equation}
For the purpose of illustrating these stabilizers, we choose our unit cell on the square lattice such that the $H$ qubit assigned to position $e_{(i,j)}$ is slightly to the left of vertex $v_{(i,j)}$ and the $V$ qubit assigned to position $e_{(i,j)}$ is slightly below $v_{(i,j)}$. Likewise, a plaquette check is specified by associating each monomial in $g(\overline{x},\overline{y})$ with an $H$ qubit and each monomial in $f(\overline{x},\overline{y})$ with a $V$ qubit, represented by the parity check matrix $H_Z=\left(g(\bar{x},\bar{y})|f(\bar{x},\bar{y}) \right)$. For the same polynomials $f(x,y)$ and $g(x,y)$ as Eq.~\eqref{eq:BBstabX}, the $Z$ stabilizer is given by:

\begin{equation}\label{eq:BBstabZ}
H_Z = \begin{tikzpicture}[scale = 0.5, baseline = {([yshift=-.5ex]current bounding box.center)}]
        \foreach \i in {0, 1, 2, 3, 4}{
            \draw[color = lightgray] (\i, -0.5) -- (\i, 4.5);}
        \foreach \i in {0, 1, 2, 3, 4}{
            \draw[color = lightgray] (-0.5, \i) -- (4.5, \i);}  
        \node at (1, 1.5) {\small $Z$};
        \node at (2, 1.5) {\small $Z$};
        \node at (1.5, 2) {\small $Z$};
        \node at (1.5, 1) {\small $Z$};
        \node at (3, 0.5) {\small $Z$};
        \node at (3.5, 3) {\small $Z$};
\end{tikzpicture}
\end{equation}
Concretely, we can think of these codes as copies of the toric code (originating from the $1+x$ term and $1+y$ term of the $f(x,y)$ and $g(x,y)$ polynomials) where each edge of the square lattice participates in an additional third vertex check specified by $\alpha,\beta$ and plaquette check specified by $a,b$. Two copies of these BB codes (which we again label \BandC) have a transversal $\mathsf{CZ}$ gate given by:
\begin{equation}\label{eq:BBSymm}
    U_{\mathsf{CZ}}=\prod_{(i,j)}(-1)^{\textcolor{dodgerblue}{n^B_{(i,j),H}}\textcolor{orange(ryb)}{n^C_{-(i,j),V}}+\textcolor{dodgerblue}{n^B_{(i,j),V}}\textcolor{orange(ryb)}{n^C_{-(i,j),H}}}
\end{equation}
If we flip the orientation of the \textcolor{dodgerblue}{B} code relative to the $\textcolor{orange(ryb)}{C}$ code about a chosen origin, this $\mathsf{CZ}$ gate can be represented as:

\newcommand{\bb}[2]{
\begin{scope}[xshift=#1]
    \def\Nx{2}
    \def\Ny{2}
    \def\Nyy{3}
    \def\Nxx{3}
    \def\Nz{1} 
    \def\bottomshift{-0.25} 

    \coordinate (eone) at (2,0);
    \coordinate (etwo) at (0.8,0.6);
    \coordinate (ez)   at (0,2.8);

    \foreach \k in {0,...,\Nz} {
        \ifnum\k=0 \def\sx{\bottomshift} \else \def\sx{0} \fi
        
        \foreach \i in {0,...,\Nxx} {
            \foreach \j in {0,...,\Nyy} {
                \coordinate (V-\i-\j-\k) at ($\i*(eone)+\j*(etwo)+\k*(ez)+0.5*(eone) + (\sx, 0)$);
            }
        }
    }

\foreach \i in {0,...,\Nx} {
    \foreach \j in {1,...,\Nyy} {

        \ifnum\i>0
            \pgfmathtruncatemacro{\im}{\i-1}
        \fi

        \coordinate (bottom) at (V-\i-\j-0);
        \coordinate (top)    at ($(V-\i-\j-1) + 0.5*(eone) - .5*(etwo)$);
        \draw[black] (bottom) -- (top);
    }
}

\foreach \i in {0,...,\Nx} {
    \foreach \j in {1,...,\Nyy} {
        \coordinate (bottom) at ($(V-\i-\j-0) + 0.5*(eone) - .5*(etwo)$);
        \coordinate (top)    at (V-\i-\j-1);
        \draw[black] (bottom) -- (top);
    }
}

    \foreach \k in {0,...,\Nz} {
        
        \ifnum\k=\Nz
            \def\col{orange(ryb)}
            \def\sx{0}            
        \else
            \def\col{#2}
            \def\sx{\bottomshift} 
        \fi

        \foreach \i in {0,...,\Nx} {
            \foreach \j in {0,...,\Ny} {
                \coordinate (A) at ($\i*(eone)+\j*(etwo)+\k*(ez) + (\sx, 0)$);
                \coordinate (B) at ($(A)+(etwo)$);
                \coordinate (C) at ($(A)+(eone)+(etwo)$);
                \coordinate (D) at ($(A)+(eone)$);

                \draw[\col] (A)--(B)--(C)--(D)--cycle;

                \coordinate (Eh) at ($(A)!0.5!(D)$);
                \coordinate (Ev) at ($(A)!0.5!(B)$);
                \filldraw[\col] (Eh) circle (2.3pt);
                \filldraw[\col] (Ev) circle (2.3pt);
            }
        }

        \foreach \j in {0,...,\Ny} {
            \coordinate (A) at ($\Nxx*(eone)+\j*(etwo)+\k*(ez) + (\sx, 0)$);
            \coordinate (B) at ($(A)+(etwo)$);
            \coordinate (Ev) at ($(A)!0.5!(B)$);
            \filldraw[\col] (Ev) circle (2.3pt);
        }

        \foreach \i in {0,...,\Nx} {
            \coordinate (A) at ($\i*(eone)+\Nyy*(etwo)+\k*(ez) + (\sx, 0)$);
            \coordinate (D) at ($(A)+(eone)$);
            \coordinate (Eh) at ($(A)!0.5!(D)$);
            \filldraw[\col] (Eh) circle (2.3pt);
        }
    }
    
    \node at (4, -.8) {\small $\textcolor{dodgerblue}{(x,y)\leftrightarrow(-x,-y)}$};
\end{scope}
}
\begin{equation}
\begin{tikzpicture}[
    scale=0.7,
    baseline={([yshift=-.05ex]current bounding box.center)}
]

\bb{0cm}{dodgerblue} 

\end{tikzpicture}
\end{equation}
To verify that $U_{\mathsf{CZ}}$ indeed a symmetry of the codespace, consider its action on a generic vertex stabilizer $A_{v_{(i,j)}}$. Under conjugation by $U_{\mathsf{CZ}}$, $A_{v_{(i,j)}}$ is mapped to itself multiplied by a $Z$ parity check specified by the polynomial $g(\overline{x},\overline{y})$ for the $H$ qubits and polynomial $f(\overline{x},\overline{y})$ for the $V$ qubits. These are precisely the $Z$ stabilizers of the original code, meaning $U_{\mathsf{CZ}}$ preserves the ground state subspace. 

We wish to gauge this symmetry to produce a non-Abelian code using our graph gauging construction. In principle, any graph satisfying our requirements in Sec.~\ref{sec:graphygraph} will work. As an illustrative example, we consider defining our symmetry enrichment operator as:
\begin{equation}\label{eq:BBSymmtik}
    \Omega_{\aP{a}\bB{B}\oC{C}}=\prod_{(i,j)}(-1)^{\textcolor{goodpurple}{n^a_{v_{(i,j)}}}(\textcolor{dodgerblue}{n^B_{(i,j),H}}\textcolor{orange(ryb)}{n^C_{-(i,j),V}}+\textcolor{dodgerblue}{n^B_{(i,j),V}}\textcolor{orange(ryb)}{n^C_{-(i,j),H}})}
\end{equation}
We choose an ancilla complex such that the ancilla vertices are a copy of $\textcolor{goodpurple}{a}$ and the edges of the ancilla graph are formed by connecting each vertex $\mathbf{v}_{(i,j)}$ with every $\mathbf{v}_{(i',j')}$ for which $\mathbf{e}_{(i',j')}\in \exd^a_0 \mathbf{v}_{(i,j)}$. For the example stabilizers of Eq.~\eqref{eq:BBstabX} and Eq.~\eqref{eq:BBstabZ}, this graph is:
\begin{equation}
 \begin{tikzpicture}[scale = 1, baseline = {([yshift=-.5ex]current bounding box.center)}]

        \foreach \i in {0, 1, 2,3}{
            \draw[color = lightcrimson] (\i, -0.5) -- (\i, 3.5);}
        \foreach \i in {0, 1, 2, 3}{
            \draw[color = lightcrimson] (-0.5, \i) -- (3.5, \i);}

        \foreach \i in {0,...,3} {
            \foreach \j in {0,...,3} {

                \filldraw[color = goodpurple] (\i,\j) circle (2 pt);
  }
}
            \draw[color = lightcrimson, line width = 0.5pt]  (2, 3) to[out=260,in=6=30] (0, 2);
            \draw[color = lightcrimson, line width = 0.5pt]  (3, 3) to[out=260,in=6=30] (1, 2);
            \draw[color = lightcrimson, line width = 0.5pt]  (4, 3) to[out=260,in=6=30] (2, 2);

            \draw[color = lightcrimson, line width = 0.5pt]  (2, 2) to[out=260,in=6=30] (0, 1);
            \draw[color = lightcrimson, line width = 0.5pt]  (3, 2) to[out=260,in=6=30] (1, 1);
            \draw[color = lightcrimson, line width = 0.5pt]  (4, 2) to[out=260,in=6=30] (2, 1);

            \draw[color = lightcrimson, line width = 0.5pt]  (2, 1) to[out=260,in=6=30] (0, 0);
            \draw[color = lightcrimson, line width = 0.5pt]  (3, 1) to[out=260,in=6=30] (1, 0);
            \draw[color = lightcrimson, line width = 0.5pt]  (4, 1) to[out=260,in=6=30] (2, 0);

            
        
        \draw[color = lightcrimson, line width = 0.5pt]  (1, 2) to[out=110,in=0] (0, 3);
        \draw[color = lightcrimson, line width = 0.5pt]  (2, 2) to[out=110,in=0] (1, 3);
        \draw[color = lightcrimson, line width = 0.5pt]  (3, 2) to[out=110,in=0] (2, 3);

        \draw[color = lightcrimson, line width = 0.5pt]  (1, 1) to[out=110,in=0] (0, 2);
        \draw[color = lightcrimson, line width = 0.5pt]  (2, 1) to[out=110,in=0] (1, 2);
        \draw[color = lightcrimson, line width = 0.5pt]  (3, 1) to[out=110,in=0] (2, 2);

        \draw[color = lightcrimson, line width = 0.5pt]  (1, 0) to[out=110,in=0] (0, 1);
        \draw[color = lightcrimson, line width = 0.5pt]  (2, 0) to[out=110,in=0] (1, 1);
        \draw[color = lightcrimson, line width = 0.5pt]  (3, 0) to[out=110,in=0] (2, 1);

\end{tikzpicture}
\end{equation}
After the gauging procedure, the example $X$ stabilizer of Eq.~\eqref{eq:BBstabX} on the $\textcolor{dodgerblue}{B}$ sublattice is transformed to:

\begin{equation}
 \begin{tikzpicture}[scale = 1, baseline = {([yshift=-.5ex]current bounding box.center)}]
        \foreach \i in {0, 1,2, 3}{
            \draw[color = dodgerblue] (\i, -0.5) -- (\i, 3.5);}
        \foreach \i in {0, 1, 2, 3}{
            \draw[color = dodgerblue] (-1, \i) -- (3.5, \i);}  

        \foreach \i in {0, 1,2, 3}{
            \draw[color = orange(ryb)] (\i-.2, -0.5) -- (\i-.2, 3.5);}
        \foreach \i in {0, 1, 2, 3}{
            \draw[color = orange(ryb)] (-1, \i-.2) -- (3.5, \i-.2);}
        \foreach \i in {0, 1, 2,3}{
            \draw[color = lightcrimson] (\i-.4, -0.5) -- (\i-.4, 3.5);}
        \foreach \i in {0, 1, 2, 3}{
            \draw[color = lightcrimson] (-1, \i-.4) -- (3.5, \i-.4);}  

        

        \draw[color = black, line width = 0.5pt]  (1.6, 3.2-1) -- (1.8-.5, 2.3-.5+1);
        \filldraw[color = orange(ryb)] (1.8-.5, 2.3-.5+1) circle (2 pt);
        \filldraw[color = red] (1.6, 3.2-1) circle (2 pt);

        \draw[color = black, line width = 0.5pt]  (1.6+.5, 2.1-1+.5) -- (2.8, 2.3-1);
        \filldraw[color = orange(ryb)] (2.8, 2.3-1) circle (2 pt);
        
        \filldraw[color = red] (1.6+.5, 2.1-1+.5) circle (2 pt);

        \draw[color = lightcrimson, line width = 0.5pt]  (2-.4, 2-.4) to[out=110,in=0] (1-.4, 3-.4);

        \draw[color = black, line width = 0.5pt]  (1.8+1-2, +2+2.3-2) -- (1.6-.47, 4-1.65);
        \filldraw[color = orange(ryb)] (1.8+1-2, +2+2.3-2) circle (2 pt);
        \filldraw[color = red] (1.6-.47, 4-1.65) circle (2 pt);

        \draw[color = lightcrimson, line width = 0.5pt]  (2-.4, 2-.4) to[out=260,in=6=30] (-.4, 1-.4);

        \draw[color = black, line width = 0.5pt]  (-.7, 1-.4+.2) -- (2.26-1, 1.05);
        \filldraw[color = orange(ryb)] (-.7, 1-.4+.2) circle (2 pt);
        \filldraw[color = red] (2.26-1, 1.05) circle (2 pt);

        \node at (1.5, 2) {\small $\textcolor{dodgerblue}{X}$};
        \node at (2.5, 2) {\small $\textcolor{dodgerblue}{X}$};
        \node at (2, 2.5) {\small $\textcolor{dodgerblue}{X}$};
        \node at (2, 1.5) {\small $\textcolor{dodgerblue}{X}$};
        \node at (0.5, 3) {\small $\textcolor{dodgerblue}{X}$};
        \node at (0, 0.5) {\small $\textcolor{dodgerblue}{X}$};

\end{tikzpicture}
\end{equation}
Here we have flipped $(i,j)\rightarrow(-i,-j)$ on the $\textcolor{orange(ryb)}{C}$ sublattice for the purposes of illustrating the stabilizer. The plaquette stabilizers are diagonal and are unchanged by the gauging procedure.

\section{Properties of Non-Abelian qLDPC Codes}\label{sec:non-Abelian-properties}

Having illustrated concrete examples of our non-Abelian qLDPC codes, the rest of this work focuses on understanding this class of codes from two complementary perspectives.
First, in this section, we understand the properties of these codes through the lens of condensed matter physics ---understanding the structure of their ground states as well as physical properties of their excitations such as non-Abelian braiding.
In the next section, we discuss how these non-Abelian codes can be used to facilitate universal quantum computation on their input qLDPC codes---enabling the preparation of logical magic states on the input codes. 

\subsection{Ground States and Logical Operators}\label{sec:groundstates}

We first analyze the structure of the ground states of our non-Abelian qLDPC codes obtained via either the homological or graph gauging procedures described in Sec.~\ref{sec:gauging-theory}.
In both constructions, the logical operators of the non-Abelian code are obtained by ``pushing'' the logical operators of the input Abelian \BandC codes through our gauging construction. 
To obtain the ground states in this way, we initialize the input codes in a configuration that is symmetric under the transversal gate we aim to gauge---e.g. for a transversal $\mathsf{CZ}$ gate, we initialize our codes in their global logical $\ket{0}$ state \footnote{We do this since any non-trivial symmetry charge of a state prior to gauging becomes a gauge charge in the non-Abelian code after gauging.}. 
The ground states of the gauged theory are then obtained by tracking the logicals through the steps of the gauging procedures of Sec.~\ref{sec:addressable} and Sec.~\ref{sec:gauging-graphs}.
In this section, we present explicit expressions for the non-Abelian ground states and point out their characteristic properties. A full derivation of the resulting ground states for both the homological and graph gauging formalism is given in Appendix~\ref{App:log}. Before writing explicit forms of the gauged logical operators, we must first briefly review some relevant concepts from homology needed to express them.

Recall that every 0-chain vertex is trivially a cycle since the boundary of a vertex is 0. Consequently, any $v\in\aP{a}$ can be expressed as a sum over some representative non-trivial elements of $H_0$ up to a boundary term. Explicitly, for any $v\in\aP{a}$, we may write:
\begin{equation}\label{eq:vmu}
    v=\partial b(v) + \sum_{\bm{\mu}\in\{\bm{\mu}\}}v_{\bm{\mu}}\, \bm{\mu}(v)
\end{equation}
Here, $\{\bm{\mu}\}$ denotes the set of representative cocycles whose cohomology classes generate $H^0$ and $\{v_{\bm{\mu}}\}$ denotes a corresponding set of representative 0-cycles which generate $H_0$, such that we have $\int_{v_{\bm{\mu'}}}\bm{\mu}=\delta_{\bm{\mu},\bm{\mu}'}$. We have also introduced a $1$-chain $b(v)\in \AR{A}$, which is fixed for each vertex once we have chosen our representative 0-cycles $\{v_{\bm{\mu}}\}$.

When the underlying chain complex of the \BandC codes describes a two dimensional lattice, this decomposition of vertices has a particularly simple interpretation. In this case, the only nontrivial generator $\bm{\mu}\in\mathsf{ker}(d^0)$ is the sum of all vertex cochains. We may then choose any single vertex $v_0$ as the dual representative $v_{\bm{\mu}}=v_0$. 
Said differently, in 2D, any vertex $v$ can be expressed in the terms of Eq.~\eqref{eq:vmu} as $v=\partial b(v)+v_0$ where $b(v)$ is the string of edges connecting $v$ and $v_0$. 
In a general chain complex, however, one must choose a reference ``base-point" $v_{\bm{\mu}}\in D_0$ for each generator of $H^0$ per Eq.~\eqref{eq:vmu}.

With the definition $\partial b(v)=v+ \sum_{\bm{\mu}\in\{\bm{\mu}\}}v_{\bm{\mu}}\, \bm{\mu}(v)$, we can now give explicit forms for our gauged logical $\tilde{X}$ operators:
\begin{align}
    \bB{\widetilde{X}^B_{\bgamma}} &= \bB{X^B_{\bgamma}} \prod_{v, e} \left(\AR{Z^A_{b(v)}} \right)^{\oC{n_{e}^C} \int_{\mathcal{M}}\textbf{v}\cupp (\bgamma\cupp \mathbf{e})} \\
    \oC{\widetilde{X}^C_{\bgamma}} &= \oC{X^C_{\bgamma}} \prod_{v, e} \left(\AR{Z^A_{b(v)}} \right)^{\bB{n_{e}^B} \int_{\mathcal{M}}\textbf{v}\cupp (\mathbf{e}\cupp \boldsymbol{\gamma})} \\
    \AR{\widetilde{X}^A_{\bgamma}} &= \AR{X^A_{\bgamma}} \prod_{v, e} \left(\bB{Z^B_{b(v)}} \right)^{\oC{n_{e}^C} \int_{\mathcal{M}}\bgamma\cupp (\textbf{v}\cupp \mathbf{e})}
\end{align}
defined for any $\bgamma\in\mathsf{ker}(\exd_{1})$. The logical $\widetilde{Z}$ operators are unchanged in the gauging procedure and are identical to those of the original \BandC codes. We note the similarity of the above expressions to the ground states of $\mathcal{D}(D_4)$ topological order on the torus \cite{margaritaknots}, with the crucial distinction that our expressions implicitly involve a sum over elements of $H^0$ in referencing $b(v)$. As we will see, this property is closely tied to the resulting non-Abelian character of the ground states.

We can similarly derive the form of ground states of a non-Abelian code which was gauged via our graph gauging procedure. However, an important difference arises in the structure of the resulting ground states, originating from the choice of ancilla complex. In the homological gauging construction, the ancilla complex had the same structure as the original \BandC codes, whereas in the graph case, the ancilla complex generally has a different structure from codes \BandC. Consequently, we expect that the ground states generated by \textcolor{red}{$\tilde{X}^A$} will have distinctive properties from the \BandC ground states.

Since the explicit form of \textcolor{red}{$\tilde{X}^A$} depends on a choice of underlying ancilla graph and in general is quite complicated, we will restrict our discussion to the ground states which can be obtained by gauging logical operators on codes \BandC. In keeping with Sec.~\ref{sec:gauging-graphs}, we consider the logical operators for the case of gauging transversal $\mathsf{SWAP}$ gates. In this case, the gauged logical $\tilde{X}$ operators on the \BandC codes are:
\begin{align}
   \bB{\tilde{X}^B}&= \bB{X^B_{\gamma}}\prod_{\mathbf{e}\in \gamma}\left(\bB{X^B_{\mathbf{e}}} \textcolor{orange(ryb)}{X^C_{\mathbf{e}}}\right)^{\AR{n^{A}_{\phii(e),\mathbf{v}_0}}} \\
   \textcolor{orange(ryb)}{\tilde{X}^C}&= \textcolor{orange(ryb)}{X^C_{\gamma}}\prod_{\mathbf{e}\in \gamma}\left(\bB{X^B_{\mathbf{e}}} \textcolor{orange(ryb)}{X^C_{\mathbf{e}}}\right)^{\AR{n^{A}_{\phii(e),\mathbf{v}_0}}} 
\end{align}
whereas the gauged logical $\tilde{Z}$ operators on the \BandC codes are:
\begin{align}
   \bB{\tilde{Z}^B} &= \bB{Z^B_{\eta}}\prod_{e\in \eta}\left(\bB{Z^B_{e}} \textcolor{orange(ryb)}{Z^C_{e}}\right)^{\AR{n^{A}_{\phii(e),\mathbf{v}_0}}}  \\
   \textcolor{orange(ryb)}{\tilde{Z}^C}&= \textcolor{orange(ryb)}{Z^C_{\eta}}\prod_{e\in \eta}\left(\bB{Z^B_{e}} \textcolor{orange(ryb)}{Z^C_{e}}\right)^{\AR{n^{A}_{\phii(e),\mathbf{v}_0}}} 
\end{align}
defined for any $\gamma\in\mathsf{ker}(d_1)$ and $\eta\in\mathsf{ker}(\partial_1)$. Here we note a few essential differences from the homological case. Firstly, whereas the homological operators involved a sum over representatives of $H^0$, the graph expressions only reference a single base-point vertex $v_0$, corresponding to the single generator of $H^0$ for connected graphs. 

We also note the different roles played by the gauged logical operators in the non-Abelian code as a consequence of gauging the $\mathsf{SWAP}$ rather than $\mathsf{CZ}$ symmetry; whereas in the $\mathsf{CZ}$ case, the $\tilde{Z}$ logicals were unchanged by the gauging procedure, in the $\mathsf{SWAP}$ case both the $\tilde{Z}$ and $\tilde{X}$ logicals are transformed. It is only the $\mathsf{SWAP}$-invariant \textit{products} of logical operators $\bB{\tilde{X^B_\gamma}}\textcolor{orange(ryb)}{\tilde{X^C_\gamma}}$ and $\bB{\tilde{Z^B_\eta}}\textcolor{orange(ryb)}{\tilde{Z^C_\eta}}$ that are invariant under the gauging map. The physical result of this is the anyons $\bB{e_A}\textcolor{orange(ryb)}{e_B}$ and $\bB{m_B}\textcolor{orange(ryb)}{m_C}$ correspond to the Abelian super-selection sectors whereas $\bB{e_B}$ and $\bB{m_C}$ correspond to the non-Abelian super-selection sectors of the gauged code.

\subsection{Non-Abelian Braiding: Single ``Gauge Charge'' on a Subsystem Symmetry}
Having derived explicit expressions for the ground states of the gauged qLDPC codes in our homological construction, we now use these states to establish the non-Abelian nature of the gauged codes. Specifically, we show that acting two logical $\tilde{X}$ operators on the ground state subspace of the code produces unpaired anyon excitations, a hallmark of non-Abelian topological order in two dimensions. For the case of our homological gauging constructions, we will see an additional richness imbued by the multiple distinct fundamental classes of the underlying chain complex in the anyon structure, in that anyons will be tied to the subsystem symmetries $\bm{\mu}$. We will also demonstrate the non-Abelian nature of the ground states obtained from our graph gauging constructions, which will have properties similar to the 2D case.

\subsubsection{Homological case}
In the more familiar setting of topological order in 2D, a signature of \textit{non-Abelian} topological order is the ability to realize single gauge-charged excitations on a torus. This single-excitation state can be prepared with a two-step procedure: first, we instantiate in the logical $\ket{0}$ state and wind a pair of anyons along a non-contractible loop of the torus. This winding is equivalent to acting with a non-trivial logical operator on the $\ket{0}$ state.
Then, we apply a second loop operator along the other non-contractible loop of the torus; crucially, this second operator intersects \textit{only once} with the first logical operator, allowing for nontrivial braiding between the underlying anyons. Physically, the second loop operator creates a pair of non-Abelian anyons, which wind around the non-contractible loop and fuse to a single anyon. 
Note that it is impossible to create a single unpaired anyon in the Abelian toric code since flipping any qubit will affect an even number of stabilizers.

Here, we confirm the non-Abelian nature of the states produced by our homological gauging construction by showing they may host isolated gauge-charged excitations. In the qLDPC context, we will see the set of subsystem symmetries present in general qLDPC codes give rise to additional super-selection sectors beyond those of conventional 2D topological order. We will show that single-excitation states are intrinsically tied to these underlying subsystem symmetries. Without loss of generality, we consider creating a single gauge charge in the $\AR{A}$ sublattice. 

To detect the presence of this single charged excitation in a state $\ket{\Psi}$, we can apply the global charge operators $\AR{\mathcal{A}^{A}_{\bm\mu}}$ defined for each $\bm{\mu}\in H^0$ on $\ket{\Psi}$, which will measure the charge parity\footnote{This global charge operator is well-defined for $D_4$ code, since the single anyon excitation has to be Abelian because of the nilpotent property of the group. Therefore, we can measure the Abelian anyon by a global charge operator defined for one of the sublattices (one of the underlying toric code) as a twisted ${\mathbb{Z}_2}^3$ theory.} of $\ket{\Psi}$.  $\AR{\mathcal{A}^{A}_{\bm\mu}}$ is defined as:
\begin{equation}
    \AR{A^A_{\bm{\mu}}}=\prod_{\mathbf{w}\in\bm{\mu}}\textcolor{red}{\mathcal{A}^A_{\mathbf{w}}}
\end{equation}

We highlight a difference from the case of conventional 2D non-Abelian topological order; in the 2D toric code, there is a single global charge parity operator associated to each code. By contrast, here there are distinct parity operators for each subsystem symmetry $\bm{\mu}$; in other words, the charges detected by each global operator are associated with a set of underlying subsystem symmetries of the code. To see this explicitly, we consider the state:
\begin{equation}
    \ket{\Psi}=\bB{\tilde{X}^B_\gamma}\bB{\ket{\overline{0}}_B}\otimes\oC{\widetilde{X}^C_{\bgamma'}}\textcolor{orange(ryb)}{\ket{\overline{0}}_C}
\end{equation}

where $\bB{\tilde{X}^B_\gamma}$ and $\oC{\widetilde{X}^C_{\bgamma'}}$ are logical operators acting along the $\bB{\gamma}$ and $\oC{\gamma'}$ loops respectively and $\ket{\overline{0}}$ represents the \textit{logical} $\ket{00}$ state of the \BandC LDPC codes. We want to compute the action of $\AR{\mathcal{A}^A_{\mathbf{\mu}}}$ on $\ket{\Psi}$ for an arbitrary subsystem symmetry $\bm{\mu}$, i.e. we want to measure the charge content associated with a ``subsystem symmetry sector" $\bm{\mu}$. To do this, we first compute the group commutator of $\bB{\tilde{X}^B_\gamma}$ with a single vertex stabilizer $\AR{\mathcal{A}^{A}_{\mathbf{w}}}$:
\begin{equation}
    \left[\AR{\mathcal{A}^{A}_{\mathbf{w}}},\bB{\tilde{X}^B_\gamma}\right]  = \prod_{\bm{\mu}\in\{\bm{\mu}\}}\prod_{\oC{e}} \left(-1\right)^{\mathbf{w}(v_{\bm{\mu}})\oC{n_{e}^C} \int_{\mathcal{M}}\bm{\mu}\cupp (\bgamma\cupp \mathbf{e})}
\end{equation}

From the form of the equation above, we see that every vertex stabilizer $\AR{A_{\mathbf{w}}}$ for which $\mathbf{w}(v_{\bm{\mu}})=0$ commutes through the first logical operator $\bB{\tilde{X}_\gamma}$, but for $\mathbf{w}(v_{\bm{\mu}})=1$, the commutator will result in a $\oC{\mathcal{Z}}$ logical operator:

\begin{equation}
\begin{split}
     \left[\AR{\mathcal{A}^{A}_{\bm{\mu}}}, \bB{\tilde{X}^B_\gamma}\right] = \oC{\mathcal{Z}^C_{(\mathcal{M}\capp \boldsymbol{\mu})\capp\bm{\gamma}}}
\end{split}
\end{equation}
%
The above commutator must then  be ``pushed through" the remaining $\oC{\widetilde{X}^C_{\bgamma'}}$ logical operator. The resulting final action on $\ket{\Psi}$ is:
\begin{equation}
    \AR{\mathcal{A}^A_{\bm{\mu}}}\ket{\Psi}=(-1)^{\int_\mathcal{M} \bm{\mu}\cupp(\bm{\gamma}\cupp\bm{\gamma'})}\ket{\Psi}
\end{equation}
due to the nontrivial commutation relation between $\oC{\mathcal{Z}^C_{(\mathcal{M}\capp \boldsymbol{\mu})\capp\bm{\gamma}}}$ and $\oC{\widetilde{X}^C_{\bgamma'}}$. We thus have seen that $\ket{\Psi}$ carries an unpaired gauge charge associated to the sector specified by $\AR{\mathcal{A}^A_{\mathbf{\mu}}}$. We emphasize again the difference to the two dimensional case; in addition to labels $\AR{A}$, $\bB{B}$ and $\textcolor{orange(ryb)}{C}$, the charge is associated with a collection of subsystem symmetries $\bm{\mu}$. 

We note that similar to the 2D case, such an excitation is not possible in an Abelian qLDPC code, even when there are multiple subsystem symmetries $\bm{\mu}$. Flipping a single qubit will always ``upset" an even number of vertex stabilizers \textit{within each $\bm{\mu}$} (if an edge were in the coboundary of an odd number of stabilizers in a given $\bm{\mu}$, then we would not have $\exd \bm{\mu}=0$). Therefore, the charge operators $A_{\bm{\mu}}$ will always measure an even parity in the Abelian case.

\subsubsection{Graph case}
Finally, we repeat this demonstration of unpaired charges for the graph gauging construction. We emphasize at the outset that the relevant difference from the homological gauging construction is the lack of multiple subsystem symmetries $\bm{\mu}$; in this case there is a single generator for $H^0$ of our (connected) graph ancilla complex, such that the unpaired gauge charges can only be associated with a unique generator $\bm{\mu}$. In keeping with Sec.~\ref{sec:gauging-graphs}, we will again consider the transversal $\mathsf{SWAP}$ gate which is always present between any two copies of a qLDPC code, using the ground states derived earlier in the section. Similarly to the homological non-Abelian construction, we consider a state:
\begin{equation}
    \ket{\Psi}=\bB{\tilde{X}^B_\gamma}\bB{\ket{\overline{0}^S}_B}\otimes \oC{\widetilde{Z}^C_{\eta}}\textcolor{orange(ryb)}{\ket{\overline{0}^S}_C}
\end{equation}
where $\ket{\overline{0}^S}$ denotes the code state symmetric under the transversal $\mathsf{SWAP}$ gate. Note that we have here chosen to act with the operators $\bB{\tilde{X}_\gamma}$ and $\oC{\widetilde{Z}_{\eta}}$ as these operators are both associated with non-Abelian anyons of the code as discussed in Sec.\ref{sec:groundstates}.

We can now repeat the same analysis but here for a single global charge operator $\AR{\mathcal{A}^A}=\prod_{\mathbf{v}\in \aP{a}}\AR{\mathcal{A}^A_{\mathbf{v}}}$ (we omit the subscript $\bm{\mu}$ here as there is a single ``subsystem symmetry", or equivalently, generator of $H^0$ on a connected graph). For a single vertex stabilizer:
\begin{equation}
    \left[\bB{\tilde{X^B_\gamma}},\textcolor{red}{\mathcal{A}^A_{\mathbf{w}}}\right] = \prod_{e\in\gamma}\left(\bB{X^B_e}\textcolor{orange(ryb)}{X^C_e}\right)^{\mathbf{w}(v_{0})}
\end{equation}

where $v_0$ is now the ``basepoint" vertex representative of the single transversal symmetry associated with the vertices of the graph. With this, we can compute for the global charge operator:
\begin{equation}
    \prod_{w}\left[\bB{\tilde{X^B_\gamma}},\textcolor{red}{\mathcal{A}^A_{\mathbf{w}}}\right]=\bB{X^B_{\gamma}}\textcolor{orange(ryb)}{X^C_{\gamma}}
\end{equation}
Note the above is a logical operator associated to an Abelian anyon of the code and will anticommute with $\oC{\tilde{Z}_\eta}$ if $\eta$ and $\gamma$ have a nontrivial overlap, i.e. if $\gamma(\eta)=+1$. Thus, we have the final result:
\begin{equation}
    \AR{\mathcal{A}^A}\ket{\Psi}=(-1)^{\gamma(\eta)}\ket{\Psi}
\end{equation}
We conclude that provided there is a nontrivial cycle in the \BandC codes, there can be a state supporting a single unpaired gauge charge in our graph construction as well.

\section{Quantum Computation on qLDPC Codes via Gauging}\label{sec:non-Abelian-computation}

Having explored the properties of our non-Abelian qLDPC codes from the perspective of many-body physics, we now turn to an interpretation of these constructions through the lens of quantum computation. As we will see, the same gauging procedure which produces a non-Abelian qLDPC code is equivalent to performing a logical symmetry measurement on the two input codes. More precisely, given a Clifford symmetry $G$, gauging this symmetry in a general qLDPC code using either of the methods of Sec.~\ref{sec:gauging-theory}, followed by an un-gauging procedure to decouple the ancilla qubits will result in a measurement of the logical action of $G$ on the input codes \BandC. 
Crucially, depending on the logical initial states of the codes, this provides a route to magic state preparation for level-$(2 + 1)$ dimensional qLDPC codes.

\subsection{Logical Clifford Measurement from Gauging and Ungauging}
We now consider the gauging and ungauging procedure \cite{williamson_tanner,ungauging,sajithcodes} for the case of a transversal $\mathsf{CZ}$ gate. Let us start by considering the global wavefunction of the \BandC codes prior to gauging.
In particular, we can express the state of our code in the logical $Z$ basis as: 
\begin{equation} \label{eq-BCSwavefunction}
    \ket{\Psi_{\bB{B}, \oC{C}}} = \sum_{\bB{\bgamma_B}, \oC{\bgamma_C}} \Psi_{\bB{\bgamma_B}, \oC{\bgamma_C}} \bB{X_{\bB{\bgamma_B}}} \textcolor{dodgerblue}{\textcolor{dodgerblue}{\ket{\overline{0}}_B}}\otimes\oC{X_{\oC{\bgamma_C}}}\textcolor{orange(ryb)}{\ket{\overline{0}}_C}
\end{equation}
where $\bB{\bgamma_B}, \oC{\bgamma_C}$ are representatives of elements in $H^1$ and $\ket{\overline{0}}$ represents the logical $\ket{00}$ state of the \BandC LDPC codes.

We begin by demonstrating that first gauging then ungauging our qLDPC codes using the homological gauging procedure of Sec.~\ref{subsubsec:d4-homological} will implement a (potentially addressable) logical $\mathsf{CZ}$ measurement.
To do so, let us consider the wavefunction of our \BandC codes combined with our ancilla cluster state: 
\begin{equation} \label{eq-wfwholeshebang}
    \ket{\Psi_{\aP{a}\AR{A}\bB{B}\oC{C}}} = \ket{\Psi_{\bB{B}, \oC{C}}} \otimes U\left( \ket{\AR{0_{A}},\aP{+_a}} \right)
\end{equation}
where $U$ is the unitary that prepares the qLDPC cluster state associated with the chain complex of the \BandC codes.

After the symmetry enrichment and measurement steps of the gauging map, the wavefunction of our system is transformed into:
\begin{align}
    &\ket{\Psi_{\aP{a}\AR{A}\bB{B}\oC{C}}} \to \prod_{\mathbf{v}} \left(\frac{1 + x_{\mathbf{v}} \aP{X_{\mathbf{v}}}}{2} \right) \Omega_{\aP{a}\bB{B}\oC{C}}\ket{\Psi_{\aP{a}\AR{A}\bB{B}\oC{C}}} \\
    & =  U \prod_{\mathbf{v}} \left(\frac{1 + x_{\mathbf{v}} \aP{X_{\mathbf{v}}}\AR{X_{\exd \mathbf{v}}}}{2}  \right)\Omega_{\aP{a}\bB{B}\oC{C}} \ket{\Psi_{\bB{B}, \oC{C}}}\ket{\AR{0_A}, \aP{+_a}}
\end{align}
where $x_{\mathbf{v}}$ is the measurement outcome of the $\aP{a}$ qubit at $\mathbf{v}$.
It will be convenient moving forward to expand the product of projectors in the equation above as: 
\begin{equation}
    \propto U \left(\sum_{\boldsymbol{\phi} \in D^0} x_{\boldsymbol{\phi}} \aP{X_{\boldsymbol{\phi}}} \AR{X_{\exd \boldsymbol{\phi}}}\right) \Omega_{\aP{a}\bB{B}\oC{C}} \ket{\Psi_{\bB{B}, \oC{C}}} \ket{\AR{0_A}, \aP{+_a}}
\end{equation}
where $x_{\boldsymbol{\phi}} = \prod_{v} x_{v}^{\boldsymbol{\phi}(v)}$.
Subsequently, to \textit{ungauge} the system, we first undo the cluster state entangler with $U^{\dagger}$. The next step is to simply measure all of the $\AR{A}$ qubits in the $Z$ basis to restore the qubit in the $\AR{A}$ sublattice back to the $\ket{\AR{0_A}}$ state (up to measurement outcome).
This results in: 
\begin{equation}
        \propto  \prod_{e} \left( 1 + z_{e} \AR{Z_{e}}\right)\left(\sum_{\boldsymbol{\phi} \in D^0} x_{\boldsymbol{\phi}} \aP{X_{\boldsymbol{\phi}}} \AR{X_{\exd \boldsymbol{\phi}}}\right) \Omega_{\aP{a}\bB{B}\oC{C}} \ket{\Psi_{\bB{B}, \oC{C}}} \ket{\AR{0_A}, \aP{+_a}}
\end{equation}
We now note that the only measurements that will result in a nonzero projection of our state are those for which $z_{e}=(-1)^{\exd\boldsymbol{\phi}(e)}$ for some $\mathbf{\boldsymbol{\phi}}\in D^0$. To see this, note that for a single co-chain $\exd\mathbf{\phi}$ in the above sum, $\AR{X_{\exd \boldsymbol{\phi}}}$ will flip all of the qubits in the $\textcolor{red}{\ket{0}_A}$ state in $\boldsymbol{\phi}$.
Therefore, the measurement in $\textcolor{red}{Z^A_e}$ will only result in a nonzero projection if the measurement outcomes are such that $-1$ is measured for every qubit in $\exd\boldsymbol{\phi}$ and $+1$ for every qubit outside of it.
We therefore know that our measurement will result in a nonzero projection such that the measurement operator takes the form $z_{e}=(-1)^{\exd\boldsymbol{\phi}_0(e)}$ for some $\boldsymbol{\phi}_0\in D^0$. We trivially insert $\AR{X^2_{\exd \phi_0}}$ to the left of $\AR{Z_e}$, and use the commutation relation $X_{\exd \phi_0} Z_e = (-1)^{\exd \phi_0(e)} Z_e X_{\exd \phi_0} = z_e Z_e X_{\exd \phi_0}$ to get
\begin{align}
        \propto \AR{X_{\exd \boldsymbol{\phi_0}}} & \prod_{e} \left( 1 + z_{e}^2 \AR{Z_{e}}\right) \AR{X_{\exd \boldsymbol{\phi_0}}}\left(\sum_{\boldsymbol{\phi} \in D^0} x_{\boldsymbol{\phi}} \aP{X_{\boldsymbol{\phi}}} \AR{X_{\exd \boldsymbol{\phi}}}\right) \\
        & \times \Omega_{\aP{a}\bB{B}\oC{C}} \ket{\Psi_{\bB{B}, \oC{C}}} \ket{\AR{0_A}, \aP{+_a}} \nonumber
\end{align}

We can simplify above with $z_e^2=+1$. By linearity, the left multiplication of $\AR{X_{\exd \boldsymbol{\phi_0}}}$ modifies the labeling for the $\AR{X_{\exd \boldsymbol{\phi}}}$ term in the parentheses to:
\begin{align}
        \propto  \AR{X_{\exd \boldsymbol{\phi}_0}}& \prod_{e} \left( 1 +  \AR{Z_{e}}\right)\left(\sum_{\boldsymbol{\phi} \in D^0} x_{\boldsymbol{\phi}} \aP{X_{\boldsymbol{\phi}}} \AR{X_{\exd (\boldsymbol{\phi} + \boldsymbol{\phi}_0)}}\right) \\ 
        &\times \Omega_{\aP{a}\bB{B}\oC{C}} \ket{\Psi_{\bB{B}, \oC{C}}}\otimes \ket{\AR{0_A}, \aP{+_a}} \nonumber
\end{align}
Now, repeating the trick from earlier to constrain the $z_{e}$ measurement outcomes, we note that the only terms in the sum over $\boldsymbol{\phi}$ that will survive the projection are those for which $\exd (\boldsymbol{\phi} + \boldsymbol{\phi}_0)=0$; these are precisely the terms corresponding to the subsystem symmetries $\bm{\mu}\in\text{ker}(\exd_{0}^a)$. Therefore, the $\AR{X_{\exd \boldsymbol{\phi}}}$ term vanishes upon projection, and we can then rewrite the projected wavefunction sum over all such $\bm{\mu}$ (and relabel $\phi \to \phi + \phi_0$):
\begin{equation}
        \propto  \AR{X_{\exd \boldsymbol{\phi}_0}} \left(\sum_{\bm{\mu} } x_{\bm{\mu} + \boldsymbol{\phi}_0} \aP{X_{\bm{\mu} + \boldsymbol{\phi}_0}}\right) \Omega_{\aP{a}\bB{B}\oC{C}} \ket{\Psi_{\bB{B}, \oC{C}}}\otimes \ket{\AR{0_A}, \aP{+_a}}
\end{equation}

Again, by linearity, $x_{\bm{\mu} + \boldsymbol{\phi}_0} \aP{X_{\bm{\mu} + \boldsymbol{\phi}_0}} = x_{\boldsymbol{\phi}_0} \aP{X_{\boldsymbol{\phi}_0}} \left(x_{\bm{\mu}} \aP{X_{\bm{\mu}}} \right)$, allowing us to factor out a global factor of $x_{\boldsymbol{\phi}_0} \aP{X_{\boldsymbol{\phi}_0}}$. We then use one round of feedforward correction to get rid of this global factor: when the measurement outcome is $x_{\phi_0}=-1$, we apply another $\aP{X_{\boldsymbol{\phi}_0}}$ to cancel out (and we can omit the global phase $x_{\boldsymbol{\phi}_0}$) and obtain
\begin{equation}
        \propto \AR{X_{\exd \boldsymbol{\phi}_0}} \left(\sum_{\bm{\mu} } x_{\bm{\mu}} \aP{X_{\bm{\mu}} }\right) \Omega_{\aP{a}\bB{B}\oC{C}} \ket{\Psi_{\bB{B}, \oC{C}}}\otimes \ket{\AR{0_A}, \aP{+_a}}
\end{equation}

Now, if we pick a basis for $\mathsf{ker}(\exd)$, which we label as $\{\boldsymbol{\mu}_i\}$, we can write this sum as a product of projectors:
\begin{equation}
        \propto  \AR{X_{\exd \boldsymbol{\phi}_0}} \prod_{\bm{\mu}\in\{\bm{\mu}_i\} }\left(1+ x_{\bm{\mu}} \aP{X_{\bm{\mu}}} \right) \Omega_{\aP{a}\bB{B}\oC{C}} \ket{\Psi_{\bB{B}, \oC{C}}}\otimes \ket{\AR{0_A}, \aP{+_a}}
\end{equation}
Now to see the action of this projector on our original qLDPC codes $\ket{\Psi_{\bB{B}, \oC{C}}}$, we simply undo the symmetry enrichment operator by applying $\Omega_{\aP{a}\bB{B}\oC{C}}^{\dagger}$ and peel off the disentangled product state on the $\AR{A}$ sublattice, yielding:
\begin{equation}
        \propto \prod_{\bm{\mu}\in\{\bm{\mu}_i\} }\left(1+ x_{\bm{\mu}} \Omega_{\aP{a}\bB{B}\oC{C}}^{\dagger}\aP{X_{\bm{\mu}}}\Omega_{\aP{a}\bB{B}\oC{C}} \right)\ket{\Psi_{\bB{B}, \oC{C}}}\otimes \ket{\aP{+_a}}
\end{equation}
Now, by the definition of the symmetry enrichment operator, $\Omega_{\aP{a}\bB{B}\oC{C}}^{\dagger}\aP{X_{\bm{\mu}}}\Omega_{\aP{a}\bB{B}\oC{C}} = \aP{X_{\boldsymbol{\mu}}} U_{\mathsf{CZ}}[\boldsymbol{\mu}]$, where $U_{\mathsf{CZ}}[\boldsymbol{\mu}]$ is a symmetry of the \BandC codes.
Indeed, in the homological framework, 
\begin{equation}
    U_{\mathsf{CZ}}[\boldsymbol{\mu}] = \prod_{e,e'} \mathsf{CZ}_{\bB{e}, \oC{e'}}^{ \int_{\mathcal{M} \capp \boldsymbol{\mu}} \mathbf{e} \cupp \mathbf{e}'}
\end{equation}
whereas in the graph framework, there is only one $\boldsymbol{\mu}$ and it is tied to an assumed symmetry of the \BandC code.
Consequently, the resulting state on the \BandC codes is: 
\begin{equation}
        \propto     \prod_{\bm{\mu}\in\{\bm{\mu}_i\} }\left(1+ x_{\boldsymbol{\mu}}U_{\mathsf{CZ}}[\boldsymbol{\mu}] \right)\ket{\Psi_{\bB{B}, \oC{C}}}
\end{equation}
The result is that gauging followed by ungauging measures the $U_{\mathsf{CZ}}[\boldsymbol{\mu}]$ symmetry on the \BandC codes. 
Moreover, the above makes manifestly clear that the measurement outcomes $x_{\boldsymbol{\mu}} = \prod_{v} x_{\mathbf{v}}^{\boldsymbol{\mu}(v)}$ are precisely the outcomes of the transversal $\mathsf{CZ}$ gate. We remark that although we chose the transversal $\mathsf{CZ}$ gate in our gauging and ungauging procedure in the example above, the same result holds for any Clifford gate.

\section{Discussion and Outlook}

In this work, we provided a systematic measurement-based gauging protocol for constructing and preparing non-Abelian qLDPC codes by gauging any transversal Clifford gate of Abelian CSS qLDPC code.
Our constructions relied on introducing an ancilla chain complex to gauge a transversal gate, subject only to very general symmetry constraints. While many ancilla complexes can in principle be used to produce a non-Abelian code, we focused on two natural approaches.
In our homological gauging construction, the presence of a generalized type of Poincar\'e duality in a quantum code gave rise to families of $\mathsf{CZ}$ and $\mathsf{SWAP}$ gates whose structure depends on the redundancies of the stabilizers underlying the code.
Complementary to this homological approach, we also explored an alternative graph-based approach.
In this approach, an ancilla graph is introduced to gauge a transversal gate of the input qLDPC code, where the structure of the ancilla graph is generally distinct from the original chain complex.
This construction applies broadly to any qLDPC code with a transversal gate, including qLDPC code with no underlying notion of Poincar\'e duality.
We illustrated both approaches concretely by presenting several concrete lattice models for non-Abelian qLDPC codes.
Notably, we provided an example of a non-Abelian bivariate bicycle code and showed that it could be concretely realized using our graph construction to gauge one of its transversal $\mathsf{CZ}$ gates.

Beyond constructing these codes, we take first steps in characterizing their physical properties as phases of matter and their utility for quantum computation.
In particular, regarding their physical properties, for both homological and graph gauging, the states realized by the gauging procedure have a non-Abelian character reminiscent of non-Abelian topological order in two dimensions.
In direct analogy to the ``single anyon on a torus" hallmark of 2D non-Abelian topological order~\cite{bombin_family_2008,Iqbal_2024}, we show that it is possible to create a single violated stabilizer in a non-Abelian qLDPC code using the global redundancy of stabilizers in its parent Abelian code.
Moreover, in the context of quantum computation, we discussed how our construction can be used to enable non-Clifford operations in qLDPC codes.
Remarkably, we demonstrated that one can prepare a magic state from any code with a logical Clifford gate. Our gauging procedure results in a measurement of the action of Clifford gate on the logical qubits of the code.  

At a broader level, our work advances our understanding of the physics of qLDPC codes and, accompanied by this understanding, we gain insight into how to practically manipulate the logical information that they store.
Consequently, our work opens the door to numerous interesting future directions in both quantum information science and condensed matter physics.

On the quantum information side, an immediate future direction of this work is to try to realize our protocol in near-term experiments capable of the required long-range connectivity~\cite{Periwal_2021, bluvstein_logical_2024,chiu_continuous_2025,sinclair_fault-tolerant_2025,bluvstein_fault-tolerant_2026,moses_race-track_2023,ransford_helios_2025}.
To do so,  the largest unaddressed question in this work is the question of fault tolerance.
In particular, to practically apply the logical $\mathsf{CZ}$ measurement protocol of Section~\ref{sec:non-Abelian-computation}, it would be necessary to see if it is possible to develop a decoder to correct against errors made during our gauging procedure.
In Ref.~\cite{williamson_tanner}, where they used a similar graph gauging protocol to measure logical Pauli operators, a decoder is developed but relies on the ancilla complex being an expander graph.
This would suggest that, for fault tolerance in our graph gauging protocol, we would need to engineer the graph carefully to have expansion properties and that our homological gauging protocol would only be fault tolerant for expander codes.
However, in Ref.~\cite{margaritaknots}, it was shown that by using a bespoke ``just-in-time'' decoder, it was possible to make the gauging and ungauging protocol fault tolerant in 2D, where these expansion properties are missing.
Developing a generalization of this just-in-time decoder to the qLDPC setting is an interesting future direction.

In addition, it would be interesting to see if the non-Abelian qLDPC codes we construct could be used in their own right for fault-tolerant quantum computation, similar to the paradigm of topological quantum computation.
On this front, there are several interesting theoretical questions to iron out:
(1) how to encode logical information in the excitations of a non-Abelian qLDPC code? (2) how to braid these excitations and implement a universal gate set? (3) how to make these procedures fault-tolerant? (4) what advantages these non-Abelian qLDPC codes pose over the conventional 2D non-Abelian codes?
These are especially relevant given recent experimental demonstrations of a universal gate set with the $\mathcal{D}(S_3)$ quantum double \cite{lo2026universaltopologicalgatesbraiding}.%

Beyond quantum computation, it would be interesting to understand these non-Abelian qLDPC codes as phases of matter.
In particular, several natural questions emerge in this directions.
As a basic question, it would be interesting to better understand the phenomenology of the codes we construct in this work---their quasi-particle content, the statistics and fusion rules of their excitations, etc.
Some of these questions are explored in detail in a forthcoming work \cite{homologicalorder}.
Beyond this, we remark that, while our work provides one construction of non-Abelian qLDPC codes, this construction is far from exhaustive.
An interesting future direction would be to characterize the landscape of non-Abelian qLDPC codes, and further, to understand what mathematical structure can be used to organize this landscape.
The key challenge here will be to efficiently describe this data especially given the issue of UV/IR mixing that is pervasive throughout these codes\cite{rakovszky2023physicsgoodldpccodes, rakovszky2024physicsgoodldpccodes, Tan_2025}.

In addition to describing these codes at their fixed point, from a phase of matter standpoint, finding ways to understand the behavior of these codes beyond their fixed point, would be an interesting question.
Indeed, several recent works have provided evidence that qLDPC codes are stable to perturbations \cite{yin_low-density_2025, de_roeck_low-density_2025}.
However, to access and compute properties of these codes away from their fixed point, a natural approach would be to study their behavior under wavefunction deformation, which has a long history in the literature for the study of quantum phases and their transitions\cite{Freedman2003, Freedman04, Freedman05a, Freedman05b, castelnovo2008, Fendley_2008, Fidkowski2009, Fendley13, haegeman_2015_gauging, zhu2019gapless, schotte2019tensornetwork, xu2020tensor, xu2021characterization, sala_2025_decoherence, stability_2025_sala, sahay2025enforcedgaplessnessstatesexponentially} and have recently been studied in the context of 2D non-Abelian topological codes \cite{sala_2025_decoherence, stability_2025_sala}.

\textit{Note Added}: During the completion of this work, we became aware of related results in \cite{zhu_non-abelian_2026} and \cite{williamson_fast_2026}.
In particular, Ref.~\cite{zhu_non-abelian_2026} discusses a construction of a non-Abelian qLDPC code and logical Clifford measurement protocol  that shares similarities with the homological approach to gauging discussed in this paper.
However, unlike our approach, the results of Ref.~\cite{zhu_non-abelian_2026} were derived to apply to codes described by Poincar\'e complexes and to certain classes of qLDPC codes mapped to these complexes via a code-to-manifold mapping.
Ref.~\cite{williamson_fast_2026} discusses the gauging of Clifford gates for logical Clifford measurement in qLDPC codes.
However, the focus of Ref.~\cite{williamson_fast_2026} was on gauging higher-form logical gates such as the flexible $\mathsf{CZ}$ gates that appears in the 3D color code, as opposed to the $0$-form $\mathsf{CZ}$ gates that we study.
Where our results overlap, they agree.

\section{Acknowledgements}

We are especially grateful to Fiona Burnell, David Long, and Tibor Rakovszky for several insightful discussions and for close collaborations on related work.
We would also like to acknowledge helpful discussions with Jeongwan Haah, Rohith Sajith, Charles Stahl, Ruben Verresen, and Ashvin Vishwanath.
M.C. acknowledges funding from Amazon Web Services, AWS Quantum Program National Science Foundation (PHY-2317110). 
The Institute for Quantum Information and Matter is an NSF Physics Frontiers Center. C.F.B.L. acknowledge support from the National Science Foundation Graduate Research Fellowship Program (NSF GRFP) and support from the Simons Collaboration on Ultra Quantum Matter, which is a grant from the Simons Foundation (618615).
V.K. acknowledges support from the Office of Naval Research Young Investigator Program (ONR YIP) under Award Number N000142412098 and from the Packard Foundation through a Packard Fellowship in Science and Engineering.
\bibliography{refs}

@article{Hierarchy,
  title = {Hierarchy of Topological Order From Finite-Depth Unitaries, Measurement, and Feedforward},
  author = {Tantivasadakarn, Nathanan and Vishwanath, Ashvin and Verresen, Ruben},
  journal = {PRX Quantum},
  volume = {4},
  issue = {2},
  pages = {020339},
  numpages = {25},
  year = {2023},
  month = {Jun},
  publisher = {American Physical Society},
  doi = {10.1103/PRXQuantum.4.020339},
  url = {https://link.aps.org/doi/10.1103/PRXQuantum.4.020339}
}

@online{verresen_efficiently_2022,
  title = {Efficiently Preparing {{Schr}}\textbackslash "odinger's Cat, Fractons and Non-{{Abelian}} Topological Order in Quantum Devices},
  author = {Verresen, Ruben and Tantivasadakarn, Nathanan and Vishwanath, Ashvin},
  date = {2022-01-06},
  eprint = {2112.03061},
  eprinttype = {arXiv},
  eprintclass = {cond-mat, physics:physics, physics:quant-ph},
  url = {http://arxiv.org/abs/2112.03061},
  urldate = {2023-12-25},
  pubstate = {prepublished}
}

@article{Tantivasadakarn_2024_LRE,
  title = {Long-Range Entanglement from Measuring Symmetry-Protected Topological Phases},
  author = {Tantivasadakarn, Nathanan and Thorngren, Ryan and Vishwanath, Ashvin and Verresen, Ruben},
  journal = {Phys. Rev. X},
  volume = {14},
  issue = {2},
  pages = {021040},
  numpages = {22},
  year = {2024},
  month = {Jun},
  publisher = {American Physical Society},
  doi = {10.1103/PhysRevX.14.021040},
  url = {https://link.aps.org/doi/10.1103/PhysRevX.14.021040}
}

@article{tantivasadakarn_shortest_2023,
  title = {Shortest {{Route}} to {{Non-Abelian Topological Order}} on a {{Quantum Processor}}},
  author = {Tantivasadakarn, Nathanan and Verresen, Ruben and Vishwanath, Ashvin},
  year = 2023,
  month = aug,
  journal = {Physical Review Letters},
  volume = {131},
  number = {6},
  pages = {060405},
  publisher = {American Physical Society},
  doi = {10.1103/PhysRevLett.131.060405},
  urldate = {2023-12-25}
}

@misc{margaritaknots,
      title={Universal fault tolerant quantum computation in 2D without getting tied in knots}, 
      author={Margarita Davydova and Andreas Bauer and Julio C. Magdalena de la Fuente and Mark Webster and Dominic J. Williamson and Benjamin J. Brown},
      year={2025},
      eprint={2503.15751},
      archivePrefix={arXiv},
      primaryClass={quant-ph},
      url={https://arxiv.org/abs/2503.15751}, 
}

@article{CalderBankShor,
  title = {Good quantum error-correcting codes exist},
  author = {Calderbank, A. R. and Shor, Peter W.},
  journal = {Phys. Rev. A},
  volume = {54},
  issue = {2},
  pages = {1098--1105},
  numpages = {0},
  year = {1996},
  month = {Aug},
  publisher = {American Physical Society},
  doi = {10.1103/PhysRevA.54.1098},
  url = {https://link.aps.org/doi/10.1103/PhysRevA.54.1098}
}

@article{Steane, volume={452},
   ISSN={1471-2946},
   url={http://dx.doi.org/10.1098/rspa.1996.0136},
   DOI={10.1098/rspa.1996.0136},
   number={1954},
   journal={Proceedings of the Royal Society of London. Series A: Mathematical, Physical and Engineering Sciences},
   publisher={The Royal Society},
   year={1996},
   month=nov, pages={2551–2577} }

@article{Kitaev_2003,
   title={Fault-tolerant quantum computation by anyons},
   volume={303},
   ISSN={0003-4916},
   url={http://dx.doi.org/10.1016/S0003-4916(02)00018-0},
   DOI={10.1016/s0003-4916(02)00018-0},
   number={1},
   journal={Annals of Physics},
   publisher={Elsevier BV},
   author={Kitaev, A.Yu.},
   year={2003},
   month=jan, pages={2–30} }

@article{barkeshli2019,
  title = {Symmetry Fractionalization, Defects, and Gauging of Topological Phases},
  author = {Barkeshli, Maissam and Bonderson, Parsa and Cheng, Meng and Wang, Zhenghan},
  year = 2019,
  month = sep,
  journal = {Physical Review B},
  volume = {100},
  number = {11},
  pages = {115147},
  issn = {2469-9950, 2469-9969},
  doi = {10.1103/PhysRevB.100.115147},
  urldate = {2024-01-06},
  langid = {english}
}

@misc{williamson_tanner,
      title={Low-overhead fault-tolerant quantum computation by gauging logical operators}, 
      author={Dominic J. Williamson and Theodore J. Yoder},
      year={2024},
      eprint={2410.02213},
      archivePrefix={arXiv},
      primaryClass={quant-ph},
      url={https://arxiv.org/abs/2410.02213}, 
}

@article{Bravyi_2024,
   title={High-threshold and low-overhead fault-tolerant quantum memory},
   volume={627},
   ISSN={1476-4687},
   url={http://dx.doi.org/10.1038/s41586-024-07107-7},
   DOI={10.1038/s41586-024-07107-7},
   number={8005},
   journal={Nature},
   publisher={Springer Science and Business Media LLC},
   author={Bravyi, Sergey and Cross, Andrew W. and Gambetta, Jay M. and Maslov, Dmitri and Rall, Patrick and Yoder, Theodore J.},
   year={2024},
   month=mar, pages={778–782} }

@article{Chen_2023,
  title = {Higher Cup Products on Hypercubic Lattices: {{Application}} to Lattice Models of Topological Phases},
  shorttitle = {Higher Cup Products on Hypercubic Lattices},
  author = {Chen, Yu-An and Tata, Sri},
  year = 2023,
  month = sep,
  journal = {Journal of Mathematical Physics},
  volume = {64},
  number = {9},
  pages = {091902},
  issn = {0022-2488},
  doi = {10.1063/5.0095189},
  urldate = {2026-01-30}
}

@misc{homologicalorder,
      author={Sahay, Rahul and Burnell, Fiona and Rakovszky, Tibor and Khemani, Vedika},
      year={2026},
      note={to appear},
      archivePrefix={arXiv},
      primaryClass={quant-ph}
}

@misc{addressablegates, 
      author={Sahay, Rahul and Long, David and Burnell, Fiona and Khemani, Vedika},
      year={2026},
      note={to appear},
      archivePrefix={arXiv},
      primaryClass={quant-ph}
}

@misc{rakovszky2023physicsgoodldpccodes,
      title={The Physics of (good) LDPC Codes I. Gauging and dualities}, 
      author={Tibor Rakovszky and Vedika Khemani},
      year={2023},
      eprint={2310.16032},
      archivePrefix={arXiv},
      primaryClass={quant-ph},
      url={https://arxiv.org/abs/2310.16032}, 
}

@misc{rakovszky2024physicsgoodldpccodes,
      title={The Physics of (good) LDPC Codes II. Product constructions}, 
      author={Tibor Rakovszky and Vedika Khemani},
      year={2024},
      eprint={2402.16831},
      archivePrefix={arXiv},
      primaryClass={quant-ph},
      url={https://arxiv.org/abs/2402.16831}, 
}

@misc{breuckmann2025cupsgatesicohomology,
      title={Cups and Gates I: Cohomology invariants and logical quantum operations}, 
      author={Nikolas P. Breuckmann and Margarita Davydova and Jens N. Eberhardt and Nathanan Tantivasadakarn},
      year={2025},
      eprint={2410.16250},
      archivePrefix={arXiv},
      primaryClass={quant-ph},
      url={https://arxiv.org/abs/2410.16250}, 
}

@misc{lin2024transversalnoncliffordgatesquantum,
      title={Transversal non-Clifford gates for quantum LDPC codes on sheaves}, 
      author={Ting-Chun Lin},
      year={2024},
      eprint={2410.14631},
      archivePrefix={arXiv},
      primaryClass={quant-ph},
      url={https://arxiv.org/abs/2410.14631}, 
}

@article{menon2025magic,
  title={Magic tricycles: Efficient magic state generation with finite block-length quantum LDPC codes},
  author={Menon, Varun and Bonilla-Ataides, J Pablo and Mehta, Rohan and Gu, Andi and Tan, Daniel Bochen and Lukin, Mikhail D},
  journal={arXiv preprint arXiv:2508.10714},
  year={2025}
}

@misc{zhu2025topological,
  title = {A Topological Theory for {{qLDPC}}: Non-{{Clifford}} Gates and Magic State Fountain on Homological Product Codes with Constant Rate and beyond the \${{N}}\textasciicircum\textbraceleft 1/3\textbraceright\$ Distance Barrier},
  shorttitle = {A Topological Theory for {{qLDPC}}},
  author = {Zhu, Guanyu},
  year = 2025,
  month = feb,
  number = {arXiv:2501.19375},
  eprint = {2501.19375},
  primaryclass = {quant-ph},
  publisher = {arXiv},
  doi = {10.48550/arXiv.2501.19375},
  urldate = {2025-03-02},
  archiveprefix = {arXiv}
}

@article{guo_toward_2026,
  title = {Toward {{Self-Correcting Quantum Codes}} for {{Neutral Atom Arrays}}},
  author = {Guo, Jinkang and Hong, Yifan and Kaufman, Adam and Lucas, Andrew},
  year = 2026,
  month = jan,
  journal = {PRX Quantum},
  volume = {7},
  number = {1},
  pages = {010301},
  publisher = {American Physical Society},
  doi = {10.1103/mfmt-fwkg},
  urldate = {2026-02-02}
}

@misc{ungauging,
      title={Ungauging quantum error-correcting codes}, 
      author={Aleksander Kubica and Beni Yoshida},
      year={2018},
      eprint={1805.01836},
      archivePrefix={arXiv},
      primaryClass={quant-ph},
      url={https://arxiv.org/abs/1805.01836}, 
}

@misc{sajithcodes,
      title={Non-Clifford gates between stabilizer codes via non-Abelian topological order}, 
      author={Rohith Sajith and Zijian Song and Brenden Roberts and Varun Menon and Yabo Li},
      year={2025},
      eprint={2505.18265},
      archivePrefix={arXiv},
      primaryClass={quant-ph},
      url={https://arxiv.org/abs/2505.18265}, 
}

@article{bluvstein_fault-tolerant_2026,
  title = {A Fault-Tolerant Neutral-Atom Architecture for Universal Quantum Computation},
  author = {Bluvstein, Dolev and Geim, Alexandra A. and Li, Sophie H. and Evered, Simon J. and Bonilla Ataides, J. Pablo and Baranes, Gefen and Gu, Andi and Manovitz, Tom and Xu, Muqing and Kalinowski, Marcin and Majidy, Shayan and Kokail, Christian and Maskara, Nishad and Trapp, Elias C. and Stewart, Luke M. and Hollerith, Simon and Zhou, Hengyun and Gullans, Michael J. and Yelin, Susanne F. and Greiner, Markus and Vuleti{\'c}, Vladan and Cain, Madelyn and Lukin, Mikhail D.},
  year = 2026,
  month = jan,
  journal = {Nature},
  volume = {649},
  number = {8095},
  pages = {39--46},
  publisher = {Nature Publishing Group},
  issn = {1476-4687},
  doi = {10.1038/s41586-025-09848-5},
  urldate = {2026-02-05},
  copyright = {2025 The Author(s)},
  langid = {english}
}

@article{sinclair_fault-tolerant_2025,
  title = {Fault-Tolerant Optical Interconnects for Neutral-Atom Arrays},
  author = {Sinclair, Josiah and Ramette, Joshua and Grinkemeyer, Brandon and Bluvstein, Dolev and Lukin, Mikhail D. and Vuleti{\'c}, Vladan},
  year = 2025,
  month = mar,
  journal = {Physical Review Research},
  volume = {7},
  number = {1},
  pages = {013313},
  publisher = {American Physical Society},
  doi = {10.1103/PhysRevResearch.7.013313},
  urldate = {2026-02-05}
}

@article{chiu_continuous_2025,
  title = {Continuous Operation of a Coherent 3,000-Qubit System},
  author = {Chiu, Neng-Chun and Trapp, Elias C. and Guo, Jinen and Abobeih, Mohamed H. and Stewart, Luke M. and Hollerith, Simon and Stroganov, Pavel L. and Kalinowski, Marcin and Geim, Alexandra A. and Evered, Simon J. and Li, Sophie H. and Lyu, Xingjian and Peters, Lisa M. and Bluvstein, Dolev and Wang, Tout T. and Greiner, Markus and Vuleti{\'c}, Vladan and Lukin, Mikhail D.},
  year = 2025,
  month = oct,
  journal = {Nature},
  volume = {646},
  number = {8087},
  pages = {1075--1080},
  publisher = {Nature Publishing Group},
  issn = {1476-4687},
  doi = {10.1038/s41586-025-09596-6},
  urldate = {2026-02-05},
  copyright = {2025 The Author(s)},
  langid = {english}
}

@article{bluvstein_logical_2024,
  title = {Logical Quantum Processor Based on Reconfigurable Atom Arrays},
  author = {Bluvstein, Dolev and Evered, Simon J. and Geim, Alexandra A. and Li, Sophie H. and Zhou, Hengyun and Manovitz, Tom and Ebadi, Sepehr and Cain, Madelyn and Kalinowski, Marcin and Hangleiter, Dominik and Bonilla Ataides, J. Pablo and Maskara, Nishad and Cong, Iris and Gao, Xun and Sales Rodriguez, Pedro and Karolyshyn, Thomas and Semeghini, Giulia and Gullans, Michael J. and Greiner, Markus and Vuleti{\'c}, Vladan and Lukin, Mikhail D.},
  year = 2024,
  month = feb,
  journal = {Nature},
  volume = {626},
  number = {7997},
  pages = {58--65},
  publisher = {Nature Publishing Group},
  issn = {1476-4687},
  doi = {10.1038/s41586-023-06927-3},
  urldate = {2026-02-05},
  copyright = {2023 The Author(s)},
  langid = {english}
}

@article{google2025quantum,
  title = {Quantum Error Correction below the Surface Code Threshold},
  author = {Acharya et al., Rajeev and {Google Quantum AI and Collaborators}},
  year = 2025,
  month = feb,
  journal = {Nature},
  volume = {638},
  number = {8052},
  pages = {920--926},
  publisher = {Nature Publishing Group},
  issn = {1476-4687},
  doi = {10.1038/s41586-024-08449-y},
  urldate = {2026-02-11},
  copyright = {2024 The Author(s)},
  langid = {english}
}

@article{moses_race-track_2023,
  title = {A {{Race-Track Trapped-Ion Quantum Processor}}},
  author = {Moses et al., S. A.},
  year = 2023,
  month = dec,
  journal = {Physical Review X},
  volume = {13},
  number = {4},
  pages = {041052},
  publisher = {American Physical Society},
  doi = {10.1103/PhysRevX.13.041052},
  urldate = {2026-01-25}
}

@misc{ransford_helios_2025,
  title = {Helios: {{A}} 98-Qubit Trapped-Ion Quantum Computer},
  shorttitle = {Helios},
  author = {Ransford et al., Anthony},
  year = 2025,
  month = nov,
  number = {arXiv:2511.05465},
  eprint = {2511.05465},
  primaryclass = {quant-ph},
  publisher = {arXiv},
  doi = {10.48550/arXiv.2511.05465},
  urldate = {2026-02-05},
  archiveprefix = {arXiv}
}

@inproceedings{panteleev_asymptotically_2022,
  title = {Asymptotically Good {{Quantum}} and Locally Testable Classical {{LDPC}} Codes},
  booktitle = {Proceedings of the 54th {{Annual ACM SIGACT Symposium}} on {{Theory}} of {{Computing}}},
  author = {Panteleev, Pavel and Kalachev, Gleb},
  year = 2022,
  month = jun,
  series = {{{STOC}} 2022},
  pages = {375--388},
  publisher = {Association for Computing Machinery},
  address = {New York, NY, USA},
  doi = {10.1145/3519935.3520017},
  urldate = {2026-02-05},
  isbn = {978-1-4503-9264-8}
}

@article{de_roeck_low-density_2025,
  title = {Low-{{Density Parity-Check Stabilizer Codes}} as {{Gapped Quantum Phases}}: {{Stability}} under {{Graph-Local Perturbations}}},
  shorttitle = {Low-{{Density Parity-Check Stabilizer Codes}} as {{Gapped Quantum Phases}}},
  author = {De Roeck, Wojciech and Khemani, Vedika and Li, Yaodong and O'Dea, Nicholas and Rakovszky, Tibor},
  year = 2025,
  month = aug,
  journal = {PRX Quantum},
  volume = {6},
  number = {3},
  pages = {030330},
  issn = {2691-3399},
  doi = {10.1103/7x71-8j7k},
  urldate = {2025-11-08},
  langid = {english}
}

@article{yin_low-density_2025,
  title = {Low-{{Density Parity-Check Codes}} as {{Stable Phases}} of {{Quantum Matter}}},
  author = {Yin, Chao and Lucas, Andrew},
  year = 2025,
  month = aug,
  journal = {PRX Quantum},
  volume = {6},
  number = {3},
  pages = {030329},
  issn = {2691-3399},
  doi = {10.1103/361k-nj4b},
  urldate = {2026-02-05},
  langid = {english}
}

@misc{williamson_fast_2026,
  title = {Fast Magic State Preparation by Gauging Higher-Form Transversal Gates in Parallel},
  author = {Williamson, Dominic J.},
  year = 2026,
  month = jan,
  number = {arXiv:2601.22939},
  eprint = {2601.22939},
  primaryclass = {quant-ph},
  publisher = {arXiv},
  doi = {10.48550/arXiv.2601.22939},
  urldate = {2026-02-02},
  archiveprefix = {arXiv}
}

@misc{li_poincare_2025,
  title = {Poincar\'e {{Duality}} and {{Multiplicative Structures}} on {{Quantum Codes}}},
  author = {Li, Yiming and Li, Zimu and Liu, Zi-Wen and Nguyen, Quynh T.},
  year = 2025,
  month = dec,
  number = {arXiv:2512.21922},
  eprint = {2512.21922},
  primaryclass = {quant-ph},
  publisher = {arXiv},
  doi = {10.48550/arXiv.2512.21922},
  urldate = {2025-12-31},
  archiveprefix = {arXiv}
}

@misc{kobayashi_clifford_2025,
  title = {Clifford {{Hierarchy Stabilizer Codes}}: {{Transversal Non-Clifford Gates}} and {{Magic}}},
  shorttitle = {Clifford {{Hierarchy Stabilizer Codes}}},
  author = {Kobayashi, Ryohei and Zhu, Guanyu and Hsin, Po-Shen},
  year = 2025,
  month = dec,
  number = {arXiv:2511.02900},
  eprint = {2511.02900},
  primaryclass = {quant-ph},
  publisher = {arXiv},
  doi = {10.48550/arXiv.2511.02900},
  urldate = {2025-12-24},
  archiveprefix = {arXiv}
}

@misc{hsin_non-abelian_2025,
  title = {Non-{{Abelian Self-Correcting Quantum Memory}} and {{Single-shot Non-Clifford Gate}} beyond the \$n\textasciicircum\textbraceleft 1/3\textbraceright\$ {{Distance Barrier}}},
  author = {Hsin, Po-Shen and Kobayashi, Ryohei and Zhu, Guanyu},
  year = 2025,
  month = sep,
  number = {arXiv:2405.11719},
  eprint = {2405.11719},
  primaryclass = {quant-ph},
  publisher = {arXiv},
  doi = {10.48550/arXiv.2405.11719},
  urldate = {2025-09-26},
  archiveprefix = {arXiv}
}

@misc{zhu_non-abelian_2026,
  title = {Non-{{Abelian qLDPC}}: {{TQFT Formalism}}, {{Addressable Gauging Measurement}} and {{Application}} to {{Magic State Fountain}} on {{2D Product Codes}}},
  shorttitle = {Non-{{Abelian qLDPC}}},
  author = {Zhu, Guanyu and Kobayashi, Ryohei and Hsin, Po-Shen},
  year = 2026,
  month = jan,
  number = {arXiv:2601.06736},
  eprint = {2601.06736},
  primaryclass = {quant-ph},
  publisher = {arXiv},
  doi = {10.48550/arXiv.2601.06736},
  urldate = {2026-01-13},
  archiveprefix = {arXiv}
}

@misc{placke_topological_2024,
  title = {Topological {{Quantum Spin Glass Order}} and Its Realization in {{qLDPC}} Codes},
  author = {Placke, Benedikt and Rakovszky, Tibor and Breuckmann, Nikolas P. and Khemani, Vedika},
  year = 2024,
  month = dec,
  number = {arXiv:2412.13248},
  eprint = {2412.13248},
  primaryclass = {quant-ph},
  publisher = {arXiv},
  doi = {10.48550/arXiv.2412.13248},
  urldate = {2025-05-13},
  archiveprefix = {arXiv}
}

@misc{huang_hybrid_2025,
  title = {Hybrid {{Lattice Surgery}}: {{Non-Clifford Gates}} via {{Non-Abelian Surface Codes}}},
  shorttitle = {Hybrid {{Lattice Surgery}}},
  author = {Huang, Sheng-Jie and Warman, Alison and {Schafer-Nameki}, Sakura and Chen, Yanzhu},
  year = 2025,
  month = oct,
  number = {arXiv:2510.20890},
  eprint = {2510.20890},
  primaryclass = {quant-ph},
  publisher = {arXiv},
  doi = {10.48550/arXiv.2510.20890},
  urldate = {2025-10-29},
  archiveprefix = {arXiv}
}

@misc{warman_transversal_2025,
  title = {Transversal {{Clifford-Hierarchy Gates}} via {{Non-Abelian Surface Codes}}},
  author = {Warman, Alison and {Schafer-Nameki}, Sakura},
  year = 2025,
  month = dec,
  number = {arXiv:2512.13777},
  eprint = {2512.13777},
  primaryclass = {quant-ph},
  publisher = {arXiv},
  doi = {10.48550/arXiv.2512.13777},
  urldate = {2025-12-18},
  archiveprefix = {arXiv}
}

@article{Shirley_2019,
   title={Foliated fracton order from gauging subsystem symmetries},
   volume={6},
   ISSN={2542-4653},
   url={http://dx.doi.org/10.21468/SciPostPhys.6.4.041},
   DOI={10.21468/scipostphys.6.4.041},
   number={4},
   journal={SciPost Physics},
   publisher={Stichting SciPost},
   author={Shirley, Wilbur and Slagle, Kevin and Chen, Xie},
   year={2019},
   month=apr }

@article{You_2018,
   title={Subsystem symmetry protected topological order},
   volume={98},
   ISSN={2469-9969},
   url={http://dx.doi.org/10.1103/PhysRevB.98.035112},
   DOI={10.1103/physrevb.98.035112},
   number={3},
   journal={Physical Review B},
   publisher={American Physical Society (APS)},
   author={You, Yizhi and Devakul, Trithep and Burnell, F. J. and Sondhi, S. L.},
   year={2018},
   month=jul }

@article{Burnell_2022,
   title={Anomaly inflow for subsystem symmetries},
   volume={106},
   ISSN={2469-9969},
   url={http://dx.doi.org/10.1103/PhysRevB.106.085113},
   DOI={10.1103/physrevb.106.085113},
   number={8},
   journal={Physical Review B},
   publisher={American Physical Society (APS)},
   author={Burnell, Fiona J. and Devakul, Trithep and Gorantla, Pranay and Lam, Ho Tat and Shao, Shu-Heng},
   year={2022},
   month=aug }

@article{Bulmash_2019,
   title={Gauging fractons: Immobile non-Abelian quasiparticles, fractals, and position-dependent degeneracies},
   volume={100},
   ISSN={2469-9969},
   url={http://dx.doi.org/10.1103/PhysRevB.100.155146},
   DOI={10.1103/physrevb.100.155146},
   number={15},
   journal={Physical Review B},
   publisher={American Physical Society (APS)},
   author={Bulmash, Daniel and Barkeshli, Maissam},
   year={2019},
   month=oct }

@article{Prem_2019,
   title={Gauging permutation symmetries as a route to non-Abelian fractons},
   volume={7},
   ISSN={2542-4653},
   url={http://dx.doi.org/10.21468/SciPostPhys.7.5.068},
   DOI={10.21468/scipostphys.7.5.068},
   number={5},
   journal={SciPost Physics},
   publisher={Stichting SciPost},
   author={Prem, Abhinav and Williamson, Dominic},
   year={2019},
   month=nov }

@article{Wang_2020,
   title={Non-Abelian gauged fracton matter field theory: Sigma models, superfluids, and vortices},
   volume={2},
   ISSN={2643-1564},
   url={http://dx.doi.org/10.1103/PhysRevResearch.2.043219},
   DOI={10.1103/physrevresearch.2.043219},
   number={4},
   journal={Physical Review Research},
   publisher={American Physical Society (APS)},
   author={Wang, Juven and Yau, Shing-Tung},
   year={2020},
   month=nov }

@article{Tu_2021,
   title={Non-Abelian fracton order from gauging a mixture of subsystem and global symmetries},
   volume={3},
   ISSN={2643-1564},
   url={http://dx.doi.org/10.1103/PhysRevResearch.3.043084},
   DOI={10.1103/physrevresearch.3.043084},
   number={4},
   journal={Physical Review Research},
   publisher={American Physical Society (APS)},
   author={Tu, Yi-Ting and Chang, Po-Yao},
   year={2021},
   month=oct }

@article{Williamson_2023,
   title={Designer non-Abelian fractons from topological layers},
   volume={107},
   ISSN={2469-9969},
   url={http://dx.doi.org/10.1103/PhysRevB.107.035103},
   DOI={10.1103/physrevb.107.035103},
   number={3},
   journal={Physical Review B},
   publisher={American Physical Society (APS)},
   author={Williamson, Dominic J. and Cheng, Meng},
   year={2023},
   month=jan }

@article{Vijay_2016,
   title={Fracton topological order, generalized lattice gauge theory, and duality},
   volume={94},
   ISSN={2469-9969},
   url={http://dx.doi.org/10.1103/PhysRevB.94.235157},
   DOI={10.1103/physrevb.94.235157},
   number={23},
   journal={Physical Review B},
   publisher={American Physical Society (APS)},
   author={Vijay, Sagar and Haah, Jeongwan and Fu, Liang},
   year={2016},
   month=dec }

@article{MacKay_2004,
   title={Sparse-Graph Codes for Quantum Error Correction},
   volume={50},
   ISSN={0018-9448},
   url={http://dx.doi.org/10.1109/TIT.2004.834737},
   DOI={10.1109/tit.2004.834737},
   number={10},
   journal={IEEE Transactions on Information Theory},
   publisher={Institute of Electrical and Electronics Engineers (IEEE)},
   author={MacKay, D.J.C. and Mitchison, G. and McFadden, P.L.},
   year={2004},
   month=oct, pages={2315–2330} }

@article{Kovalev_2013,
   title={Quantum Kronecker sum-product low-density parity-check codes with finite rate},
   volume={88},
   ISSN={1094-1622},
   url={http://dx.doi.org/10.1103/PhysRevA.88.012311},
   DOI={10.1103/physreva.88.012311},
   number={1},
   journal={Physical Review A},
   publisher={American Physical Society (APS)},
   author={Kovalev, Alexey A. and Pryadko, Leonid P.},
   year={2013},
   month=jul }

@inproceedings{Leverrier_2015,
   title={Quantum Expander Codes},
   url={http://dx.doi.org/10.1109/FOCS.2015.55},
   DOI={10.1109/focs.2015.55},
   booktitle={2015 IEEE 56th Annual Symposium on Foundations of Computer Science},
   publisher={IEEE},
   author={Leverrier, Anthony and Tillich, Jean-Pierre and Zemor, Gilles},
   year={2015},
   month=oct, pages={810–824} }

@inproceedings{leverrier2022quantumtannercodes,
  title = {Quantum {{Tanner}} Codes},
  booktitle = {2022 {{IEEE}} 63rd {{Annual Symposium}} on {{Foundations}} of {{Computer Science}} ({{FOCS}})},
  author = {Leverrier, Anthony and Z{\'e}mor, Gilles},
  year = 2022,
  month = oct,
  pages = {872--883},
  issn = {2575-8454},
  doi = {10.1109/FOCS54457.2022.00117},
  urldate = {2026-02-08}
}

@article{Breuckmann_2021,
   title={Balanced Product Quantum Codes},
   volume={67},
   ISSN={1557-9654},
   url={http://dx.doi.org/10.1109/TIT.2021.3097347},
   DOI={10.1109/tit.2021.3097347},
   number={10},
   journal={IEEE Transactions on Information Theory},
   publisher={Institute of Electrical and Electronics Engineers (IEEE)},
   author={Breuckmann, Nikolas P. and Eberhardt, Jens N.},
   year={2021},
   month=oct, pages={6653–6674} }

@article{Panteleev_2021,
   title={Degenerate Quantum LDPC Codes With Good Finite Length Performance},
   volume={5},
   ISSN={2521-327X},
   url={http://dx.doi.org/10.22331/q-2021-11-22-585},
   DOI={10.22331/q-2021-11-22-585},
   journal={Quantum},
   publisher={Verein zur Forderung des Open Access Publizierens in den Quantenwissenschaften},
   author={Panteleev, Pavel and Kalachev, Gleb},
   year={2021},
   month=nov, pages={585} }

@article{Breuckmann_2021_rev,
  title = {Quantum {{Low-Density Parity-Check Codes}}},
  author = {Breuckmann, Nikolas P. and Eberhardt, Jens Niklas},
  year = 2021,
  month = oct,
  journal = {PRX Quantum},
  volume = {2},
  number = {4},
  pages = {040101},
  publisher = {American Physical Society},
  doi = {10.1103/PRXQuantum.2.040101},
  urldate = {2024-09-27}
}

@article{RevModPhys.80.1083,
  title = {Non-Abelian anyons and topological quantum computation},
  author = {Nayak, Chetan and Simon, Steven H. and Stern, Ady and Freedman, Michael and Das Sarma, Sankar},
  journal = {Rev. Mod. Phys.},
  volume = {80},
  issue = {3},
  pages = {1083--1159},
  numpages = {0},
  year = {2008},
  month = {Sep},
  publisher = {American Physical Society},
  doi = {10.1103/RevModPhys.80.1083},
  url = {https://link.aps.org/doi/10.1103/RevModPhys.80.1083}
}

@article{Iqbal_2024,
   title={Non-Abelian topological order and anyons on a trapped-ion processor},
   volume={626},
   ISSN={1476-4687},
   url={http://dx.doi.org/10.1038/s41586-023-06934-4},
   DOI={10.1038/s41586-023-06934-4},
   number={7999},
   journal={Nature},
   publisher={Springer Science and Business Media LLC},
   author={Iqbal, Mohsin and Tantivasadakarn, Nathanan and Verresen, Ruben and Campbell, Sara L. and Dreiling, Joan M. and Figgatt, Caroline and Gaebler, John P. and Johansen, Jacob and Mills, Michael and Moses, Steven A. and Pino, Juan M. and Ransford, Anthony and Rowe, Mary and Siegfried, Peter and Stutz, Russell P. and Foss-Feig, Michael and Vishwanath, Ashvin and Dreyer, Henrik},
   year={2024},
   month=feb, pages={505–511} }

@misc{lo2026universaltopologicalgatesbraiding,
      title={Universal Topological Gates from Braiding and Fusing Anyons on Quantum Hardware}, 
      author={Chiu Fan Bowen Lo and Anasuya Lyons and Dan Gresh and Michael Mills and Peter E. Siegfried and Maxwell D. Urmey and Nathanan Tantivasadakarn and Henrik Dreyer and Ashvin Vishwanath and Ruben Verresen and Mohsin Iqbal},
      year={2026},
      eprint={2601.20956},
      archivePrefix={arXiv},
      primaryClass={quant-ph},
      url={https://arxiv.org/abs/2601.20956}, 
}

@article{Tillich_2014,
   title={Quantum LDPC Codes With Positive Rate and Minimum Distance Proportional to the Square Root of the Blocklength},
   volume={60},
   ISSN={1557-9654},
   url={http://dx.doi.org/10.1109/TIT.2013.2292061},
   DOI={10.1109/tit.2013.2292061},
   number={2},
   journal={IEEE Transactions on Information Theory},
   publisher={Institute of Electrical and Electronics Engineers (IEEE)},
   author={Tillich, Jean-Pierre and Zemor, Gilles},
   year={2014},
   month=feb, pages={1193–1202} }

@article{wen_topological_1990,
  title = {Topological Orders in Rigid States},
  author = {Wen, X. G.},
  year = 1990,
  month = feb,
  journal = {International Journal of Modern Physics B},
  volume = {04},
  number = {02},
  pages = {239--271},
  publisher = {World Scientific Publishing Co.},
  issn = {0217-9792},
  doi = {10.1142/S0217979290000139},
  urldate = {2024-09-27}
}

@article{RevModPhys.89.041004,
  title = {Colloquium: Zoo of quantum-topological phases of matter},
  author = {Wen, Xiao-Gang},
  journal = {Rev. Mod. Phys.},
  volume = {89},
  issue = {4},
  pages = {041004},
  numpages = {17},
  year = {2017},
  month = {Dec},
  publisher = {American Physical Society},
  doi = {10.1103/RevModPhys.89.041004},
  url = {https://link.aps.org/doi/10.1103/RevModPhys.89.041004}
}

@inproceedings{Hastings_2021, series={STOC ’21},
   title={Fiber bundle codes: breaking the
            n
            1/2
            polylog(
            n
            ) barrier for Quantum LDPC codes},
   url={http://dx.doi.org/10.1145/3406325.3451005},
   DOI={10.1145/3406325.3451005},
   booktitle={Proceedings of the 53rd Annual ACM SIGACT Symposium on Theory of Computing},
   publisher={ACM},
   author={Hastings, Matthew B. and Haah, Jeongwan and O’Donnell, Ryan},
   year={2021},
   month=jun, pages={1276–1288},
   collection={STOC ’21} }

@article{Haah_2011,
   title={Local stabilizer codes in three dimensions without string logical operators},
   volume={83},
   ISSN={1094-1622},
   url={http://dx.doi.org/10.1103/PhysRevA.83.042330},
   DOI={10.1103/physreva.83.042330},
   number={4},
   journal={Physical Review A},
   publisher={American Physical Society (APS)},
   author={Haah, Jeongwan},
   year={2011},
   month=apr }

@article{bravyi_2010,
  title = {Topological Quantum Order: Stability under Local Perturbations},
  shorttitle = {Topological Quantum Order},
  author = {Bravyi, Sergey and Hastings, Matthew and Michalakis, Spyridon},
  year = 2010,
  month = sep,
  journal = {Journal of Mathematical Physics},
  volume = {51},
  number = {9},
  pages = {093512},
  issn = {0022-2488, 1089-7658},
  doi = {10.1063/1.3490195},
  urldate = {2026-01-29}
}

@ARTICLE{SipserSpiel,
  author={Sipser, M. and Spielman, D.A.},
  journal={IEEE Transactions on Information Theory}, 
  title={Expander codes}, 
  year={1996},
  volume={42},
  number={6},
  pages={1710-1722},
  doi={10.1109/18.556667}}

@article{Bombin_2007,
   title={Homological error correction: Classical and quantum codes},
   volume={48},
   ISSN={1089-7658},
   url={http://dx.doi.org/10.1063/1.2731356},
   DOI={10.1063/1.2731356},
   number={5},
   journal={Journal of Mathematical Physics},
   publisher={AIP Publishing},
   author={Bombin, H. and Martin-Delgado, M. A.},
   year={2007},
   month=may }

@inproceedings{dinur_good_2023,
  title = {Good {{Quantum LDPC Codes}} with {{Linear Time Decoders}}},
  booktitle = {Proceedings of the 55th {{Annual ACM Symposium}} on {{Theory}} of {{Computing}}},
  author = {Dinur, Irit and Hsieh, Min-Hsiu and Lin, Ting-Chun and Vidick, Thomas},
  year = 2023,
  month = jun,
  series = {{{STOC}} 2023},
  pages = {905--918},
  publisher = {Association for Computing Machinery},
  address = {New York, NY, USA},
  doi = {10.1145/3564246.3585101},
  urldate = {2026-02-07},
  isbn = {978-1-4503-9913-5}
}

@article{GappedLDPC,
  title = {Low-Density Parity-Check Stabilizer Codes as Gapped Quantum Phases: Stability under Graph-Local Perturbations},
  author = {De Roeck, Wojciech and Khemani, Vedika and Li, Yaodong and O'Dea, Nicholas and Rakovszky, Tibor},
  journal = {PRX Quantum},
  volume = {6},
  issue = {3},
  pages = {030330},
  numpages = {31},
  year = {2025},
  month = {Aug},
  publisher = {American Physical Society},
  doi = {10.1103/7x71-8j7k},
  url = {https://link.aps.org/doi/10.1103/7x71-8j7k}
}

@article{Tan_2025,
  title = {Fracton Models from Product Codes},
  author = {Tan, Yi and Roberts, Brenden and Tantivasadakarn, Nathanan and Yoshida, Beni and Yao, Norman Y.},
  year = 2025,
  month = sep,
  journal = {Physical Review Research},
  volume = {7},
  number = {3},
  pages = {L032062},
  issn = {2643-1564},
  doi = {10.1103/f48m-rlh3},
  urldate = {2026-02-08},
  langid = {english}
}

@article{bravyi_universal_2006,
  title = {Universal Quantum Computation with the {$\nu$} = 5 / 2 Fractional Quantum {{Hall}} State},
  author = {Bravyi, Sergey},
  year = 2006,
  month = apr,
  journal = {Physical Review A},
  volume = {73},
  number = {4},
  pages = {042313},
  issn = {1050-2947, 1094-1622},
  doi = {10.1103/PhysRevA.73.042313},
  urldate = {2026-02-08},
  copyright = {http://link.aps.org/licenses/aps-default-license},
  langid = {english}
}

@article{cui_universal_2015,
  title = {Universal Quantum Computation with Weakly Integral Anyons},
  author = {Cui, Shawn X. and Hong, Seung-Moon and Wang, Zhenghan},
  year = 2015,
  month = may,
  journal = {Quantum Information Processing},
  volume = {14},
  number = {8},
  eprint = {1401.7096},
  primaryclass = {quant-ph},
  pages = {2687--2727},
  issn = {1570-0755, 1573-1332},
  doi = {10.1007/s11128-015-1016-y},
  urldate = {2024-06-07},
  archiveprefix = {arXiv},
  langid = {english}
}

@article{cong_universal_2017,
  title = {Universal {{Quantum Computation}} with {{Gapped Boundaries}}},
  author = {Cong, Iris and Cheng, Meng and Wang, Zhenghan},
  year = 2017,
  month = oct,
  journal = {Physical Review Letters},
  volume = {119},
  number = {17},
  pages = {170504},
  issn = {0031-9007, 1079-7114},
  doi = {10.1103/PhysRevLett.119.170504},
  urldate = {2025-04-15},
  copyright = {https://link.aps.org/licenses/aps-default-license},
  langid = {english}
}

@article{laubscher_universal_2019,
  title = {Universal Quantum Computation in the Surface Code Using Non-{{Abelian}} Islands},
  author = {Laubscher, Katharina and Loss, Daniel and Wootton, James R.},
  year = 2019,
  month = jul,
  journal = {Physical Review A},
  volume = {100},
  number = {1},
  pages = {012338},
  issn = {2469-9926, 2469-9934},
  doi = {10.1103/PhysRevA.100.012338},
  urldate = {2024-12-18}
}

@article{levaillant_universal_2015,
  title = {Universal Gates via Fusion and Measurement Operations on {{SU}} ( 2 ) 4 Anyons},
  author = {Levaillant, Claire and Bauer, Bela and Freedman, Michael and Wang, Zhenghan and Bonderson, Parsa},
  year = 2015,
  month = jul,
  journal = {Physical Review A},
  volume = {92},
  number = {1},
  pages = {012301},
  issn = {1050-2947, 1094-1622},
  doi = {10.1103/PhysRevA.92.012301},
  urldate = {2025-10-28},
  copyright = {http://link.aps.org/licenses/aps-default-license},
  langid = {english}
}

@article{bonderson_2008,
  title = {Measurement-{{Only Topological Quantum Computation}}},
  author = {Bonderson, Parsa and Freedman, Michael and Nayak, Chetan},
  year = 2008,
  month = jun,
  journal = {Physical Review Letters},
  volume = {101},
  number = {1},
  pages = {010501},
  issn = {0031-9007, 1079-7114},
  doi = {10.1103/PhysRevLett.101.010501},
  urldate = {2026-01-16}
}

@article{freedman_modular_2002,
  title = {A {{Modular Functor Which}} Is {{Universal}} for {{Quantum Computation}}},
  author = {Freedman, Michael H. and Larsen, Michael and Wang, Zhenghan},
  year = 2002,
  month = jun,
  journal = {Communications in Mathematical Physics},
  volume = {227},
  number = {3},
  pages = {605--622},
  issn = {1432-0916},
  doi = {10.1007/s002200200645},
  urldate = {2024-08-29},
  langid = {english}
}

@inproceedings{Anshu_2023, series={STOC ’23},
   title={NLTS Hamiltonians from Good Quantum Codes},
   url={http://dx.doi.org/10.1145/3564246.3585114},
   DOI={10.1145/3564246.3585114},
   booktitle={Proceedings of the 55th Annual ACM Symposium on Theory of Computing},
   publisher={ACM},
   author={Anshu, Anurag and Breuckmann, Nikolas P. and Nirkhe, Chinmay},
   year={2023},
   month=jun, pages={1090–1096},
   collection={STOC ’23} }

@article{Gorantla_2023,
   title={Gapped lineon and fracton models on graphs},
   volume={107},
   ISSN={2469-9969},
   url={http://dx.doi.org/10.1103/PhysRevB.107.125121},
   DOI={10.1103/physrevb.107.125121},
   number={12},
   journal={Physical Review B},
   publisher={American Physical Society (APS)},
   author={Gorantla, Pranay and Lam, Ho Tat and Seiberg, Nathan and Shao, Shu-Heng},
   year={2023},
   month=mar }

@article{bombin_family_2008,
  title = {Family of Non-{{Abelian Kitaev}} Models on a Lattice: {{Topological}} Condensation and Confinement},
  shorttitle = {Family of Non-{{Abelian Kitaev}} Models on a Lattice},
  author = {Bombin, H. and {Martin-Delgado}, M. A.},
  year = 2008,
  month = sep,
  journal = {Physical Review B},
  volume = {78},
  number = {11},
  pages = {115421},
  publisher = {American Physical Society},
  doi = {10.1103/PhysRevB.78.115421},
  urldate = {2024-02-07}
}

@article{chen_universal_2025,
  title = {A Universal Circuit Set Using the {{S3}} Quantum Double},
  author = {Chen, Liyuan and Ren, Yuanjie and Fan, Ruihua and Jaffe, Arthur},
  year = 2025,
  month = jul,
  journal = {npj Quantum Information},
  volume = {11},
  number = {1},
  pages = {112},
  publisher = {Nature Publishing Group},
  issn = {2056-6387},
  doi = {10.1038/s41534-025-01063-4},
  urldate = {2026-02-08},
  copyright = {2025 The Author(s)},
  langid = {english}
}

@misc{lo_universal_2025,
  title = {Universal {{Quantum Computation}} with the \${{S}}\_3\$ {{Quantum Double}}: {{A Pedagogical Exposition}}},
  shorttitle = {Universal {{Quantum Computation}} with the \${{S}}\_3\$ {{Quantum Double}}},
  author = {Lo, Chiu Fan Bowen and Lyons, Anasuya and Verresen, Ruben and Vishwanath, Ashvin and Tantivasadakarn, Nathanan},
  year = 2025,
  month = feb,
  number = {arXiv:2502.14974},
  eprint = {2502.14974},
  primaryclass = {quant-ph},
  publisher = {arXiv},
  doi = {10.48550/arXiv.2502.14974},
  urldate = {2025-03-01},
  archiveprefix = {arXiv}
}

@article{Haah_2013,
   title={Commuting Pauli Hamiltonians as Maps between Free Modules},
   volume={324},
   ISSN={1432-0916},
   url={http://dx.doi.org/10.1007/s00220-013-1810-2},
   DOI={10.1007/s00220-013-1810-2},
   number={2},
   journal={Communications in Mathematical Physics},
   publisher={Springer Science and Business Media LLC},
   author={Haah, Jeongwan},
   year={2013},
   month=oct, pages={351–399} }

@misc{yoder2025tourgrossmodularquantum,
      title={Tour de gross: A modular quantum computer based on bivariate bicycle codes}, 
      author={Theodore J. Yoder and Eddie Schoute and Patrick Rall and Emily Pritchett and Jay M. Gambetta and Andrew W. Cross and Malcolm Carroll and Michael E. Beverland},
      year={2025},
      eprint={2506.03094},
      archivePrefix={arXiv},
      primaryClass={quant-ph},
      url={https://arxiv.org/abs/2506.03094}, 
}

@misc{eberhardt2024logicaloperatorsfoldtransversalgates,
      title={Logical Operators and Fold-Transversal Gates of Bivariate Bicycle Codes}, 
      author={Jens Niklas Eberhardt and Vincent Steffan},
      year={2024},
      eprint={2407.03973},
      archivePrefix={arXiv},
      primaryClass={quant-ph},
      url={https://arxiv.org/abs/2407.03973}, 
}

@misc{cross2025improvedqldpcsurgerylogical,
      title={Improved QLDPC Surgery: Logical Measurements and Bridging Codes}, 
      author={Andrew W. Cross and Zhiyang He and Patrick J. Rall and Theodore J. Yoder},
      year={2025},
      eprint={2407.18393},
      archivePrefix={arXiv},
      primaryClass={quant-ph},
      url={https://arxiv.org/abs/2407.18393}, 
}

@misc{xu2025batchedhighratelogicaloperations,
      title={Batched high-rate logical operations for quantum LDPC codes}, 
      author={Qian Xu and Hengyun Zhou and Dolev Bluvstein and Madelyn Cain and Marcin Kalinowski and John Preskill and Mikhail D. Lukin and Nishad Maskara},
      year={2025},
      eprint={2510.06159},
      archivePrefix={arXiv},
      primaryClass={quant-ph},
      url={https://arxiv.org/abs/2510.06159}, 
}

@misc{xu2023constantoverheadfaulttolerantquantumcomputation,
      title={Constant-Overhead Fault-Tolerant Quantum Computation with Reconfigurable Atom Arrays}, 
      author={Qian Xu and J. Pablo Bonilla Ataides and Christopher A. Pattison and Nithin Raveendran and Dolev Bluvstein and Jonathan Wurtz and Bane Vasic and Mikhail D. Lukin and Liang Jiang and Hengyun Zhou},
      year={2023},
      eprint={2308.08648},
      archivePrefix={arXiv},
      primaryClass={quant-ph},
      url={https://arxiv.org/abs/2308.08648}, 
}

@misc{xu2024fastparallelizablelogicalcomputation,
      title={Fast and Parallelizable Logical Computation with Homological Product Codes}, 
      author={Qian Xu and Hengyun Zhou and Guo Zheng and Dolev Bluvstein and J. Pablo Bonilla Ataides and Mikhail D. Lukin and Liang Jiang},
      year={2024},
      eprint={2407.18490},
      archivePrefix={arXiv},
      primaryClass={quant-ph},
      url={https://arxiv.org/abs/2407.18490}, 
}

@misc{zheng2025highratesurgeryconstantoverheadlogical,
      title={High-Rate Surgery: towards constant-overhead logical operations}, 
      author={Guo Zheng and Liang Jiang and Qian Xu},
      year={2025},
      eprint={2510.08523},
      archivePrefix={arXiv},
      primaryClass={quant-ph},
      url={https://arxiv.org/abs/2510.08523}, 
}

@misc{berthusen2025automorphismgadgetshomologicalproduct,
      title={Automorphism gadgets in homological product codes}, 
      author={Noah Berthusen and Michael J. Gullans and Yifan Hong and Maryam Mudassar and Shi Jie Samuel Tan},
      year={2025},
      eprint={2508.04794},
      archivePrefix={arXiv},
      primaryClass={quant-ph},
      url={https://arxiv.org/abs/2508.04794}, 
}

@misc{swaroop2025universaladaptersquantumldpc,
      title={Universal adapters between quantum LDPC codes}, 
      author={Esha Swaroop and Tomas Jochym-O'Connor and Theodore J. Yoder},
      year={2025},
      eprint={2410.03628},
      archivePrefix={arXiv},
      primaryClass={quant-ph},
      url={https://arxiv.org/abs/2410.03628}, 
}

@misc{he2025extractorsqldpcarchitecturesefficient,
      title={Extractors: QLDPC Architectures for Efficient Pauli-Based Computation}, 
      author={Zhiyang He and Alexander Cowtan and Dominic J. Williamson and Theodore J. Yoder},
      year={2025},
      eprint={2503.10390},
      archivePrefix={arXiv},
      primaryClass={quant-ph},
      url={https://arxiv.org/abs/2503.10390}, 
}

@article{yoshida_topological_2015,
  title = {Topological Color Code and Symmetry-Protected Topological Phases},
  author = {Yoshida, Beni},
  year = 2015,
  month = jun,
  journal = {Physical Review B},
  volume = {91},
  number = {24},
  pages = {245131},
  issn = {1098-0121, 1550-235X},
  doi = {10.1103/PhysRevB.91.245131},
  urldate = {2026-02-10},
  copyright = {http://link.aps.org/licenses/aps-default-license},
  langid = {english}
}

@article{kesselring_anyon_2024,
  title = {Anyon {{Condensation}} and the {{Color Code}}},
  author = {Kesselring, Markus S. and Magdalena De La Fuente, Julio C. and Thomsen, Felix and Eisert, Jens and Bartlett, Stephen D. and Brown, Benjamin J.},
  year = 2024,
  month = mar,
  journal = {PRX Quantum},
  volume = {5},
  number = {1},
  pages = {010342},
  issn = {2691-3399},
  doi = {10.1103/PRXQuantum.5.010342},
  urldate = {2024-04-10},
  langid = {english}
}

@article{bridgeman_tensor_2017,
  title = {Tensor Networks with a Twist: {{Anyon-permuting}} Domain Walls and Defects in Projected Entangled Pair States},
  shorttitle = {Tensor Networks with a Twist},
  author = {Bridgeman, Jacob C. and Bartlett, Stephen D. and Doherty, Andrew C.},
  year = 2017,
  month = dec,
  journal = {Physical Review B},
  volume = {96},
  number = {24},
  pages = {245122},
  publisher = {American Physical Society},
  doi = {10.1103/PhysRevB.96.245122},
  urldate = {2026-02-10}
}

@article{beverland_protected_2016,
  title = {Protected Gates for Topological Quantum Field Theories},
  author = {Beverland, Michael E. and Buerschaper, Oliver and Koenig, Robert and Pastawski, Fernando and Preskill, John and Sijher, Sumit},
  year = 2016,
  month = feb,
  journal = {Journal of Mathematical Physics},
  volume = {57},
  number = {2},
  publisher = {AIP Publishing},
  issn = {0022-2488, 1089-7658},
  doi = {10.1063/1.4939783},
  urldate = {2025-07-09},
  langid = {english}
}

@article{kesselring_boundaries_2018,
  title = {The Boundaries and Twist Defects of the Color Code and Their Applications to Topological Quantum Computation},
  author = {Kesselring, Markus S. and Pastawski, Fernando and Eisert, Jens and Brown, Benjamin J.},
  year = 2018,
  month = oct,
  journal = {Quantum},
  volume = {2},
  eprint = {1806.02820},
  primaryclass = {quant-ph},
  pages = {101},
  issn = {2521-327X},
  doi = {10.22331/q-2018-10-19-101},
  urldate = {2025-05-15},
  archiveprefix = {arXiv},
  langid = {english}
}

@article{davydova_quantum_2024,
  title = {Quantum Computation from Dynamic Automorphism Codes},
  author = {Davydova, Margarita and Tantivasadakarn, Nathanan and Balasubramanian, Shankar and Aasen, David},
  year = 2024,
  month = aug,
  journal = {Quantum},
  volume = {8},
  pages = {1448},
  publisher = {Verein zur F\"orderung des Open Access Publizierens in den Quantenwissenschaften},
  doi = {10.22331/q-2024-08-27-1448},
  urldate = {2026-02-10},
  langid = {british}
}

@article{koenig_quantum_2010,
  title = {Quantum Computation with {{Turaev}}--{{Viro}} Codes},
  author = {Koenig, Robert and Kuperberg, Greg and Reichardt, Ben W.},
  year = 2010,
  month = dec,
  journal = {Annals of Physics},
  volume = {325},
  number = {12},
  pages = {2707--2749},
  issn = {0003-4916},
  doi = {10.1016/j.aop.2010.08.001},
  urldate = {2025-05-02}
}

@article{breuckmann_hyperbolic_2017,
  title = {Hyperbolic and {{Semi-Hyperbolic Surface Codes}} for {{Quantum Storage}}},
  author = {Breuckmann, Nikolas P. and Vuillot, Christophe and Campbell, Earl and Krishna, Anirudh and Terhal, Barbara M.},
  year = 2017,
  month = sep,
  journal = {Quantum Science and Technology},
  volume = {2},
  number = {3},
  eprint = {1703.00590},
  primaryclass = {quant-ph},
  pages = {035007},
  issn = {2058-9565},
  doi = {10.1088/2058-9565/aa7d3b},
  urldate = {2025-03-21},
  archiveprefix = {arXiv}
}

@article{lavasani_universal_2019,
  title = {Universal Logical Gates with Constant Overhead: Instantaneous {{Dehn}} Twists for Hyperbolic Quantum Codes},
  shorttitle = {Universal Logical Gates with Constant Overhead},
  author = {Lavasani, Ali and Zhu, Guanyu and Barkeshli, Maissam},
  year = 2019,
  month = aug,
  journal = {Quantum},
  volume = {3},
  eprint = {1901.11029},
  primaryclass = {quant-ph},
  pages = {180},
  issn = {2521-327X},
  doi = {10.22331/q-2019-08-26-180},
  urldate = {2025-08-12},
  archiveprefix = {arXiv},
  langid = {english}
}

@article{zhu_instantaneous_2020,
  title = {Instantaneous Braids and {{Dehn}} Twists in Topologically Ordered States},
  author = {Zhu, Guanyu and Lavasani, Ali and Barkeshli, Maissam},
  year = 2020,
  month = aug,
  journal = {Physical Review B},
  volume = {102},
  number = {7},
  eprint = {1806.06078},
  primaryclass = {cond-mat},
  pages = {075105},
  issn = {2469-9950, 2469-9969},
  doi = {10.1103/PhysRevB.102.075105},
  urldate = {2025-08-12},
  archiveprefix = {arXiv}
}

@misc{first2024good2querylocallytestable,
      title={On Good $2$-Query Locally Testable Codes from Sheaves on High Dimensional Expanders}, 
      author={Uriya A. First and Tali Kaufman},
      year={2024},
      eprint={2208.01778},
      archivePrefix={arXiv},
      primaryClass={math.CO},
      url={https://arxiv.org/abs/2208.01778}, 
}

@misc{panteleev2024maximallyextendablesheafcodes,
      title={Maximally Extendable Sheaf Codes}, 
      author={Pavel Panteleev and Gleb Kalachev},
      year={2024},
      eprint={2403.03651},
      archivePrefix={arXiv},
      primaryClass={cs.IT},
      url={https://arxiv.org/abs/2403.03651}, 
}

@book{hatcher2002algebraic,
  title     = {Algebraic Topology},
  author    = {Hatcher, Allen},
  year      = {2002},
  publisher = {Cambridge University Press},
  url       = {https://pi.math.cornell.edu/~hatcher/AT/AT.pdf}
}

@article{xu2004strong,
  title = {Strong-{{Weak Coupling Self-Duality}} in the {{Two-Dimensional Quantum Phase Transition}} of \$p+ip\$ {{Superconducting Arrays}}},
  author = {Xu, Cenke and Moore, J. E.},
  year = 2004,
  month = jul,
  journal = {Physical Review Letters},
  volume = {93},
  number = {4},
  pages = {047003},
  publisher = {American Physical Society},
  doi = {10.1103/PhysRevLett.93.047003},
  urldate = {2026-02-10}
}

@article{Periwal_2021,
   title={Programmable interactions and emergent geometry in an array of atom clouds},
   volume={600},
   ISSN={1476-4687},
   url={http://dx.doi.org/10.1038/s41586-021-04156-0},
   DOI={10.1038/s41586-021-04156-0},
   number={7890},
   journal={Nature},
   publisher={Springer Science and Business Media LLC},
   author={Periwal, Avikar and Cooper, Eric S. and Kunkel, Philipp and Wienand, Julian F. and Davis, Emily J. and Schleier-Smith, Monika},
   year={2021},
   month=dec, pages={630–635} }

@article{sala_2025_decoherence,
  title = {Decoherence and Wave-Function Deformation of ${D}_{4}$ Non-Abelian Topological Order},
  author = {Sala, Pablo and Alicea, Jason and Verresen, Ruben},
  journal = {Phys. Rev. X},
  volume = {15},
  issue = {3},
  pages = {031002},
  numpages = {43},
  year = {2025},
  month = {Jul},
  publisher = {American Physical Society},
  doi = {10.1103/5ywn-6d3q},
  url = {https://link.aps.org/doi/10.1103/5ywn-6d3q}
}

@article{stability_2025_sala,
  title = {Stability and Loop Models from Decohering Non-Abelian Topological Order},
  author = {Sala, Pablo and Verresen, Ruben},
  journal = {Phys. Rev. Lett.},
  volume = {134},
  issue = {25},
  pages = {250403},
  numpages = {10},
  year = {2025},
  month = {Jun},
  publisher = {American Physical Society},
  doi = {10.1103/fy9r-hpcw},
  url = {https://link.aps.org/doi/10.1103/fy9r-hpcw}
}

@Article{Freedman2003,
author={Freedman, Michael H.},
title={A Magnetic Model with a Possible Chern-Simons Phase},
journal={Communications in Mathematical Physics},
year={2003},
month={Mar},
day={01},
volume={234},
number={1},
pages={129-183},
issn={1432-0916},
doi={10.1007/s00220-002-0785-1},
url={https://doi.org/10.1007/s00220-002-0785-1}
}

@article{Freedman04,
title = {A class of P,T-invariant topological phases of interacting electrons},
journal = {Annals of Physics},
volume = {310},
number = {2},
pages = {428-492},
year = {2004},
issn = {0003-4916},
doi = {https://doi.org/10.1016/j.aop.2004.01.006},
url = {https://www.sciencedirect.com/science/article/pii/S0003491604000260},
author = {Michael Freedman and Chetan Nayak and Kirill Shtengel and Kevin Walker and Zhenghan Wang},
}

@article{Freedman05a,
  title = {Line of Critical Points in $2+1$ Dimensions: Quantum Critical Loop Gases and Non-Abelian Gauge Theory},
  author = {Freedman, Michael and Nayak, Chetan and Shtengel, Kirill},
  journal = {Phys. Rev. Lett.},
  volume = {94},
  issue = {14},
  pages = {147205},
  numpages = {4},
  year = {2005},
  month = {Apr},
  publisher = {American Physical Society},
  doi = {10.1103/PhysRevLett.94.147205},
  url = {https://link.aps.org/doi/10.1103/PhysRevLett.94.147205}
}

@article{Freedman05b,
  title = {Extended Hubbard Model with Ring Exchange: A Route to a Non-Abelian Topological Phase},
  author = {Freedman, Michael and Nayak, Chetan and Shtengel, Kirill},
  journal = {Phys. Rev. Lett.},
  volume = {94},
  issue = {6},
  pages = {066401},
  numpages = {4},
  year = {2005},
  month = {Feb},
  publisher = {American Physical Society},
  doi = {10.1103/PhysRevLett.94.066401},
  url = {https://link.aps.org/doi/10.1103/PhysRevLett.94.066401}
}

@article{castelnovo2008,
  title = {Quantum topological phase transition at the microscopic level},
  author = {Castelnovo, Claudio and Chamon, Claudio},
  journal = {Phys. Rev. B},
  volume = {77},
  issue = {5},
  pages = {054433},
  numpages = {14},
  year = {2008},
  month = {Feb},
  publisher = {American Physical Society},
  doi = {10.1103/PhysRevB.77.054433},
  url = {https://link.aps.org/doi/10.1103/PhysRevB.77.054433}
}

@article{Fendley_2008,
   title={Topological order from quantum loops and nets},
   volume={323},
   ISSN={0003-4916},
   url={http://dx.doi.org/10.1016/j.aop.2008.04.011},
   DOI={10.1016/j.aop.2008.04.011},
   number={12},
   journal={Annals of Physics},
   publisher={Elsevier BV},
   author={Fendley, Paul},
   year={2008},
   month=dec, pages={3113–3136} }

@article{Fidkowski2009,
  title    = "From String Nets to Nonabelions",
  author   = "Fidkowski, Lukasz and Freedman, Michael and Nayak, Chetan and
              Walker, Kevin and Wang, Zhenghan",
  journal  = "Communications in Mathematical Physics",
  volume   =  287,
  number   =  3,
  pages    = "805--827",
  month    =  may,
  year     =  2009
}

@article{Fendley13,
  title = {Fibonacci topological order from quantum nets},
  author = {Fendley, Paul and Isakov, Sergei V. and Troyer, Matthias},
  journal = {Phys. Rev. Lett.},
  volume = {110},
  issue = {26},
  pages = {260408},
  numpages = {5},
  year = {2013},
  month = {Jun},
  publisher = {American Physical Society},
  doi = {10.1103/PhysRevLett.110.260408},
  url = {https://link.aps.org/doi/10.1103/PhysRevLett.110.260408}
}

@article{haegeman_2015_gauging,
  title = {Gauging Quantum States: From Global to Local Symmetries in Many-Body Systems},
  author = {Haegeman, Jutho and Van Acoleyen, Karel and Schuch, Norbert and Cirac, J. Ignacio and Verstraete, Frank},
  journal = {Phys. Rev. X},
  volume = {5},
  issue = {1},
  pages = {011024},
  numpages = {10},
  year = {2015},
  month = {Feb},
  publisher = {American Physical Society},
  doi = {10.1103/PhysRevX.5.011024},
  url = {https://link.aps.org/doi/10.1103/PhysRevX.5.011024}
}

@article{zhu2019gapless,
  title = {Gapless Coulomb State Emerging from a Self-Dual Topological Tensor-Network State},
  author = {Zhu, Guo-Yi and Zhang, Guang-Ming},
  journal = {Phys. Rev. Lett.},
  volume = {122},
  issue = {17},
  pages = {176401},
  numpages = {6},
  year = {2019},
  month = {Apr},
  publisher = {American Physical Society},
  doi = {10.1103/PhysRevLett.122.176401},
  url = {https://link.aps.org/doi/10.1103/PhysRevLett.122.176401}
}

@article{schotte2019tensornetwork,
  title = {Tensor-network approach to phase transitions in string-net models},
  author = {Schotte, Alexis and Carrasco, Jose and Vanhecke, Bram and Vanderstraeten, Laurens and Haegeman, Jutho and Verstraete, Frank and Vidal, Julien},
  journal = {Phys. Rev. B},
  volume = {100},
  issue = {24},
  pages = {245125},
  numpages = {8},
  year = {2019},
  month = {Dec},
  publisher = {American Physical Society},
  doi = {10.1103/PhysRevB.100.245125},
  url = {https://link.aps.org/doi/10.1103/PhysRevB.100.245125}
}

@article{xu2020tensor,
  title = {Tensor Network Approach to Phase Transitions of a Non-Abelian Topological Phase},
  author = {Xu, Wen-Tao and Zhang, Qi and Zhang, Guang-Ming},
  journal = {Phys. Rev. Lett.},
  volume = {124},
  issue = {13},
  pages = {130603},
  numpages = {6},
  year = {2020},
  month = {Apr},
  publisher = {American Physical Society},
  doi = {10.1103/PhysRevLett.124.130603},
  url = {https://link.aps.org/doi/10.1103/PhysRevLett.124.130603}
}

@article{xu2021characterization,
  title = {Characterization of topological phase transitions from a non-Abelian topological state and its Galois conjugate through condensation and confinement order parameters},
  author = {Xu, Wen-Tao and Schuch, Norbert},
  journal = {Phys. Rev. B},
  volume = {104},
  issue = {15},
  pages = {155119},
  numpages = {16},
  year = {2021},
  month = {Oct},
  publisher = {American Physical Society},
  doi = {10.1103/PhysRevB.104.155119},
  url = {https://link.aps.org/doi/10.1103/PhysRevB.104.155119}
}

@misc{sahay2025enforcedgaplessnessstatesexponentially,
      title={Enforced Gaplessness from States with Exponentially Decaying Correlations}, 
      author={Rahul Sahay and Curt von Keyserlingk and Ruben Verresen and Carolyn Zhang},
      year={2025},
      eprint={2503.01977},
      archivePrefix={arXiv},
      primaryClass={cond-mat.str-el},
      url={https://arxiv.org/abs/2503.01977}, 
}

@misc{hsin2025automorphismgaugetheorieshigher,
      title={Automorphism in Gauge Theories: Higher Symmetries and Transversal Non-Clifford Logical Gates}, 
      author={Po-Shen Hsin and Ryohei Kobayashi},
      year={2025},
      eprint={2511.15783},
      archivePrefix={arXiv},
      primaryClass={cond-mat.str-el},
      url={https://arxiv.org/abs/2511.15783}, 
}

\onecolumngrid
\begin{appendix}

\section{Derivation for Gauging using Cup Product to $D_4$ qLDPC code}\label{D4-cup-derivation}

Here, we provide a detailed derivation for the projectors in the non-Abelian $D_4$ qLDPC code (see Sec.~\ref{subsubsec:d4-homological}) upon gauging transversal $\mathsf{CZ}$ gate using the cup product structure of the input qLDPC code.

We require that the input qLDPC code has a well-defined cup product, which implies that these codes admit the following transversal gate: 
\begin{equation}
        U_{\mathsf{CZ}}[\mathcal{M}] = \prod_{e,e'} \exp\left( i \pi\,  \bB{n^B_{e}} \oC{n^C_{e'}} \int_{\mathcal{M}} \mathbf{e} \cupp \mathbf{e}'\right)
\end{equation}
where $\mathcal{M}$ is an arbitrary element of the top homology group $H_2$. 
These transversal gates are not necessarily symmetries of the input qLDPC code.
However, they are emergent symmetries of the ground state subspace---and are sometimes called topological symmetries. These transversal gates are the $\mathbb{Z}_2$ symmetries we will gauge to produce a non-Abelian code.
To do so, we use the same chain complex as our ancilla chain complex, such that the stabilizers are: 
\begin{equation}
      \aP{X_{\mathbf{v}}^a} \AR{X_{\exd \mathbf{v}}^A} = \AR{Z_{e}^A} \aP{Z_{\partial e}^a} = 1
\end{equation}
Having chosen an ancilla complex, we consider the transversal $\mathsf{CZ}$ gate between the \BandC complexes as an emergent symmetry in the code space, and ``symmetry enrich'' it with the unitary operator: 
\begin{equation}
    \Omega_{\aP{a}\bB{B}\oC{C}} =  \prod_{w} \prod_{e,e'} \exp\left(i \pi\,  \aP{n^{a}_w} \bB{n^B_{e}} \oC{n^{C}_{e'}}  \int_{\mathcal{M}} \mathbf{w} \cupp (\mathbf{e} \cupp \mathbf{e}')\right)
\end{equation}
This unitary has the crucial property that it ties the 0-cocycle symmetry $\aP{X_{\bm{\mu}}^a}$ for $\bm{\mu}\in \text{ker}(d^0)$ of the cluster state to the transversal $\mathsf{CZ}$ gate of the two toric codes on the \BandC complex.
Specifically:
\begin{align}
     \Omega_{\aP{a}\bB{B}\oC{C}} \aP{X^a_{\bm{\mu}}} \Omega_{\aP{a}\bB{B}\oC{C}} &= \aP{X^a_{ \bm{\mu}}}\prod_{e,e'}\exp\left(i \pi \bB{n_{e}^B} \oC{n_{e'}^C} \int_{\mathcal{M}} \bm{\mu} \cupp (\mathbf{e} \cupp \mathbf{e}') \right)\\
     &= \aP{X_{\bm{\mu}}^a}  \prod_{e,e'}(-1)^{\left( \bB{n_{e}^B} \oC{n_{e'}^C} \int_{\mathcal{M} \capp \bm{\mu}}  \mathbf{e} \cupp \mathbf{e}' \right)} = \aP{X_{\bm{\mu}}^a} U_{\mathsf{CZ}}[\mathcal{M}\capp \bm{\mu}]
\end{align}
Under this unitary, we can track how the stabilizers transform.
A representative set of these transformations is given by:
\begin{align}
    \aP{X^a_{\mathbf{v}}} \AR{X^A_{\exd \mathbf{v}}} &\to \aP{X^a_{\mathbf{v}}} \AR{X^A_{\exd \mathbf{v}}}\prod_{e,e'}(-1)^{\bB{n_{e}^B} \oC{n_{e'}^C} \int_{\mathcal{M} \capp \mathbf{v}} \mathbf{e} \cupp \mathbf{e}'}\\
    \bB{X_{\exd \mathbf{v}}^B}  &\to \bB{X_{\exd \mathbf{v}}^{B}}\prod_{w,e'} (-1)^{\aP{n_w^a} \oC{n_{e'}^C} \int_{\mathcal{M}} \mathbf{w} \cupp (\exd \mathbf{v} \cupp \mathbf{e}')}\\
    \oC{X_{\exd \mathbf{v}}^C}  &\to \oC{X_{\exd \mathbf{v}}^{C}} \prod_{w,e}(-1)^{\aP{n_w^a} \bB{n_{e}^{B}} \int_{\mathcal{M}} \mathbf{w} \cupp (\mathbf{e} \cupp \exd \mathbf{v})}
\end{align}

\subsection{Gauging the transversal $\mathsf{CZ}$ gate}

Prior to gauging the transversal $\mathsf{CZ}$ gate, we simplify the second and third term.
Starting with the second, we use the Leibniz rule to rewrite:
\begin{equation}
    \bB{X_{\exd \mathbf{v}}^{B}} \prod_{w,e'} (-1)^{\aP{n_w^a} \oC{n_{e'}^C} \int_{\mathcal{M}} \mathbf{w} \cupp (\exd \mathbf{v} \cupp \mathbf{e'})} = \bB{X_{\exd \mathbf{v}}^{B}} \prod_{w,e'} (-1)^{\aP{n_w^a} \oC{n_{e'}^C} \int_{\mathcal{M}} \exd\mathbf{w} \cupp ( \mathbf{v} \cupp \mathbf{e'})}\prod_{w,e'} (-1)^{\aP{n_w^a} \oC{n_{e'}^C} \int_{\mathcal{M}} \mathbf{w} \cupp ( \mathbf{v} \cupp \exd\mathbf{e'})}
\end{equation}
where we have used that the total derivative $\exd(\mathbf{w} \cupp ( \mathbf{v} \cupp \mathbf{e'}))$ vanishes in the integral over the fundamental class $\mathcal{M}$. We can then use Stokes theorem to rewrite the above as:
\begin{equation}
   \bB{X_{\exd \mathbf{v}}^{B}} \prod_{e'}(\aP{Z_{\partial((\mathbf{v}\cupp\mathbf{e'})\frown\mathcal{M})}^a})^{ \oC{n_{e'}^C} }\prod_{w,e'}  (-1)^{\aP{n_w^a} \oC{n_{e'}^C} \int_{\mathcal{M}} \mathbf{w} \cupp ( \mathbf{v} \cupp \exd\mathbf{e'})}
\end{equation}
Note that we have not assumed any associativity property of the cup product. We note now that the $\textcolor{goodpurple}{a}$ sublattice term in the first product can be expressed in terms of the $\textcolor{red}{A}$ sublattice variables per the cluster stabilizers. We then have:
\begin{equation}
   \bB{X_{\exd \mathbf{v}}^{B}}\prod_{e'} (\AR{Z_{(\mathbf{v}\cupp\mathbf{e'})\frown\mathcal{M}}^A})^{ \oC{n_{e'}^C} }\prod_{w,e'}  (-1)^{\aP{n_w^a} \oC{n_{e'}^C} \int_{\mathcal{M}} \mathbf{w} \cupp ( \mathbf{v} \cupp \exd\mathbf{e'})}=\bB{X_{\exd \mathbf{v}}^{B}} \prod_{e,e'} (-1)^{ \AR{n_{e}^A}\oC{n_{e'}^C} \int_{\mathcal{M}}\mathbf{e}\cupp (\mathbf{v}\cupp\mathbf{e'})}\prod_{w,e'}(-1)^{\aP{n_w^a} \oC{n_{e'}^C} \int_{\mathcal{M}} \mathbf{w} \cupp ( \mathbf{v} \cupp \exd\mathbf{e'})}
\end{equation}

Finally, we consider just the second term for some arbitrary vertex $\mathbf{w}$:
\begin{equation}
    \prod_{e'} (-1)^{\aP{n_w^a}\oC{n_{e'}^C} \int_{\mathcal{M}} \mathbf{w} \cupp ( \mathbf{v} \cupp \exd\mathbf{e'})} = \prod_{e'} \left(\oC{Z^C_{e'}}\right)^{\aP{n_w^a} \int_{\mathcal{M}} \mathbf{w} \cupp ( \mathbf{v} \cupp \exd\mathbf{e'})}
\end{equation}
The above term commutes with the stabilizer $\textcolor{orange(ryb)}{X_{\exd\mathbf{v'}}}$ for any $\oC{\mathbf{v'}}$, using the fact that $\exd^2=0$. Hence, it is a $Z$ stabilizer on the original $\textcolor{orange(ryb)}{C}$ code which is equal to $+1$ acting on the ground state. Therefore, we observe that the term:
\begin{equation}
   \prod_{w,e'} (-1)^{\aP{n_w^a} \oC{n_{e'}^C} \int_{\mathcal{M}} \mathbf{w} \cupp ( \mathbf{v} \cupp \exd\mathbf{e'})}
\end{equation}
can safely be set equal to $+1$, leaving us with the remaining simplified expression for the gauged stabilizer:
\begin{equation}
    \textcolor{dodgerblue}{\mathcal{A}^{B}}=\bB{X_{\exd \mathbf{v}}^{B}} \prod_{e,e'} (-1)^{ \AR{n_{e}^A}\oC{n_{e'}^C} \int_{\mathcal{M}}\mathbf{e}\cupp (\mathbf{v}\cupp\mathbf{e'})}
\end{equation}
Similar algebra applies to obtain $\textcolor{orange(ryb)}{\mathcal{A}^{C}}$. After projecting into $\aP{X_{\mathbf{v}}^a} = +1$, and using the identity $(-1)^{n_1n_2}=\mathsf{CZ}_{1,2}$ to simplify expressions, the final vertex stabilizers of the $D_4$ model are:
%
\begin{align}
&\AR{\mathcal{A}^{A}}=\AR{X_{\exd\mathbf{v}}^A}\prod_{e,e'} 
\left(\mathsf{CZ}_{\bB{e},\oC{e'}}\right)^{\int_{\mathcal{M}}\mathbf{v}\cupp (\mathbf{e}\cupp\mathbf{e'})}\\
&\bB{\mathcal{A}^{B}}=\bB{X_{\exd\mathbf{v}}^B}\prod_{e,e'} 
\left(\mathsf{CZ}_{\AR{e},\oC{e'}}\right)^{\int_{\mathcal{M}}\mathbf{e}\cupp (\mathbf{v}\cupp\mathbf{e'})} \\
&\oC{\mathcal{A}^{C}}=\oC{X_{\exd\mathbf{v}}^C}\prod_{e,e'} 
\left(\mathsf{CZ}_{\bB{e},\AR{e'}}\right)^{\int_{\mathcal{M}}\mathbf{e'}\cupp (\mathbf{e}\cupp\mathbf{v})}
\end{align}

confirming the expressions given in Eq.~\eqref{eq-homologicalA}, ~\eqref{eq-homologicalB}, ~\eqref{eq-homologicalC}.


\section{Derivation of Non-Abelian worked examples}

\subsection{Derivations for Non-Abelian $D_4$ Fractons by Gauging transversal $\mathsf{CZ}$ Gates}\label{append-sub:fracton-from-graph-gauging}

Here, we provide a detailed derivation for the stabilizers (see Eq.~\eqref{eq:fracton-stabilizers-graph-gauging}) of the non-Abelian fracton model. 

In the main text, we provide the initial stabilizers (Eq.~\eqref{eq:qubit-ALP-stabilizers} and \eqref{eq:cubic-graph-cluster-state}) and the symmetry enrichment (Eq.~\eqref{eq:cubic-symm-enrich}). We reproduce the symmetry enrichment (controlled on the qubits in the $\aP{a}$ sublattice, and target on the qubits on the \BandC sublattices) below for ease of the reader.
\begin{equation}\label{eq:cubic-symm-enrich}

\end{equation}
where the green thickened lines denote $\mathsf{CZ}$ gates (in different shades of green to aid later visualization). 

After the symmetry enrichment step, the $Z$ stabilizers of the cluster state remain the same, whereas the $X$ stabilizer of the cluster state is modified to
\begin{equation}
    %
\end{equation}

The $X_{\exd \mathbf{v}}$ stabilizer of the \BandC sublattices becomes:
\begin{equation}
    %
,
\end{equation}

Let's rewrite the $X_{\exd \mathbf{v}}$ stabilizer in a form that makes the later projection step easier. Take the first stabilizer as the example. 
\begin{equation}
\begin{aligned}
    %
\end{aligned}
\end{equation}

Then, we recall that we can use the $\oC{Z}_{\partial p}=+1$ to trivialize the term in the parentheses. So finally, the stabilizer is 
\begin{equation}
    %
\end{equation}

We can go through a similar derivation for the $X$ stabilizer in the $\oC{C}$ sublattice to obtain
\begin{equation}
    %
.
\end{equation}

After projecting the $\aP{a}$ sublattice to $X_{\aP{a}}=+1$, only the product of $Z$ stabilizers around plaquettes commute with this measurement, so there are three plaquette operators:
\begin{equation}
    %
.
\end{equation}

After projecting the $\aP{a}$ sublattice to $X_{\aP{a}}=+1$, the cluster state $X$ stabilizer simply reduces to
\begin{equation}
    %
\end{equation}

To sum up, the resulting stabilizer of the non-Abelian LDPC code are:
\begin{equation}
    %
,
\end{equation}

where the green lines (shaded differently for easier visualization) all represent $\mathsf{CZ}$ gates.


\section{Non-Abelian $S_3$ Fractons by Gauging Charge Conjugation}\label{app:s3-ALP}

Here, we provide a derivation of obtaining an $S_3$ fracton model by gauging the transversal charge conjugate gate (i.e. a $\mathbb{Z}_2$ charge conjugation symmetry) present in the $\mathbb{Z}_3$ generalization of the anisotropic lineon planeon (ALP) model.

Before we proceed, we comment on relevant literature. There are related examples of gauging the charge conjugation symmetry of fracton models in the literature. Ref.~\cite{Bulmash_2019} noted that their construction is applicable to gauging charge conjugation symmetry of $\mathbb{Z}_3$ X-cube model~\cite{Vijay_2016}, and ref.~\cite{Tu_2021} provides an explicit example of a model equivalent to gauging global $\mathbb{Z}_2$ charge conjugation symmetry of the $\mathbb{Z}_3$ X-cube model. We do not know of an explicit lattice construction of gauging global charge conjugation symmetry of $\mathbb{Z}_3$ ALP model to our knowledge.

We take the qutrit generalization of ALP model to be on the \bB{B} sublattices, where the stabilizers are given by
\begin{equation}\label{eq:qutrit-ALP-stabilizers}
    H = - \sum_{p \in \bB{\mathsf{B}_2}} \left(  
    \underbrace{
}_{X_{\exd \mathbf{v}}} + \text{h.c.} \right)
\end{equation}
This model enjoys a global charge conjugation symmetry $\mathcal{C}$ that maps qutrit Pauli operators to their Hermitian conjugate: $\mathcal{Z} \leftrightarrow \mathcal{Z}^\dagger$ and $\mathcal{X} \leftrightarrow \mathcal{X}^\dagger$. $\mathcal{X}$ and $\mathcal{Z}$ denote the qutrit clock matrices, and $\mathcal{C}$ is the qutrit charge-conjugation operator,
\begin{equation}\label{eq:x_and_z_and_C_def}
    \mathcal{Z} = %
\;.
\end{equation}
We use the unscripted $Z$ and $X$ to denote the Pauli matrices acting on the \textit{qubit} degrees of freedom.

Now we use a graph cluster state (see Sec.~\ref{sec:gauging-graphs}) to gauge this transversal $\mathcal{C}$ gate. Similar to the case of gauging transversal $\mathsf{CZ}$ gate (see Sec.~\ref{subsec:graph-gauging-fracton}), we use a cluster state on the cubic lattice to gauge this transversal gate (note that the ancilla degrees of freedom remains to be qubits), where the stabilizers are:
\begin{equation}
    %
.
\end{equation}

The symmetry enrichment is given by the controlled charge conjugation ($C\mathcal{C}$) gate

\begin{equation}\label{eq:cubic-symm-enrich}
    %
,
\end{equation}

where the above pattern is repeated for each vertex $\aP{\mathbf{v}}$ of the cubic lattice.

This symmetry enrichment pattern modifies the stabilizers to be

\begin{equation}
    %
\end{equation}

based on the nontrivial commutation between $\mathcal{X},X$ and $C\mathcal{C}$.

Now, let's express the stabilizers associated with the ALP model to prepare for the later projection step. For the $\bB{\mathcal{X}}$ stabilizer,

\begin{equation}
    %
.
\end{equation}

We first raise the stabilizer to the power $\aP{Z_1}$ and insert $\aP{{Z_2}^2}=+1$, then we express the vertex degrees of freedom in the $\aP{a}$ sublattice in terms of edge degrees of freedom in the $\AR{A}$ sublattice using the relation $\aP{Z}\AR{Z}\aP{Z}=+1$ along edges from the cluster state stabilizer.

Similarly, we can massage the $\bB{\mathcal{Z}}$ stabilizer to
\begin{equation}
    %
\end{equation}

Finally, we project the $\aP{a}$ sublattice to $\aP{X}=+1$. Only the product of $Z$ stabilizers around plaquettes commute with this measurement, so there are three plaquette operators:
\begin{equation}
    %
.
\end{equation}    

to obtain the precursor stabilizer
\begin{equation}
    \AR{\tilde{\mathcal{A}}^A} =
    %
\end{equation}

Finally, we symmetrize the precursor stabilizers to obtain commuting projectors for the non-Abelian $S_3$ fracton code. The resulting projectors are

\begin{equation}
     \AR{\mathcal{A}^A}= \frac{1}{3}\left(1+ \AR{\tilde{\mathcal{A}}^A} +\text{h.c.}\right), \quad 
     \bB{\mathcal{A}^B}= \frac{1}{3}\left(1+ \bB{\tilde{\mathcal{A}}^B} +\text{h.c.}\right), 
     \quad 
     \bB{\tilde{\mathcal{B}}^B} = \frac{1}{3}\left(1+ \bB{\tilde{\mathcal{B}}^B}+\text{h.c.}\right), 
\end{equation}

completing the derivation for the commuting projectors of the non-Abelian $S_3$ ALP model.

\section{Ground States of Non-Abelian qLDPC codes}\label{App:log}

\subsubsection{Non-Abelian Logical Operators from Homological Gauging}

To derive the logical operators obtained from homologically gauging, let us consider, for sake of discussion, a logical $X$ operator of the $\bB{B}$ code $\bB{X_{\boldsymbol{\gamma}}}$, where $\boldsymbol{\gamma} \in H^1$ is some (non-trivial) $1$-cocycle.
Let us note that after symmetry enrichment (Eq.~\eqref{eq-CCZSEop}), $\bB{X_{\bgamma}}$ is transformed to: 
\begin{equation} \label{eq-Xgammaenrich}
    \Omega_{\aP{a}, \bB{B}\oC{C}} \bB{X_{\bgamma}}\Omega_{\aP{a}, \bB{B}\oC{C}}^{\dagger} = \bB{X_{\bgamma}} \prod_{\aP{v}, \oC{e}}(-1)^{\aP{n^a_{v}}\textcolor{orange(ryb)}{n^C_{e}}\int_{\mathcal{M}}\textbf{v}\cupp (\bgamma\cupp \mathbf{e})}
\end{equation}
Now, to gauge the system, we must project the $\aP{a}$ sublattice into the $\ket{+}$ state.
Consequently, to update this operator after gauging, we seek to re-write it in a form that manifestly commutes with the projection.

To do so, we follow a procedure similar to how we updated the $X$-stabilizers post-projection.
In particular, note that because we initialize our $\bB{B}$ and $\oC{C}$ codes in the $\ket{0}$ state, we can safely set their $Z$ logical operator to $1$.
Our aim will be to ``pull out'' a $Z$-logical from Eq.~\eqref{eq-Xgammaenrich}.
We do so by re-writing:
\begin{equation}
    \Omega_{\aP{a}, \bB{B}\oC{C}} \bB{X_{\bgamma}}\Omega_{\aP{a}, \bB{B}\oC{C}}^{\dagger} =  \bB{X_{\bgamma}} \prod_{\aP{v}, \oC{e}}(\aP{Z_v})^{\textcolor{orange(ryb)}{n^C_{e}}\int_{\mathcal{M}}\textbf{v}\cupp (\bgamma\cupp \mathbf{e})}
\end{equation}
Subsequently, we note that any chain $v$ can be trivially re-written as a sum of elements in $H_0 \simeq D_0/\mathsf{im}(\partial_1)$ plus a term in $\mathsf{im}(\partial_1)$.
In particular, letting $\boldsymbol{\mu}$ denote a basis of elements in $H^0$ and $v_{\boldsymbol{\mu}}$ be a basis of $H_0$ such that $\int_{v_{\boldsymbol{\mu}}} \boldsymbol{\mu}' = \delta_{\boldsymbol{\mu}, \boldsymbol{\mu}'}$, we have that: 
\begin{equation}
    v = \sum_{\boldsymbol{\mu}} \left( \int_{v} \boldsymbol{\mu} \right) v_{\boldsymbol{\mu}} + \partial_1 b(v)
\end{equation}
where $b(v) \in D_1$ is some co-boundary that depends on $v$.
With this decomposition, we may re-write: 
\begin{equation}
    \aP{Z_v} =  \aP{Z_{\partial b(v)}}\prod_{\boldsymbol{\mu}}(\aP{Z_{v_{\mu}}})^{\int_{v} \boldsymbol{\mu}} = \AR{Z_{b(v)}}\prod_{\boldsymbol{\mu}}(\aP{Z_{v_{\mu}}})^{\int_{v} \boldsymbol{\mu}}
\end{equation}
Consequently, we have that:
\begin{align}
    \Omega_{\aP{a}, \bB{B}\oC{C}} \bB{X_{\bgamma}}\Omega_{\aP{a}, \bB{B}\oC{C}}^{\dagger} = &\bB{X_{\bgamma}} \prod_{v, \oC{e}} \left(\AR{Z_{b(v)}} \right)^{\oC{n_{e}^C} \int_{\mathcal{M}}\textbf{v}\cupp (\bgamma\cupp \mathbf{e})} \\
    & \times  \prod_{\oC{e}} \left((-1)^{\oC{n_{e}^C} \int_{\mathcal{M}} \boldsymbol{\mu} \cupp (\boldsymbol{\gamma} \cupp \mathbf{e})} \right)^{\aP{n_{v_{\boldsymbol{\mu}}}}} 
\end{align}
However, let us note that the term in parentheses in the second line is simply $(\oC{Z_{(\mathcal{M} \capp \boldsymbol{\mu}) \capp \boldsymbol{\gamma}}})^{\aP{n_{v_{\boldsymbol{\mu}}}}}$, which is a logical $Z$ operator of the $\oC{C}$ code raised to some power [c.f. discussion below Eq.~\eqref{eq-capproduct} for properties of the cap product]. 
Hence, after gauging our logical operators can be re-written as:
\begin{equation}
    \bB{\widetilde{X}_{\bgamma}} = \bB{X_{\bgamma}} \prod_{v, \oC{e}} \left(\AR{Z_{b(v)}} \right)^{\oC{n_{e}^C} \int_{\mathcal{M}}\textbf{v}\cupp (\bgamma\cupp \mathbf{e})}
\end{equation}
One can similarly derive an expression for the logical operators associated with the $\oC{C}$ code and $\AR{A}$ code: 
\begin{align}
    \oC{\widetilde{X}_{\bgamma}} &= \oC{X_{\bgamma}} \prod_{v, \oC{e}} \left(\AR{Z_{b(v)}} \right)^{\bB{n_{e}^B} \int_{\mathcal{M}}\textbf{v}\cupp (\mathbf{e}\cupp \boldsymbol{\gamma})} 
\end{align}

\begin{equation}
    \AR{\widetilde{X}_{\bgamma}} = \AR{X_{\bgamma}} \prod_{v, \bB{e}} \left(\bB{Z_{b(v)}} \right)^{\oC{n_{e}^C} \int_{\mathcal{M}}\bgamma\cupp (\textbf{v}\cupp \mathbf{e})}
\end{equation}

\subsubsection{Ground States for Graph State Construction}
We now turn to the ground states of the graph construction introduced in Sec.~\ref{sec:graphygraph} and construct the ground states following a similar procedure as the previous section, namely taking a logical $\overline{X}$ operator $\textcolor{dodgerblue}{\overline{X}^B}=\textcolor{dodgerblue}{X^B_{\gamma}}$ and following it through the graph gauging procedure. In keeping with Sec.~\ref{sec:graphygraph}, we will focus on the specific case of gauging a transversal $\mathsf{SWAP}$ gate. We remark that unlike the previous example where the ancilla complex was a copy of the original codes' chain complex, here the ground states on the $\textcolor{red}{A}$ sublattice will have a form distinct from those on \BandC. After symmetry enrichment, the ground state $\textcolor{dodgerblue}{\overline{X}^B}$ on the $\textcolor{dodgerblue}{B}$ code is transformed to:

\begin{equation}
   \textcolor{dodgerblue}{\overline{X}^B}\rightarrow \textcolor{dodgerblue}{X^B_{\gamma}}\prod_{\mathbf{e}\in \gamma}\left(\textcolor{dodgerblue}{X^B_{\mathbf{e}}} \textcolor{orange(ryb)}{X^C_{\mathbf{e}}}\right)^{\aP{n^{a}_{\phii(e)}}} 
\end{equation}
As in the addressable case, we now must write the above in a form that is written in terms of only the $\textcolor{red}{A}$ sublattice variables. To do so, we assume our original qLDPC codes were initialized in a $\mathsf{SWAP}$ symmetric case such that $\textcolor{dodgerblue}{X^B_{\gamma}} \textcolor{orange(ryb)}{X^C_{\gamma}}=+1$ on the ground state subspace. With this assumption, we are free to choose any vertex $\mathbf{v}_0\in\textcolor{goodpurple}{a}$ and multiply the above by $\left(\textcolor{dodgerblue}{X^B_{\gamma}} \textcolor{orange(ryb)}{X^C_{\gamma}}\right)^{\textcolor{goodpurple}{n^a_{\mathbf{v}_0}}}$. The transformed logical operator after multiplication is:

\begin{equation}
   \textcolor{dodgerblue}{\overline{X}^B}\rightarrow \textcolor{dodgerblue}{X^B_{\gamma}}\prod_{\mathbf{e}\in \gamma}\left(\textcolor{dodgerblue}{X^B_{\mathbf{e}}} \textcolor{orange(ryb)}{X^C_{\mathbf{e}}}\right)^{\AR{n^{A}_{\phii(e),\mathbf{v}_0}}} 
\end{equation}
\end{appendix}

\end{document}